\documentclass[aps,prx,10pt, twocolumn, superscriptaddress]{revtex4-2}

\usepackage{standalone, gensymb, tensor, amsmath,amsfonts,amssymb,amsthm,bbm,bm, braket}
\usepackage[T1]{fontenc}
\usepackage{scalerel}
\usepackage{wasysym}
\usepackage{comment}
\usepackage{enumerate}
\usepackage{graphicx}
\usepackage{mathtools}
\usepackage{xparse}
\usepackage{ifthen}
\usepackage{tensor}
\usepackage{tikz}
\usepackage{tikz-network}
\usetikzlibrary{patterns,decorations.pathreplacing}
\theoremstyle{break}        

\theoremstyle{break}        

\usepackage{color}
\definecolor{myred}{RGB}{232,102,102}
\definecolor{myblue}{RGB}{166,218,149}
\definecolor{mygreyblue}{RGB}{166,218,149}
\definecolor{mygreyred}{RGB}{232,102,102}
\definecolor{myvioletc}{RGB}{45,130,60}
\definecolor{mygreen}{RGB}{187,187,255}
\definecolor{myorange}{RGB}{255,145,33}
\definecolor{OliveGreen}{RGB}{85,107,47}
\definecolor{NavyBlue}{RGB}{0,0,128}
\definecolor{myviolet}{RGB}{210,145,178}
\definecolor{mygrey1}{RGB}{220,220,220}
\definecolor{mygrey2}{RGB}{211,211,211}
\definecolor{mygrey3}{RGB}{192,192,192}
\definecolor{mygrey4}{RGB}{169,169,169}
\usepackage{tensor}
\newcommand{\be}{\begin{equation}}
\newcommand{\ee}{\end{equation}}
\newcommand{\ba}{\begin{aligned}}
\newcommand{\ea}{\end{aligned}}
\newcommand{\bw}{\begin{widetext}}
\newcommand{\ew}{\end{widetext}}
\newcommand{\1}{\mathbbm{1}}
\newcommand{\limsc}{{\textstyle \lim_{k}}}

\theoremstyle{plain}

\theoremstyle{plain}

\newtheorem{conjecture}{Conjecture}
\usepackage[colorlinks,bookmarks=false,citecolor=NavyBlue,linkcolor=OliveGreen,urlcolor=blue]{hyperref}

\newcommand{\mcirc}{\mathbin{\scalerel*{\fullmoon}{G}}}

\renewcommand{\boxed}[1]{%
  \framebox{\raisebox{0pt}[0.4\baselineskip][0.025\baselineskip]{\hbox to 0.25cm{\hss#1\hss}}}}

\tikzset{
    cross/.pic = {
    \draw[thick, fill=white] (0,0) circle (0.075);

    }
}

\tikzset{
    cross_chi/.pic = {
    \draw[thick, fill=white] (0,0) circle (0.075);
    \node[scale=0.7] at (0.2,0) {$\chi$};
    
    }
}

\tikzset{
    arrowhead/.pic = {
\draw[thick, fill=white] (0,0) circle (0.075);
    }
}
\tikzset{
    U/.pic = {
\draw[thick] (-.25,0) -- (0.25,.5);
\draw[thick] (-0.25,.5) -- (0.25,0);
\draw[thick, fill=myred, rounded corners=2pt] (-0.125,0.375) rectangle (.125,0.125);
\draw[thick] (0,0.325) -- (.075,0.325) -- (.075,0.25);
}}

\tikzset{
    Ufolded/.pic = {
\draw[thick] (-.25,0) -- (0.25,.5);
\draw[thick] (-0.25,.5) -- (0.25,0);
\draw[thick, fill=myorange, rounded corners=2pt] (-0.125,0.375) rectangle (.125,0.125);
\draw[thick] (0,0.325) -- (.075,0.325) -- (.075,0.25);
}}

\tikzset{
    U_dag/.pic = {
\draw[thick] (-.25,0) -- (0.25,.5);
\draw[thick] (-0.25,.5) -- (0.25,0);
\draw[thick, fill=myblue, rounded corners=2pt] (-0.125,0.375) rectangle (.125,0.125);
\draw[thick] (0,0.325) -- (.075,0.325) -- (.075,0.25);
}}

\tikzset{
    U_dagfolded/.pic = {
\draw[thick] (-.25,0) -- (0.25,.5);
\draw[thick] (-0.25,.5) -- (0.25,0);
\draw[thick, fill=mygreen, rounded corners=2pt] (-0.125,0.375) rectangle (.125,0.125);
\draw[thick] (0,0.325) -- (.075,0.325) -- (.075,0.25);
}}

\tikzset{
    state_pair/.pic = {
    \draw[line width=4,line cap=round] (0.7,0.4) -- (-0.7,0.4);
    \draw[line width=4,line cap=round] (0.7,-0.4) -- (-0.7,-0.4);
    \draw[thick] (-0.3,0.4) -- (-0.3,-0.4);
    \draw[thick] (0.3,-0.4) -- (0.3,0.4);
    }
}

\tikzset{
    Ur/.pic = {
\draw[thick] (-.25,-0.25) -- (0.25,.25);
\draw[thick] (-0.25,.25) -- (0.25,-0.25);
\draw[thick, fill=myred, rounded corners=2pt] (-0.125,0.125) rectangle (.125,-0.125);
\draw[thick] (0,0.075) -- (.075,0.075) -- (.075,0);
}}

\tikzset{
    Ur_dag/.pic = {
\draw[thick] (-.25,-0.25) -- (0.25,.25);
\draw[thick] (-0.25,.25) -- (0.25,-0.25);
\draw[thick, fill=myblue, rounded corners=2pt] (-0.125,0.125) rectangle (.125,-.125);
\draw[thick] (0,0.075) -- (.075,0.075) -- (.075,0.0);
}}

\tikzset{
    Urfolded/.pic = {
\draw[thick] (-.25,-0.25) -- (0.25,.25);
\draw[thick] (-0.25,.25) -- (0.25,-0.25);
\draw[thick, fill=myorange, rounded corners=2pt] (-0.125,0.125) rectangle (.125,-0.125);
\draw[thick] (0,0.075) -- (.075,0.075) -- (.075,0);
}}

\tikzset{
    Ur_dagfolded/.pic = {
\draw[thick] (-.25,-0.25) -- (0.25,.25);
\draw[thick] (-0.25,.25) -- (0.25,-0.25);
\draw[thick, fill=mygreen, rounded corners=2pt] (-0.125,0.125) rectangle (.125,-.125);
\draw[thick] (0,0.075) -- (.075,0.075) -- (.075,0.0);
}}

\tikzset{
    state_dagfolded/.pic = {
\draw[thick] (-0.5,-0.1)--(0.5,-0.1);
\draw[thick, fill=mygreen, rounded corners=1pt] (-0.2,0.12) rectangle (0.2,-0.12);
\draw[thick] (-0.075,-0.07) -- (-0.15,-0.07) -- (-0.15,0.005);
}}

\tikzset{
    statefolded/.pic = {
\draw[thick] (-0.5,-0.1)--(0.5,-0.1);
\draw[thick, fill=myorange, rounded corners=1pt] (-0.2,0.12) rectangle (0.2,-0.12);
\draw[thick] (0.075,0.07) -- (0.15,0.07) -- (0.15,-0.005);
}}

\tikzset{
    state_dagnolegs/.pic = {
\draw[thick] (-0.5,-0.1)--(0.5,-0.1);
\draw[thick, fill=myblue, rounded corners=1pt] (-0.2,0.12) rectangle (0.2,-0.12);
\draw[thick] (-0.075,-0.07) -- (-0.15,-0.07) -- (-0.15,0.005);
}}

\tikzset{
    statenolegs/.pic = {
\draw[thick] (-0.5,-0.1)--(0.5,-0.1);
\draw[thick, fill=myred, rounded corners=1pt] (-0.2,0.12) rectangle (0.2,-0.12);
\draw[thick] (0.075,0.07) -- (0.15,0.07) -- (0.15,-0.005);
}}

\tikzset{
    state_dagfoldedlegs/.pic = {
\draw[thick] (0,0)--(0.35,0.21);
\draw[thick] (0,0)--(-0.35,0.21);
\draw[thick] (-0.5,-0.1)--(0.5,-0.1);
\draw[thick, fill=mygreen, rounded corners=1pt] (-0.2,0.12) rectangle (0.2,-0.12);
}}

\tikzset{
    statefoldedlegs/.pic = {
\draw[thick] (0,0)--(0.35,0.21);
\draw[thick] (0,0)--(-0.35,0.21);
\draw[thick] (-0.5,-0.1)--(0.5,-0.1);
\draw[thick, fill=myorange, rounded corners=1pt] (-0.2,0.12) rectangle (0.2,-0.12);
}}

\tikzset{
    state_daglegs/.pic = {
\draw[thick] (0,0)--(0.35,0.21);
\draw[thick] (0,0)--(-0.35,0.21);
\draw[thick] (-0.5,-0.1)--(0.5,-0.1);
\draw[thick, fill=myblue, rounded corners=1pt] (-0.2,0.12) rectangle (0.2,-0.12);
}}

\tikzset{
    statelegs/.pic = {
\draw[thick] (0,0)--(0.35,0.21);
\draw[thick] (0,0)--(-0.35,0.21);
\draw[thick] (-0.5,-0.1)--(0.5,-0.1);
\draw[thick, fill=myred, rounded corners=1pt] (-0.2,0.12) rectangle (0.2,-0.12);
}}

\tikzset{
    coloured_lines/.pic = {
\draw[thick] (-1.4,-0.7)--(-0.7,0);
\draw[thick] (-2.1,-1.4)--(-2.8,-2.1);
\draw[thick] (-3.5,-2.8)--(-3.5,-4.2);
\draw[thick] (8.4,-0.7)--(7.7,0);
\draw[thick] (9.1,-1.4)--(9.8,-2.1);
\draw[thick] (10.5,-2.8)--(10.5,-4.2);
\draw[thick] (9.8,-4.9)--(9.1,-5.6);
\draw[thick] (8.4,-6.3)--(7.7,-7);
\draw[thick] (-2.8,-4.9)--(-2.1,-5.6);
\draw[thick] (-1.4,-6.3)--(-0.7,-7);
\draw[thick,color=black, rounded corners=1pt] (0,0.7)--(0,0.4)--(-0.4,0)--(-0.7,0)--(-1.4,-0.7)--(-1.4,-1)--(-1.8,-1.4)--(-2.1,-1.4)--(-2.8,-2.1)--(-2.8,-2.4)--(-3.2,-2.8)--(-3.5,-2.8)--(-3.5,-4.2)--(-3.2,-4.2)--(-2.8,-4.6)--(-2.8,-4.9)--(-2.1,-5.6)--(-1.8,-5.6)--(-1.4,-6)--(-1.4,-6.3)--(-0.7,-7)--(-0.4,-7)--(0,-7.4)--(0,-7.7);
\foreach \i in {1,2}
{
\draw[thick,color=black, rounded corners=1pt] (1.4*\i+0,0.7)--(1.4*\i+0,0.4)--(1.4*\i+-0.4,0)--(1.4*\i+-1,0)--(1.4*\i+-1.4,-0.4)--(1.4*\i+-1.4,-1)--(1.4*\i+-1.8,-1.4)--(1.4*\i+-2.4,-1.4)--(1.4*\i+-2.8,-1.8)--(1.4*\i+-2.8,-2.4)--(1.4*\i+-3.2,-2.8)--(1.4*\i+-3.8,-2.8)--(1.4*\i+-4.2,-3.2)--(1.4*\i+-4.2,-3.8)--(1.4*\i+-3.8,-4.2)--(1.4*\i+-3.2,-4.2)--(1.4*\i+-2.8,-4.6)--(1.4*\i+-2.8,-5.2)--(1.4*\i+-2.4,-5.6)--(1.4*\i+-1.8,-5.6)--(1.4*\i+-1.4,-6)--(1.4*\i+-1.4,-6.6)--(1.4*\i+-1,-7)--(1.4*\i+-0.4,-7)--(1.4*\i+0,-7.4)--(1.4*\i+0,-7.7);
}
\draw[thick,color=black, rounded corners=1pt]
(7,0.7)--(7,0.4)--(7.4,0)--(7.7,0)--(8.4,-0.7)--(8.4,-1)--(8.8,-1.4)--(9.1,-1.4)--(9.8,-2.1)--(9.8,-2.4)--(10.2,-2.8)--(10.5,-2.8)--(10.5,-4.2)--(10.2,-4.2)--(9.8,-4.6)--(9.8,-4.9)--(9.1,-5.6)--(8.8,-5.6)--(8.4,-6)--(8.4,-6.3)--(7.7,-7)--(7.4,-7)--(7,-7.4)--(7,-7.7);
\foreach \i in {1,2}
{\draw[thick,color=black, rounded corners=1pt] 
(7-1.4*\i,0.7)--(7-1.4*\i,0.4)--(7.4-1.4*\i,0)--(8-1.4*\i,0)--(8.4-1.4*\i,-0.4)--(8.4-1.4*\i,-1)--(8.8-1.4*\i,-1.4)--(9.4-1.4*\i,-1.4)--(9.8-1.4*\i,-1.8)--(9.8-1.4*\i,-2.4)--(10.2-1.4*\i,-2.8)--(10.8-1.4*\i,-2.8)--(11.2-1.4*\i,-3.2)--(11.2-1.4*\i,-3.8)--(10.8-1.4*\i,-4.2)--(10.2-1.4*\i,-4.2)--(9.8-1.4*\i,-4.6)--(9.8-1.4*\i,-5.2)--(9.4-1.4*\i,-5.6)--(8.8-1.4*\i,-5.6)--(8.4-1.4*\i,-6)--(8.4-1.4*\i,-6.6)--(8-1.4*\i,-7)--(7.4-1.4*\i,-7)--(7-1.4*\i,-7.4)--(7-1.4*\i,-7.7);
}
\draw[thick,color=myred, rounded corners=1pt] (4.2,-0.4)--(4.2,-1)--(4.6,-1.4)--(5.2,-1.4)--(5.6,-1.8)--(5.6,-2.4)--(6,-2.8)--(6.6,-2.8)--(7,-3.2)--(7,-3.8)--(6.6,-4.2)--(6,-4.2)--(5.6,-4.6)--(5.6,-5.2)--(5.2,-5.6)--(4.6,-5.6)--(4.2,-6)--(4.2,-6.6)--(3.8,-7)--(3.2,-7)--(2.8,-6.6)--(2.8,-6)--(2.4,-5.6)--(1.8,-5.6)--(1.4,-5.2)--(1.4,-4.6)--(1,-4.2)--(0.4,-4.2)--(0,-3.8)--(0,-3.2)--(0.4,-2.8)--(1,-2.8)--(1.4,-2.4)--(1.4,-1.8)--(1.8,-1.4)--(2.4,-1.4)--(2.8,-1)--(2.8,-0.4)--(3.2,0)--(3.8,0)--(4.2,-0.4);
\draw[thick,color=myred, rounded corners=1pt]
(4.2,-1.8)--(4.2,-2.4)--(4.6,-2.8)--(5.2,-2.8)--(5.6,-3.2)--(5.6,-3.8)--(5.2,-4.2)--(4.6,-4.2)--(4.2,-4.6)--(4.2,-5.2)--(3.8,-5.6)--(3.2,-5.6)--(2.8,-5.2)--(2.8,-4.6)--(2.4,-4.2)--(1.8,-4.2)--(1.4,-3.8)--(1.4,-3.2)--(1.8,-2.8)--(2.4,-2.8)--(2.8,-2.4)--(2.8,-1.8)--(3.2,-1.4)--(3.8,-1.4)--(4.2,-1.8);
\draw[thick,color=myred, rounded corners=1pt]
(3.8,-2.8)--(4.2,-3.2)--(4.2,-3.8)--(3.8,-4.2)--(3.2,-4.2)--(2.8,-3.8)--(2.8,-3.2)--(3.2,-2.8)--(3.8,-2.8);
}
\foreach \i in {0,...,5}
{\path (1.4*\i,0.7) pic[rotate=0] {arrowhead};
\path (1.4*\i,-7.7) pic[rotate=180] {arrowhead};
}
}

\tikzset{
int_times/.pic = {
    \foreach \i in {0,...,2}
{\path (1.4*\i,0) pic[rotate=225] {Urfolded};
\path (4.2+1.4*\i,0) pic[rotate=-45] {Ur_dagfolded};
\path (4.2+1.4*\i,-7) pic[rotate=45] {Urfolded};
\path (1.4*\i,-7) pic[rotate=135] {Ur_dagfolded};}
\foreach \i in {0,...,3}
{\path (-1.4+1.4*\i,-1.4) pic[rotate=225] {Urfolded};
\path (4.2+1.4*\i,-1.4) pic[rotate=-45] {Ur_dagfolded};
\path (4.2+1.4*\i,-5.6) pic[rotate=45] {Urfolded};
\path (-1.4+1.4*\i,-5.6) pic[rotate=135] {Ur_dagfolded};}
\foreach \i in {0,...,4}
{\path (-2.8+1.4*\i,-2.8) pic[rotate=225] {Urfolded};
\path (4.2+1.4*\i,-2.8) pic[rotate=-45] {Ur_dagfolded};
\path (4.2+1.4*\i,-4.2) pic[rotate=45] {Urfolded};
\path (-2.8+1.4*\i,-4.2) pic[rotate=135] {Ur_dagfolded};
}

\draw[thick] (-1.4,-0.7)--(-0.7,0);
\draw[thick] (-2.1,-1.4)--(-2.8,-2.1);
\draw[thick] (-3.5,-2.8)--(-3.5,-4.2);
\draw[thick] (8.4,-0.7)--(7.7,0);
\draw[thick] (9.1,-1.4)--(9.8,-2.1);
\draw[thick] (10.5,-2.8)--(10.5,-4.2);
\draw[thick] (9.8,-4.9)--(9.1,-5.6);
\draw[thick] (8.4,-6.3)--(7.7,-7);
\draw[thick] (-2.8,-4.9)--(-2.1,-5.6);
\draw[thick] (-1.4,-6.3)--(-0.7,-7);

\foreach \i in {0,...,5}
{\path (1.4*\i,0.7) pic[rotate=0] {arrowhead};
\path (1.4*\i,-7.7) pic[rotate=180] {arrowhead};
}
}}

\tikzset{
    cage/.pic = {
    
\draw[thick,myred] (2.1,0)--(1.4,-0.7);
\draw[thick,myred] (0.7,-1.4)--(0,-2.1);
\draw[thick,myred] (16.1,0)--(16.8,-0.7);
\draw[thick,myred] (17.5,-1.4)--(18.2,-2.1);
\draw[thick,myred] (2.1,-7)--(1.4,-6.3);
\draw[thick,myred] (0.7,-5.6)--(0,-4.9);
\draw[thick,myred] (16.1,-7)--(16.8,-6.3);
\draw[thick,myred] (17.5,-5.6)--(18.2,-4.9);
\foreach \i in {0,...,9}
{\draw[thick] (2.8+1.4*\i,0.7)--(2.8+1.4*\i,-7.7);}
\foreach \i in {0,...,1}
{\draw[thick] (-0.7,-2.8-1.4*\i)--(18.9,-2.8-1.4*\i);}
\draw[thick,myred] (1.4,-0.7)--(1.4,-6.3);
\draw[thick,myred] (0,-2.1)--(0,-4.9);
\draw[thick,myred] (18.2,-2.1)--(18.2,-4.9);
\draw[thick,myred] (16.8,-6.3)--(16.8,-0.7);
\draw[thick,myred] (2.1,0)--(16.1,0);
\draw[thick,myred] (2.1,-7)--(16.1,-7);
\draw[thick,myred] (0.7,-1.4)--(17.5,-1.4);
\draw[thick,myred] (0.7,-5.6)--(17.5,-5.6);

\foreach \i in {2,3}
{\path (18.9,-1.4*\i) pic[rotate = 270] {arrowhead=0.1};
}
\foreach \i in {2,3}
{\path (-0.7,-1.4*\i) pic[rotate = 90] {arrowhead=0.1};
}
\foreach \i in {2,...,11}
{\path (1.4*\i,0.7) pic[rotate =0] {arrowhead=0.1};
}
\foreach \i in {2,...,11}
{\path (1.4*\i,-7.7) pic[rotate =180] {arrowhead=0.1};
}

}}

\tikzset{
    state_dag/.pic = {
\draw[thick] (0,0)--(0.25,-0.15);
\draw[thick] (0,0)--(-0.25,-0.15);
\draw[thick] (-0.5,0.1)--(0.5,0.1);
\draw[thick, fill=myblue, rounded corners=1pt] (-0.2,0.12) rectangle (0.2,-0.12);
\draw[thick] (0.075,0.07) -- (0.15,0.07) -- (0.15,-0.005);
}}

\tikzset{
    state/.pic = {
\draw[thick] (0,0)--(0.25,0.15);
\draw[thick] (0,0)--(-0.25,0.15);
\draw[thick] (-0.5,-0.1)--(0.5,-0.1);
\draw[thick, fill=myred, rounded corners=1pt] (-0.2,0.12) rectangle (0.2,-0.12);
\draw[thick] (0.075,0.07) -- (0.15,0.07) -- (0.15,-0.005);
}}

\tikzset{
    state2/.pic = {
\draw[thick] (0,0)--(0.25,0.15);
\draw[thick] (0,0)--(-0.25,0.15);
\draw[thick] (-0.5,-0.1)--(0.5,-0.1);
\draw[thick, fill=myblue, rounded corners=1pt] (-0.2,0.12) rectangle (0.2,-0.12);
\draw[thick] (-0.075,-0.07) -- (-0.15,-0.07) -- (-0.15,0.005);
}}

\tikzset{
    state_dag2/.pic = {
\draw[thick] (0,0)--(0.25,-0.15);
\draw[thick] (0,0)--(-0.25,-0.15);
\draw[thick] (-0.5,0.1)--(0.5,0.1);
\draw[thick, fill=myred, rounded corners=1pt] (-0.2,0.12) rectangle (0.2,-0.12);
\draw[thick] (-0.075,-0.07) -- (-0.15,-0.07) -- (-0.15,0.005);
}}

\tikzset{
    curl_1/.pic = {
    \draw[ thick] (0,0) arc (-45:90:0.08);
    }
}
\tikzset{
    curl_2/.pic = {
    \draw[ thick] (0,0) arc (45:-90:0.08);
    }
}
\tikzset{
    curl_3/.pic = {
    \draw[ thick] (0,0) arc (-90:-225:0.08);
    }
}
\tikzset{
    curl_4/.pic = {
    \draw[ thick] (0,0) arc (90:225:0.08);
    }
}
\tikzset{
    curl_5/.pic = {
    \draw[ thick] (0,0) arc (270:405:0.08);
    }
}
\tikzset{
    curl_6/.pic = {
    \draw[ thick] (0,0) arc (90:-45:0.08);
    }
}

\begin{document}

\title{Entanglement Barriers in Dual-Unitary Circuits}

\author{Isaac Reid}
\affiliation{Rudolf Peierls Centre for Theoretical Physics, Clarendon Laboratory, Oxford University, Parks Road, Oxford OX1 3PU, United Kingdom}

\author{Bruno Bertini}
\affiliation{Rudolf Peierls Centre for Theoretical Physics, Clarendon Laboratory, Oxford University, Parks Road, Oxford OX1 3PU, United Kingdom}

\date{\today}

\begin{abstract}

After quantum quenches in many-body systems, finite subsystems evolve non-trivially in time, eventually approaching a stationary state. In typical situations, the reduced density matrix of a given subsystem begins and ends this endeavour as a low-entangled vector in the space of operators. This means that if its entanglement in the operator space initially grows (which is generically the case), it must eventually decrease, describing a barrier-shaped curve. Understanding the shape of this ``entanglement barrier'' is interesting for three main reasons: (i) it quantifies the dynamics of entanglement in the (open) subsystem; (ii) it gives information on the approximability of the reduced density matrix by means of matrix product operators; (iii) it shows qualitative differences depending on the type of dynamics undergone by the system, signalling quantum chaos. Here we compute exactly the shape of the entanglement barriers described by different R\'enyi entropies after quantum quenches in dual-unitary circuits initialised in a class of solvable matrix product states (MPS)s. We show that, for free (SWAP-like) circuits, the entanglement entropy behaves as in rational conformal field theories (CFT)s. On the other hand, for completely chaotic dual-unitary circuits it behaves as in holographic CFTs, exhibiting a longer entanglement barrier that drops rapidly when the subsystem thermalises. Interestingly, the entanglement spectrum is non-trivial in the completely chaotic case. Higher R\'enyi entropies behave in an increasingly similar way to rational CFTs, such that the free and completely chaotic barriers are identical in the limit of infinite replicas (i.e.\ for the so called min-entropy). We also show that, upon increasing the bond dimension of the MPSs, the barrier maintains the same shape. It simply shifts to the left to accommodate for the larger initial entanglement.  
\end{abstract}

\maketitle

\tableofcontents

\section{Introduction}

Isolated quantum many-body systems with short-range interactions relax locally in space~\cite{polkovnikov2011, dalessio2016quantum, gogolin2016equilibration, eisert2015quantum, calabrese2016introduction, essler2016quench}. Namely, when one of these systems is driven out of equilibrium and then let to evolve (unitarily), its local subsystems are subject to the effective bath produced by the rest of the system, which enables them to thermalise~\cite{polkovnikov2011, dalessio2016quantum, gogolin2016equilibration, eisert2015quantum, calabrese2016introduction, essler2016quench}. This means that the reduced density matrix of a given local subsystem follows a complex non-unitary evolution that eventually brings it to a stationary value. The latter is generically a Gibbs state specified by the conservation laws with local spatial density~\cite{ilievski2016quasilocal, essler2016quench}.

Throughout this process the dynamics of the reduced density matrix are conveniently characterised by its \emph{operator entanglement}, i.e.\ the entanglement of the reduced density matrix seen as a vector in the operator space~\cite{zanardi2001entanglement}. The operator entanglement of the reduced density matrix is interesting for three main reasons. 

First, as any other entanglement related quantity, it returns universal information about the time evolution of the system, distilling out most of the observable- or system- specific details. For this reason the evolution of the entanglement is often captured by universal descriptions like conformal field theory (CFT)~\cite{calabrese2005evolution, calabrese2016quantum, asplund2015entanglement,  liu2014entanglement, calabrese2009entanglement, wang2019barrier, dubail2017entanglement} or effective statistical mechanical models in the space-time~\cite{nahum2017quantum, jonay2018coarsegrained, zhou2020entanglement}. 

Second, the operator entanglement of the reduced density matrix gives information about the efficiency of classical simulations of the quantum dynamics. More precisely, it estimates the computational cost of approximating the density matrix with a matrix product operator (MPO)~\cite{prosen2007operator,dubail2017entanglement, pizorn2009operator, Note11}. \footnotetext[11]{See the discussion in Sec.~\ref{section:osee_intro} and Appendix~\ref{app:approximability} for a precise formulation of this statement.}
Understanding its behaviour could then help the ongoing effort~\cite{prosen2007is, haegeman2011time, haegeman2016unifying, leviatan2017quantum, kloss2018time, white2018quantum, znidaric2019nonequilibrium, krumnow2019overcoming, rakovszky2020dissipationassisted, schmitt2021observations} to simulate the time-evolution of quantum systems with tensor network methods~\cite{schollwock2011density, cirac2020matrix}, overcoming the well-known obstacle posed by the linear growth of conventional (state) entanglement~\cite{calabrese2005evolution,  liu2014entanglement, fagotti2008evolution, alba2017entanglement, alba2018entanglement, laeuchli2008spreading, kim2013ballistic, pal2018entangling, nahum2017quantum, bertini2019entanglement, piroli2020exact, chan2018solution}. 

Third, the dynamics of the operator entanglement of the reduced density matrix have recently been proposed as an indicator of quantum chaos~\cite{wang2019barrier}. Indeed, they have been shown to display qualitatively different features depending on whether or not the dynamics of the system are chaotic. This complements a similar observation made concerning the entanglement of local operators~\cite{bertini2020operator, alba2019operator, alba2020diffusion}.  

Specifically, considering the time evolution generated by a standard quantum quench protocol~\cite{calabrese2006time, calabrese2007quantum} --- the system is prepared in a low-entangled state %(the ground state of a short-ranged Hamiltonian) 
and then let to evolve unitarily --- the operator entanglement of the reduced density matrix describes a barrier-shaped curve~\cite{wang2019barrier, dubail2017entanglement}. Indeed, since both the initial and the eventual thermal states are low-entangled, after an initial linear growth the operator entanglement has to decay. Interestingly, while the initial linear growth appears \emph{universal} (it is very similar in nature to the linear growth of state entanglement) the way in which it decays has been observed to depend on the chaoticity of the dynamics~\cite{wang2019barrier}.  

Up to now this behaviour has been demonstrated analytically in CFT (both rational~\cite{dubail2017entanglement} and holographic~\cite{wang2019barrier}) and for random unitary circuits with infinitely large local Hilbert space dimension  $d$~\cite{wang2019barrier}. Additionally, it has been demonstrated numerically for free fermionic systems~\cite{dubail2017entanglement}. No analytical result, however, exists concerning the operator-entanglement dynamics in specific (microscopic) models (integrable or not) with local interactions and finite $d$ or, in general, for specific clean systems.

In this paper we fill this gap and provide the first exact confirmation of the aforementioned picture for a concrete class of local quantum circuits defined for \emph{any} local Hilbert space dimension. Specifically, we focus on the so called \emph{dual-unitary} class~\cite{bertini2019exact}, which has recently emerged as a useful arena where several dynamical quantities~\cite{bertini2019exact, bertini2019entanglement, lerose2020influence, piroli2020exact, gopalakrishnan2019unitary, claeys2021ergodic, gutkin2020local, suzuki2021computational, fritzsch2021eigenstate} and quantum chaos indicators~\cite{bertini2020operator, claeys2020maximum, bertini2018exact, bertini2020operator2, flack2020statistics} can be computed exactly even in the absence of any structure related to integrability. In fact, dual-unitary circuits are generically \emph{quantum chaotic}~\cite{bertini2021random, bertini2020scrambling, bertini2020operator, bertini2018exact} but also contain some integrable points~\cite{bertini2020operator2}. 

After preparing the system in the class of ``solvable'' MPSs recently introduced in Ref.~\cite{piroli2020exact} we find that the operator entanglement of the reduced density matrix follows a universal evolution (independent of the specific gate as long as it is dual-unitary) in the growth region. In the decay region, however, we find that the behaviour of the operator entanglement depends on the specific dual-unitary gate defining the dynamics. We compute it in the extreme cases of the SWAP gate (implementing free evolution) and (in a certain scaling limit) of generic or ``completely chaotic'' dual-unitary circuits~\cite{bertini2020operator}, characterised by no additional structure besides dual-unitarity. We find that in the non-interacting case the barrier has a symmetric shape while it is elongated for generic systems. This is in quantitative agreement with the predictions of rational and holographic CFT respectively. Moreover we find that increasing the operator entanglement of the initial state results in an effective rigid shift to the left of the entanglement barrier, which maintains the same shape. Given the universality of the of the entanglement dynamics we expect these results to describe the behaviour of generic systems. For example, the ubiquitous linear growth of state entanglement for generic systems is confirmed in dual-unitary circuits by a gate-independent entanglement evolution from solvable MPSs~\cite{piroli2020exact}.

The rest of the paper is laid out as follows. In Sec.~\ref{sec:setting} we describe our setting, defining dual-unitary circuits, the class of solvable MPSs, and the observables of interest. In Sec.~\ref{sec:results} we describe and discuss our main results. In Sec.~\ref{sec:derivation} we present the main steps of our derivation, and, finally, Sec.~\ref{sec:conclusions} contains our conclusions. A number of more technical discussions and derivations are reported in the three appendices complementing the main text.

\section{Setting}
\label{sec:setting}

We consider brickwork-like local quantum circuits. These are one-dimensional chains of qudits, each one with $d$ internal states, where the time evolution is generated by discrete applications of the the operator  
\be
\mathbb U = \prod_{x \,\in \mathbb Z_L} U_{x}\prod_{x \in  \mathbb Z_L+1/2} U_{x}\,.
\label{eq:Uglob}
\ee
Here we denoted by $2L$ the number of qudits and placed them at equally spaced positions on a chain of length $L$. Moreover, we introduced
\be
U_{x} = \Pi_{2L}^{2x} (U \otimes \1^{\otimes 2L-2})\Pi_{2L}^{-2x}\,,
\ee
where the ``local gate''  $U\in U(d^2)$ is a two-site unitary matrix that defines the local interactions and $\Pi_{2L}$ is the unitary operator implementing a periodic shift on a chain of $2L$ sites, i.e.
\begin{align}
\!\!\!\Pi_{2L} &\ket{j_1}\!\otimes\!\ket{j_2}\!\otimes\! \cdots \!\otimes\!\ket{j_{2L}} \notag\\
&= \ket{j_{2L}}\!\otimes\!\ket{j_1}\!\otimes\! \cdots \!\otimes\!\ket{j_{2L-1}}, \,\, j_i\in\{0,\ldots,d-1\},
\end{align}
where 
\be
\{\ket{j}, \qquad j \in\{ 0 ,\ldots, d-1\}\},
\ee
denotes an orthonormal basis of the ``local''  Hilbert space $\mathcal{H}_j\simeq \mathbb{C}^d$.

\begin{figure}[t]
\centering
\includegraphics[width=0.45\textwidth]{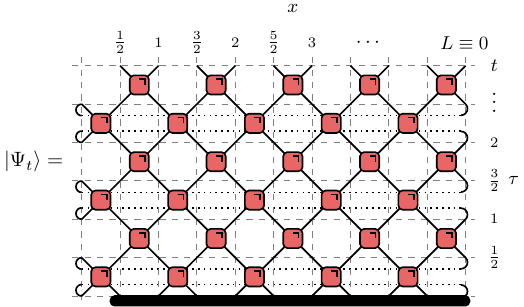}
\caption{State of a local quantum circuit after $t$ steps of evolution.}
\label{fig:statet}
\end{figure}

In this setting, we study the out-of-equilibrium dynamics generated by a quantum quench protocol~\cite{calabrese2006time, calabrese2007quantum}. Namely, the system is prepared in some (non-stationary) initial state $\ket{\Psi_0}$ at time $t=0$ and then evolved using $\mathbb U$. The state at time $t\in\mathbb N$ is therefore given by    
\be
\ket{\Psi_t} = \mathbb U^t \ket{\Psi_0}\,.
\label{eq:quench}
\ee
The time-evolution of the system is conveniently represented by the diagram in Fig.~\ref{fig:statet}, which is reminiscent of those used in tensor-network theory~\cite{cirac2020matrix}. More specifically, to produce the diagram we set   
\begin{equation}
\label{eq:U}
\bra{i}\otimes\braket{j|U|k}\otimes\ket{l}=
\begin{tikzpicture}[baseline=(current  bounding  box.center), scale=1]
\def\eps{0.5};
\draw[thick] (-2.25,0.5)node[left]{$i$}-- (-1.25,-0.5)node[right]{$l$};
\draw[thick] (-2.25,-0.5)node[left]{$k$} -- (-1.25,0.5)node[right]{$j$};
\draw[ thick, fill=myred, rounded corners=2pt] (-2,0.25) rectangle (-1.5,-0.25);
\draw[thick] (-1.75,0.15)-- (-1.6,0.15) -- (-1.6,0);
\Text[x=-2,y=-1]{}
\end{tikzpicture}\equiv U_{kl}^{ij}\,,
\end{equation}
where we added a mark to stress that $U$ is generically not symmetric under space reflection (left to right flip) and time reversal (transposition). Note that: (i) when legs of different operators are joined together a sum over the index of the corresponding local space is understood; (ii) the time direction runs from bottom to top, hence lower legs correspond to incoming indices (matrix row) and upper legs to outgoing indices (matrix column); (iii) objects with open legs without labels represent operators, rather than their matrix elements. 

Note that the quantum-circuit setting described above, besides providing an approximation of the continuous-time dynamics via Suzuki-Trotter decomposition~\cite{suzuki1991general, osborne2006efficient}, is currently implemented in real-world quantum simulators~\cite{arute2019quantum}.

\subsection{Operator entanglement} \label{section:osee_intro}

As set out in the introduction here we are interested in studying the entanglement of the reduced density matrix seen as a vector in the operator space~\cite{pizorn2009operator, zanardi2001entanglement, dubail2017entanglement, wang2019barrier}. Let us now discuss the precise definition of this quantity and the way we go about to measure it. 

\begin{figure*}[t]
\centering
\includegraphics[width=0.85\textwidth]{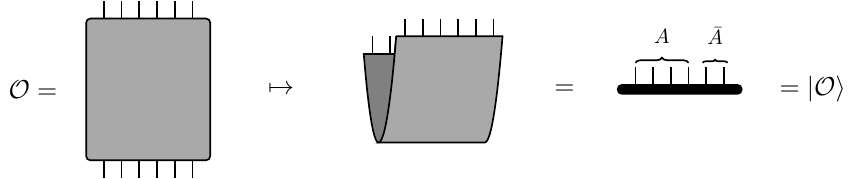}
\caption{Pictorial representation of the operator-to-state mapping and of the bipartition of $B$ in $A$ and $\bar A$. The operator entanglement entropies \eqref{eq:Renyi} measure the entanglement between $A$ and $\bar A$.}
\label{fig:sketch}
\end{figure*}

A convenient way to access the operator entanglement of a generic operator $\mathcal O$ acting non-trivially on a region $B$ ($2|B|$ contiguous qubits) is to introduce an operator-to-state mapping
\be
\mathcal O \longmapsto \ket{\mathcal O},
\label{eq:operatortostate}
\ee
that associates operators on ${\mathcal{H}_{B}=\mathbb C^{d^{2|B|}}}$ to states in ${\mathcal{H}_{B} \otimes \mathcal{H}_{B} = \mathbb C^{d^{4|B|}}}$. In particular, since we are interested in the entanglement among different connected regions in space, we consider a mapping that preserves locality
\begin{align}
\ket{\mathcal O} :=  \sum_{s_j,r_j\in\{0,\ldots,d-1\}}   \braket{\boldsymbol s |\mathcal O|\boldsymbol r}\ket{d\boldsymbol s+\boldsymbol r},
\label{eq:operatortostate2}
\end{align}
where we used the shorthand notation  
\begin{align}
&\!\!\!\!\ket{\boldsymbol s} = \ket{s_1}\!\otimes\! \cdots \!\otimes\!\ket{s_{2|B|}}\in {\mathcal H}_B\\
&\!\!\!\!\!\ket{d\boldsymbol s\!+\!\boldsymbol r}  = \ket{d s_1\!+\! r_1}\!\otimes\! \cdots \!\otimes\!\ket{d s_{2|B|}\!+\! r_{2|B|}}\in \mathcal{H}_{B} \!\otimes\! \mathcal{H}_{B}.\notag
\end{align}
The object $\ket{\mathcal O}$ is now a pure state one can estimate its entanglement by computing the R\'enyi entropies corresponding to the bipartition of $B$ into two connected regions of space~\cite{pizorn2009operator, zanardi2001entanglement} (see Fig.~\ref{fig:sketch}). Namely, considering $B = A \bar A$ we have
\be
S^{(\alpha)}(A,\mathcal O) = \frac{\log{\rm tr}( {\rm P}^\alpha_{\!A}(\mathcal O))}{1-\alpha} \,,
\label{eq:Renyi}
\ee
where we introduced the super-operatorial density matrix reduced to the subsystem $A$
\be
{\rm P}_{\!A}(\mathcal O) =  \frac{{\rm tr}_{\!\bar A}(\ket{\mathcal O}\!\bra{\mathcal O})}{\braket{\mathcal O|\mathcal O}}=\frac{{\rm tr}_{\!\bar A}(\ket{\mathcal O}\!\bra{\mathcal O})}{{\rm tr}(\mathcal O \mathcal O^\dag)}\!.
\label{eq:rhosuper}
\ee
In particular
\be
S(A,\mathcal O) = \lim_{\alpha\to1^+} S^{(\alpha)}(A,\mathcal O),
\ee
gives the so called von-Neumann entropy, which is a bona fide \emph{measure} of bipartite entanglement~\cite{amico2008entanglement}. 

To see in what sense $S^{(\alpha)}(A,\mathcal O) $ quantify the entanglement of $\ket{\mathcal O}$ (normalised) it is useful to consider its Schmidt decomposition 
\be
\ket{\mathcal O} = \sum_{j} \sigma_j \ket{A,j}\otimes\ket{\bar A, j}, 
\label{eq:Schmidt}
\ee
where the sum runs up to $d^{2\min(|A|,|\bar A|)}$, $\ket{A,j}$ ($\ket{\bar A,j}$) are orthonormal states in the subsystem $A$ ($\bar A$), and the real numbers $\sigma_j$, known as Schmidt coefficients, fulfil
\be
\sigma_j\geq\sigma_{j+1}\geq 0,\qquad \sum_j \sigma^2_j=1\,.
\ee
The second of these conditions follows from the normalisation of the state.

Substituting \eqref{eq:Schmidt} in \eqref{eq:Renyi} one finds that the R\'enyi entropies are expressed solely in terms of the Schmidt coefficients  
\be
S^{(\alpha)}(A,\mathcal O) = \frac{1}{1-\alpha} \log \sum_j \sigma_j^{2\alpha}\,.
\label{eq:RenyiSchmidt}
\ee
It is then clear that computing $\{S^{(\alpha)}(A,\mathcal O)\}_{\alpha\in\mathbb R}$ one characterises the distribution of Schmidt coefficients (the so-called entanglement spectrum~\cite{laflorencie2016quantum, li2008entanglement}). This gives an estimation of how close $\ket{\mathcal O}$ is to a product state or, upon undoing the mapping \eqref{eq:operatortostate}--\eqref{eq:operatortostate2}, how close $\mathcal O$ is to a product operator.  

Specialising to the case of interest, we can estimate the entanglement of the reduced density matrix by computing the operatorial R\'enyi entropies~\eqref{eq:Renyi} for ${\mathcal O = \rho_B(t)}$, namely 
\be
S^{(\alpha)}(A, {\rho_{A \bar A}(t)})  = \frac{\log{\rm tr}( {\rm P}^\alpha_{\!A}(\rho_{A \bar A}(t)))}{1-\alpha}\,.
\label{eq:Renyirho}
\ee
Note that the operator entanglement of the reduced density matrix is not related to the entanglement of the reduced density matrix seen as a mixed state~\cite{prosen2007operator, dubail2017entanglement}.

In the introduction, however, we also mentioned that the entanglement of $\rho_B(t)$ is often used as an indicator for the \emph{approximability} of the state, i.e., to determine whether or not the reduced density matrix can be efficiently approximated by a MPO. In this regard, as discussed in Appendix~\ref{app:approximability}, it is also useful to consider the operatorial R\'enyi entropies of the \emph{square root} of the state (note that in the language of quantum information~\cite{nielsen2011quantum} $\ket{\sqrt{\rho_{B}(t)}}$ is a \emph{purification} of $\rho_{B}(t)$). Namely 
\be
S^{(\alpha)}(A, \sqrt{\rho_{A \bar A}(t)})  = \frac{\log{\rm tr}( {\rm P}^\alpha_{\!A}( \sqrt{\rho_{A \bar A}(t)}))}{1-\alpha} \,.
\label{eq:Renyisquareroot}
\ee
To summarise the discussion in Appendix~\ref{app:approximability} we have that if $S^{(\alpha)}(A, \sqrt{\rho_{A \bar A}(t)})$ scales at most logarithmically with $|B|$ for all bipartitions $B=A\bar A$ and $\alpha\in[0,1]$, the state is approximable. Conversely, if $S^{(\alpha)}(A, {\rho_{A \bar A}(t)})$ scales faster than logarithmically with $|B|$ for some $A$ and $\alpha>1$ the state is not approximable. Incidentally, we note that the von Neumann entropy of $\sqrt{\rho_B(t)}$, also known as reflective entropy~\cite{dutta2019canonical}, is the only quantity related to the operator entanglement of ${\rho_B(t)}$ that has been accessed in holographic CFTs~\cite{wang2019barrier}.

Here, to achieve a comprehensive description including both \eqref{eq:Renyirho} and \eqref{eq:Renyisquareroot}, we follow Ref.~\cite{wang2019barrier} and look at R\'enyi entropies of generic positive powers of the reduced density matrix, i.e.   
\be
\!\!S^{(\alpha)}_{\beta}(A, t)\equiv S^{(\alpha)}(A, \rho^\beta_{A \bar A}(t))  = \frac{\log{\rm tr}({\rm P}_{\!A,\beta}^\alpha(t) )}{1-\alpha}\,,
\label{eq:Renyigen}
\ee
where $\beta>0$ and we introduced the shorthand notation
\be
{\rm P}_{\!A,\beta}(t) \equiv  {\rm P}_{\!A}(\rho_{A \bar A}^\beta(t)))\,.
\label{eq:Pbeta}
\ee
Specifically, we compute \eqref{eq:Renyigen} by means of a replica approach: first we consider $\alpha=n\geq2$ and $\beta=m\geq1$ with $n,m\in\mathbb Z$ and then analytically continue the result to 
\be
(\alpha,\beta)\in\mathcal D \!=\! \{(z,w)\in \mathbb C^2\!\!:\, {\rm Re}[z]\geq0\,, {\rm Re}[w]\geq0\}\,.
\label{eq:D}
\ee

\subsection{Dual-Unitary Circuits} 
\label{section:dual_unitary}

As discussed in the introduction, to access \eqref{eq:Renyigen} analytically here we consider a special class of local quantum circuits, called ``dual-unitary circuits'', where the local gates $U$ are \emph{dual-unitary}. This means that the tensor \eqref{eq:U} is unitary both when considered as map from bottom to top and as map from left to right~\cite{bertini2019exact}. While the first is just the standard condition enforcing the unitarity of the time evolution, the second ensures that also the evolution in space (sometimes also called time-channel evolution) is unitary. In formulae we have  
\begin{align}
&\sum_{a,b}U_{i,j}^{a,b}(U^{a,b}_{k,l})^*=\delta_{i,k}\delta_{j,l}, & & \text{unitarity in time},\\
&\sum_{a,b} U_{b,j}^{a,i}(U^{a,k}_{b,l})^*=\delta_{i,k} \delta_{j,l},  & &  \text{unitarity in space}. 
\end{align}
These relations can be expressed diagrammatically. Indeed, introducing a representation for the hermitian conjugate of $U$,
\be
U^{\dag}= \begin{tikzpicture}[baseline=(current  bounding  box.center), scale=0.45]
\draw[thick] (4.5,0) -- (6.5,2);
\draw[thick] (4.5,2) -- (6.5,0);
\draw[thick, fill=myblue, rounded corners=2pt] (5,1.5) rectangle (6,0.5);
\draw[thick] (5.5,1.3) -- (5.8,1.3) -- (5.8,1);
\Text[x=5,y=-.35]{}
\end{tikzpicture},
\ee
we have 
\begin{align} 
\begin{tikzpicture}[baseline=(current  bounding  box.center), scale=0.45]
\draw[thick] (0.5,1.5) -- (0.5,-1.5);
\draw[thick] (-0.5,-1.5) -- (-0.5,1.5);
\draw[thick] (0,1) -- (1,2);
\draw[thick] (0,1) -- (-1,2);
\draw[thick, fill=myred, rounded corners=2pt] (-0.5,1.5) rectangle (.5,0.5);
\draw[thick] (0,1.3) -- (.3,1.3) -- (.3,1);
\draw[thick] (-1,-2) -- (0,-1);
\draw[thick] (0,-1) -- (1,-2);
\draw[thick, fill=myblue, rounded corners=2pt] (-.5,-.5) rectangle (.5,-1.5);
\draw[thick] (0,-.7) -- (.3,-.7) -- (.3,-1);
\Text[x=.25,y=-2.5]{}
\end{tikzpicture}
&=
\begin{tikzpicture}[baseline=(current  bounding  box.center), scale=0.45]
\draw[thick] (4.5,1.5) -- (4.5,-1.5);
\draw[thick] (3.5,-1.5) -- (3.5,1.5);
\draw[thick] (4.5,1.5) -- (5,2);
\draw[thick] (3.5,1.5) -- (3,2);
\draw[thick] (3,-2) -- (3.5,-1.5);
\draw[thick] (4.5,-1.5) -- (5,-2);
\Text[x=3.75,y=-2.5]{}
\end{tikzpicture}
=
\begin{tikzpicture}[baseline=(current  bounding  box.center), scale=0.45]
\draw[thick] (8.5,1.5) -- (8.5,-1.5);
\draw[thick] (7.5,-1.5) -- (7.5,1.5);
\draw[thick] (8,1) -- (9,2);
\draw[thick] (8,1) -- (7,2);
\draw[thick, fill=myblue, rounded corners=2pt] (7.5,1.5) rectangle (8.5,0.5);
\draw[thick] (8,1.3) -- (8.3,1.3) -- (8.3,1);
\draw[thick] (7,-2) -- (8,-1);
\draw[thick] (8,-1) -- (9,-2);
\draw[thick, fill=myred, rounded corners=2pt] (7.5,-.5) rectangle (8.5,-1.5);
\draw[thick] (8,-.7) -- (8.3,-.7) -- (8.3,-1);
\Text[x=8.25,y=-2.5]{}
\end{tikzpicture},\label{eq:du1}\\
\notag\\
\begin{tikzpicture}[baseline=(current  bounding  box.center), scale=0.45]
\draw[thick] (-0.5,1.5) -- (2,1.5);
\draw[thick] (-0.5,.5) -- (2,.5);
\draw[thick] (0,1) -- (-1,0);
\draw[thick] (0,1) -- (-1,2);
\draw[thick, fill=myred, rounded corners=2pt] (-0.5,1.5) rectangle (.5,0.5);
\draw[thick] (0,0.7) -- (-0.3,0.7) -- (-0.3,1);
\draw[thick] (2,1) -- (3,2);
\draw[thick] (2,1) -- (3,0);
\draw[thick, fill=myblue, rounded corners=2pt] (1.5,1.5) rectangle (2.5,.5);
\draw[thick] (2,1.3) -- (2.3,1.3) -- (2.3,1);
\Text[x=1.75,y=-.35]{}
\end{tikzpicture}
& =
\begin{tikzpicture}[baseline=(current  bounding  box.center), scale=0.45]
\draw[thick] (5.5,1.5) -- (8.5,1.5);
\draw[thick] (5.5,.5) -- (8.5,.5);
\draw[thick] (5.5,.5) -- (5,0);
\draw[thick] (5.5,1.5) -- (5,2);
\draw[thick] (8.5,1.5) -- (9,2);
\draw[thick] (8.5,.5) -- (9,0);
\Text[x=8.5,y=-.35]{}
\end{tikzpicture}
=
\begin{tikzpicture}[baseline=(current  bounding  box.center), scale=0.45]
\draw[thick] (11.5,1.5) -- (14,1.5);
\draw[thick] (11.5,.5) -- (14,.5);
\draw[thick] (12,1) -- (11,0);
\draw[thick] (12,1) -- (11,2);
\draw[thick, fill=myblue, rounded corners=2pt] (11.5,1.5) rectangle (12.5,0.5);
\draw[thick] (12,1.3) -- (12.3,1.3) -- (12.3,1.0);
\draw[thick] (14,1) -- (15,2);
\draw[thick] (14,1) -- (15,0);
\draw[thick, fill=myred, rounded corners=2pt] (13.5,1.5) rectangle (14.5,.5);
\draw[thick] (13.7,1) -- (13.7,0.7) -- (14,0.7);
\Text[x=14,y=-.35]{}
\end{tikzpicture}\,.\label{eq:du2}
\end{align}
For chains of qubits ($d=2$) one can find all local gates fulfilling these conditions~\cite{bertini2019exact}. Specifically, they can be parametrised as follows
\begin{equation} \label{eq:parametrisation}
U=e^{i \phi} (u_{+} \otimes u_{-}) V[J] (v_{+} \otimes v_{-}), 
\end{equation}
where $\phi , J \in \mathbb{R}$, $u_{\pm}, v_{\pm} \in {\rm SU}(2)$ and
\begin{equation} 
\!\!\!\!V[J]\!=\!\exp\!\left[-i\left(\frac{\pi}{4} \sigma_1 \otimes \sigma_1 \!+\! \frac{\pi}{4} \sigma_2 \otimes \sigma_2 \!+\! J \sigma_3 \otimes \sigma_3\right)\right]\!,
\end{equation}
where $\{\sigma_\alpha\}_{\alpha=1,2,3}$ are Pauli matrices. For ${d>2}$ a complete parameterisation is yet unknown, although parameterisations for certain sub-families have been presented in Refs.~\cite{claeys2021ergodic, gutkin2020local, bertini2021random}. 

The simplest example of a dual-unitary gate, defined for any local Hilbert space dimension, is the SWAP gate. This is the gate that simply exchanges the quantum states at each of its input sites with no interaction among the two, i.e. 
\begin{equation}
[{\textrm{SWAP}}]_{ab}^{cd} = \delta_{ad} \delta_{bc},
\label{eq:SWAP}
\end{equation}
or, in tensor-network notation,
\begin{equation}
{\textrm{SWAP}}=\begin{tikzpicture}[baseline=(current  bounding  box.center), scale=0.5]
\draw[thick] (-1,-1) .. controls (-0.9,0) and (0.9,0) .. (1,1);
\draw[thick, white, fill=white] (0,0) circle (0.1);
\draw[thick] (-1,1) .. controls (-0.9,0) and (0.9,0) .. (1,-1);

\end{tikzpicture}.
\ee
It is immediate to see that this fulfils the dual-unitarity conditions \eqref{eq:du1}--\eqref{eq:du2}. 

Here we will use the SWAP gate as the representative example of non-interacting dual-unitary circuits, and contrast the dynamics generated by ${\textrm{SWAP}}$ with those of  ``generic'' dual-unitary circuits. As a representative of the latter we will use the class of ``completely chaotic'' dual-unitary circuits introduced in Ref.~\cite{bertini2020operator}. In essence, the completely chaotic dual-unitary gates are those subject to no other constraints but \eqref{eq:du1} and \eqref{eq:du2}. A more precise definition can be given by introducing the following $d^{4x} \times d^{4x}$ dimensional ``transfer matrices'' 
\begin{align}
{\mathcal T}_{1,x} &= \begin{tikzpicture}[baseline=(current  bounding  box.center), scale=0.75]
\foreach \i in {0,...,2}
{\path (1.4+1.4*\i,0) pic[rotate=315, scale = 1.5] {Urfolded};
}
\foreach \i in {0,...,2}
{\path (5.6+1.4*\i,0) pic[rotate=225, scale = 1.5] {Ur_dagfolded};
}
\path (0.7,0) pic[rotate = 90] {arrowhead=0.1};
\path (9.1,0) pic[rotate = 270] {arrowhead=0.1};
\draw [thick,decorate,decoration={brace}]
(8.55,-.75) -- (1.25,-.75);
\Text[x=3,y=1.25]{}
\node[scale=1] at (5,-1.25) {$2x$};
\end{tikzpicture},\\
{\mathcal T}_{2,x} &= \begin{tikzpicture}[baseline=(current  bounding  box.center), scale=0.75]
\foreach \i in {0,...,2}
{\path (1.4+1.4*\i,0) pic[rotate=135, scale = 1.5] {Ur_dagfolded};
}
\foreach \i in {0,...,2}
{\path (5.6+1.4*\i,0) pic[rotate=45, scale = 1.5] {Urfolded};
}
\path (0.7,0) pic[rotate = 90] {arrowhead=0.1};
\path (9.1,0) pic[rotate = 270] {arrowhead=0.1};
\draw [thick,decorate,decoration={brace}]
(8.55,-.75) -- (1.25,-.75);
\Text[x=3,y=1.25]{}
\node[scale=1] at (5,-1.25) {$2x$};
\end{tikzpicture},
\end{align}
which are constructed in terms of folded (or doubled) gates  
\begin{align}
&\begin{tikzpicture}[baseline=(current  bounding  box.center), scale=0.7]
\def\eps{0.5};
\draw[ thick] (-4.25,0.5) -- (-3.25,-0.5);
\draw[ thick] (-4.25,-0.5) -- (-3.25,0.5);
\draw[ thick, fill=myorange, rounded corners=2pt] (-4,0.25) rectangle (-3.5,-0.25);
\draw[thick] (-3.75,0.15) -- (-3.6,0.15) -- (-3.6,0);
\Text[x=-3.95,y=-.725]{}
\end{tikzpicture}
=
\begin{tikzpicture}[baseline=(current  bounding  box.center), scale=0.7]
\draw[thick] (-1.65,0.65) -- (-0.65,-0.35);
\draw[thick] (-1.65,-0.35) -- (-0.65,0.65);
\draw[ thick, fill=myblue, rounded corners=2pt] (-1.4,0.4) rectangle (-.9,-0.1);
\draw[thick] (-1.15,0) -- (-1,0) -- (-1,0.15);
\draw[thick] (-2.25,0.5) -- (-1.25,-0.5);
\draw[thick] (-2.25,-0.5) -- (-1.25,0.5);
\draw[ thick, fill=myred, rounded corners=2pt] (-2,0.25) rectangle (-1.5,-0.25);
\draw[thick] (-1.75,0.15) -- (-1.6,0.15) -- (-1.6,0);
\Text[x=-1.95,y=-.725]{}
%\Text[x=-4.8,y=0.05]{$W=$}
\end{tikzpicture}= U\otimes U^\dag\,, \label{eq:doublegates1}\\
& \begin{tikzpicture}[baseline=(current  bounding  box.center), scale=0.7]
\def\eps{0.5};
\draw[ thick] (-4.25,0.5) -- (-3.25,-0.5);
\draw[ thick] (-4.25,-0.5) -- (-3.25,0.5);
\draw[ thick, fill=mygreen, rounded corners=2pt] (-4,0.25) rectangle (-3.5,-0.25);
\draw[thick] (-3.75,0.15) -- (-3.6,0.15) -- (-3.6,0);
\Text[x=-3.95,y=-.725]{}
\end{tikzpicture}
=
\begin{tikzpicture}[baseline=(current  bounding  box.center), scale=0.7]
\draw[thick] (-1.65,0.65) -- (-0.65,-0.35);
\draw[thick] (-1.65,-0.35) -- (-0.65,0.65);
\draw[ thick, fill=myred, rounded corners=2pt] (-1.4,0.4) rectangle (-.9,-0.1);
\draw[thick] (-1.15,0) -- (-1,0) -- (-1,0.15);
\draw[thick] (-2.25,0.5) -- (-1.25,-0.5);
\draw[thick] (-2.25,-0.5) -- (-1.25,0.5);
\draw[ thick, fill=myblue, rounded corners=2pt] (-2,0.25) rectangle (-1.5,-0.25);
\draw[thick] (-1.75,0.15) -- (-1.6,0.15) -- (-1.6,0);
%\Text[x=-4.8,y=0.05]{$W=$}
\Text[x=-1.5,y=-.725]{}
\end{tikzpicture}= U^\dag\otimes U,
\label{eq:doublegates2}
\end{align}
and folded wires 
\begin{equation}
\begin{tikzpicture}[baseline=(current  bounding  box.center), scale=0.8]
\draw[very thick] (-0.15,0.25) -- (-0.15,-0.251);
\draw[thick, fill=white] (-.15,-0.25) circle (0.1cm); 
\Text[x=-0.15,y=-0.65]{}
\end{tikzpicture} =
\frac{1}{\sqrt{d}}\,\,
\begin{tikzpicture}[baseline=(current  bounding  box.center), scale=0.8]
\draw[thick] (-2,0.25) -- (-2,-0.251);
\draw[thick] (-1.5,0.4) -- (-1.5,-0.101);
\draw[thick] (-2,-0.25) to[out=-85,in=-80] ( (-1.505,-0.1);
\Text[x=-1.75,y=-0.65]{}
\end{tikzpicture}\equiv \ket{\mcirc}. \label{eq:foldedwire}
\end{equation}
Note that the legs of double gates carry a $d^2$-dimensional Hilbert space and that, because of \eqref{eq:du1} and \eqref{eq:du2}, the folded gates fulfil
\begin{align}
&\begin{tikzpicture}[baseline=(current  bounding  box.center), scale=.75]
\def\eps{0.5};
\draw[very thick] (-4.25,0.5) -- (-3.25,-0.5);
\draw[very thick] (-4.25,-0.5) -- (-3.25,0.5);
\draw[ thick, fill=myorange, rounded corners=2pt] (-4,0.25) rectangle (-3.5,-0.25);
\draw[thick] (-3.75,0.15) -- (-3.6,0.15) -- (-3.6,0);
\Text[x=-2.75,y=0.0, anchor = center]{$=$}
\draw[thick, fill=white] (-4.25,-0.5) circle (0.1cm); 
\draw[thick, fill=white] (-3.25,-0.5) circle (0.1cm); 
\draw[very thick] (-2.25,0.5) -- (-2.25,-0.5);
\draw[very thick] (-1.25,-0.5) -- (-1.25,0.5);
\draw[thick, fill=white] (-2.25,-0.5) circle (0.1cm); 
\draw[thick, fill=white] (-1.25,-0.5) circle (0.1cm); 
\Text[x=-1,y=0.0, anchor = center]{,}
\end{tikzpicture}
\quad\qquad \begin{tikzpicture}[baseline=(current  bounding  box.center), scale=.75]
\def\eps{0.5};
\draw[very thick] (-4.25,0.5) -- (-3.25,-0.5);
\draw[very thick] (-4.25,-0.5) -- (-3.25,0.5);
\draw[ thick, fill=myorange, rounded corners=2pt] (-4,0.25) rectangle (-3.5,-0.25);
\draw[thick] (-3.75,0.15) -- (-3.6,0.15) -- (-3.6,0);
\Text[x=-2.75,y=0.0, anchor = center]{$=$}
\draw[thick, fill=white] (-4.25,0.5) circle (0.1cm); 
\draw[thick, fill=white] (-3.25,0.5) circle (0.1cm); 
\draw[very thick] (-2.25,0.5) -- (-2.25,-0.5);
\draw[very thick] (-1.25,-0.5) -- (-1.25,0.5);
\draw[thick, fill=white] (-2.25,0.5) circle (0.1cm); 
\draw[thick, fill=white] (-1.25,0.5) circle (0.1cm); 
\Text[x=-1,y=0.0, anchor = center]{,}
\end{tikzpicture}\\
&\notag\\
&\begin{tikzpicture}[baseline=(current  bounding  box.center), scale=.75]
\def\eps{0.5};
\draw[very thick] (-4.25,0.5) -- (-3.25,-0.5);
\draw[very thick] (-4.25,-0.5) -- (-3.25,0.5);
\draw[ thick, fill=mygreen, rounded corners=2pt] (-4,0.25) rectangle (-3.5,-0.25);
\draw[thick] (-3.75,0.15) -- (-3.6,0.15) -- (-3.6,0);
\Text[x=-2.75,y=0.0, anchor = center]{$=$}
\draw[thick, fill=white] (-4.25,-0.5) circle (0.1cm); 
\draw[thick, fill=white] (-3.25,-0.5) circle (0.1cm); 
\draw[very thick] (-2.25,0.5) -- (-2.25,-0.5);
\draw[very thick] (-1.25,-0.5) -- (-1.25,0.5);
\draw[thick, fill=white] (-2.25,-0.5) circle (0.1cm); 
\draw[thick, fill=white] (-1.25,-0.5) circle (0.1cm); 
\Text[x=-1,y=0.0, anchor = center]{,}
\end{tikzpicture}
\quad\qquad \begin{tikzpicture}[baseline=(current  bounding  box.center), scale=.75]
\def\eps{0.5};
\draw[very thick] (-4.25,0.5) -- (-3.25,-0.5);
\draw[very thick] (-4.25,-0.5) -- (-3.25,0.5);
\draw[ thick, fill=mygreen, rounded corners=2pt] (-4,0.25) rectangle (-3.5,-0.25);
\draw[thick] (-3.75,0.15) -- (-3.6,0.15) -- (-3.6,0);
\Text[x=-2.75,y=0.0, anchor = center]{$=$}
\draw[thick, fill=white] (-4.25,0.5) circle (0.1cm); 
\draw[thick, fill=white] (-3.25,0.5) circle (0.1cm); 
\draw[very thick] (-2.25,0.5) -- (-2.25,-0.5);
\draw[very thick] (-1.25,-0.5) -- (-1.25,0.5);
\draw[thick, fill=white] (-2.25,0.5) circle (0.1cm); 
\draw[thick, fill=white] (-1.25,0.5) circle (0.1cm); 
\Text[x=-1,y=0.0, anchor = center]{,}
\end{tikzpicture}\\
&\notag\\
&\begin{tikzpicture}[baseline=(current  bounding  box.center), scale=.75]
\def\eps{0.5};
\draw[very thick] (-4.25,0.5) -- (-3.25,-0.5);
\draw[very thick] (-4.25,-0.5) -- (-3.25,0.5);
\draw[ thick, fill=myorange, rounded corners=2pt] (-4,0.25) rectangle (-3.5,-0.25);
\draw[thick] (-3.75,0.15) -- (-3.6,0.15) -- (-3.6,0);
\Text[x=-2.75,y=0.0, anchor = center]{$=$}
\draw[thick, fill=white] (-4.25,0.5) circle (0.1cm); 
\draw[thick, fill=white] (-4.25,-0.5) circle (0.1cm); 
\draw[very thick] (-2.25,0.5) -- (-1.25,0.5);
\draw[very thick] (-2.25,-0.5) -- (-1.25,-0.5);
\draw[thick, fill=white] (-2.25,-0.5) circle (0.1cm); 
\draw[thick, fill=white] (-2.25,0.5) circle (0.1cm);
\Text[x=-1,y=0.0, anchor = center]{,}
\end{tikzpicture}
\quad\qquad\begin{tikzpicture}[baseline=(current  bounding  box.center), scale=.75]
\def\eps{0.5};
\draw[very thick] (-4.25,0.5) -- (-3.25,-0.5);
\draw[very thick] (-4.25,-0.5) -- (-3.25,0.5);
\draw[ thick, fill=myorange, rounded corners=2pt] (-4,0.25) rectangle (-3.5,-0.25);
\draw[thick] (-3.75,0.15) -- (-3.6,0.15) -- (-3.6,0);
\Text[x=-2.75,y=0.0, anchor = center]{$=$}
\draw[thick, fill=white] (-3.25,-0.5) circle (0.1cm); 
\draw[thick, fill=white] (-3.25, 0.5) circle (0.1cm); 
\draw[very thick] (-1.25,0.5) -- (-2.25, 0.5);
\draw[very thick] (-1.25,-0.5) -- (-2.25,-0.5);
\draw[thick, fill=white] (-1.25, 0.5) circle (0.1cm); 
\draw[thick, fill=white] (-1.25,-0.5) circle (0.1cm); 
\Text[x=-1,y=0.0, anchor = center]{,}
\end{tikzpicture}\\
&\notag\\
&\begin{tikzpicture}[baseline=(current  bounding  box.center), scale=.75]
\def\eps{0.5};
\draw[very thick] (-4.25,0.5) -- (-3.25,-0.5);
\draw[very thick] (-4.25,-0.5) -- (-3.25,0.5);
\draw[ thick, fill=mygreen, rounded corners=2pt] (-4,0.25) rectangle (-3.5,-0.25);
\draw[thick] (-3.75,0.15) -- (-3.6,0.15) -- (-3.6,0);
\Text[x=-2.75,y=0.0, anchor = center]{$=$}
\draw[thick, fill=white] (-4.25,0.5) circle (0.1cm); 
\draw[thick, fill=white] (-4.25,-0.5) circle (0.1cm); 
\draw[very thick] (-2.25,0.5) -- (-1.25,0.5);
\draw[very thick] (-2.25,-0.5) -- (-1.25,-0.5);
\draw[thick, fill=white] (-2.25,-0.5) circle (0.1cm); 
\draw[thick, fill=white] (-2.25,0.5) circle (0.1cm);
\Text[x=-1,y=0.0, anchor = center]{,}
\end{tikzpicture}
\quad\qquad\begin{tikzpicture}[baseline=(current  bounding  box.center), scale=.75]
\def\eps{0.5};
\draw[very thick] (-4.25,0.5) -- (-3.25,-0.5);
\draw[very thick] (-4.25,-0.5) -- (-3.25,0.5);
\draw[ thick, fill=mygreen, rounded corners=2pt] (-4,0.25) rectangle (-3.5,-0.25);
\draw[thick] (-3.75,0.15) -- (-3.6,0.15) -- (-3.6,0);
\Text[x=-2.75,y=0.0, anchor = center]{$=$}
\draw[thick, fill=white] (-3.25,-0.5) circle (0.1cm); 
\draw[thick, fill=white] (-3.25, 0.5) circle (0.1cm); 
\draw[very thick] (-1.25,0.5) -- (-2.25, 0.5);
\draw[very thick] (-1.25,-0.5) -- (-2.25,-0.5);
\draw[thick, fill=white] (-1.25, 0.5) circle (0.1cm); 
\draw[thick, fill=white] (-1.25,-0.5) circle (0.1cm); 
\Text[x=-1,y=0.0, anchor = center]{,}
\end{tikzpicture}\\
&\notag\\
&\begin{tikzpicture}[baseline=(current  bounding  box.center), scale=0.4]
\draw[thick] (0.5,1.5) -- (0.5,-1.5);
\draw[thick] (-0.5,-1.5) -- (-0.5,1.5);
\draw[thick] (0,1) -- (1,2);
\draw[thick] (0,1) -- (-1,2);
\draw[thick, fill=myorange, rounded corners=2pt] (-0.5,1.5) rectangle (.5,0.5);
\draw[thick] (0,1.3) -- (.3,1.3) -- (.3,1);
\draw[thick] (-1,-2) -- (0,-1);
\draw[thick] (0,-1) -- (1,-2);
\draw[thick, fill=mygreen, rounded corners=2pt] (-.5,-.5) rectangle (.5,-1.5);
\draw[thick] (0,-.7) -- (.3,-.7) -- (.3,-1);
\Text[x=.25,y=-2.5]{}
\end{tikzpicture}
=
\begin{tikzpicture}[baseline=(current  bounding  box.center), scale=0.4]
\draw[thick] (4.5,1.5) -- (4.5,-1.5);
\draw[thick] (3.5,-1.5) -- (3.5,1.5);
\draw[thick] (4.5,1.5) -- (5,2);
\draw[thick] (3.5,1.5) -- (3,2);
\draw[thick] (3,-2) -- (3.5,-1.5);
\draw[thick] (4.5,-1.5) -- (5,-2);
\Text[x=3.75,y=-2.5]{}
\end{tikzpicture}
=
\begin{tikzpicture}[baseline=(current  bounding  box.center), scale=0.4]
\draw[thick] (8.5,1.5) -- (8.5,-1.5);
\draw[thick] (7.5,-1.5) -- (7.5,1.5);
\draw[thick] (8,1) -- (9,2);
\draw[thick] (8,1) -- (7,2);
\draw[thick, fill=mygreen, rounded corners=2pt] (7.5,1.5) rectangle (8.5,0.5);
\draw[thick] (8,1.3) -- (8.3,1.3) -- (8.3,1);
\draw[thick] (7,-2) -- (8,-1);
\draw[thick] (8,-1) -- (9,-2);
\draw[thick, fill=myorange, rounded corners=2pt] (7.5,-.5) rectangle (8.5,-1.5);
\draw[thick] (8,-.7) -- (8.3,-.7) -- (8.3,-1);
\Text[x=8.25,y=-2.5]{}
\end{tikzpicture}, \\
&\begin{tikzpicture}[baseline=(current  bounding  box.center), scale=0.4]
\draw[thick] (-0.5,1.5) -- (2,1.5);
\draw[thick] (-0.5,.5) -- (2,.5);
\draw[thick] (0,1) -- (-1,0);
\draw[thick] (0,1) -- (-1,2);
\draw[thick, fill=myorange, rounded corners=2pt] (-0.5,1.5) rectangle (.5,0.5);
\draw[thick] (0,0.7) -- (-0.3,0.7) -- (-0.3,1);
\draw[thick] (2,1) -- (3,2);
\draw[thick] (2,1) -- (3,0);
\draw[thick, fill=mygreen, rounded corners=2pt] (1.5,1.5) rectangle (2.5,.5);
\draw[thick] (2,1.3) -- (2.3,1.3) -- (2.3,1);
\Text[x=1.75,y=-.35]{}
\end{tikzpicture}
=
\begin{tikzpicture}[baseline=(current  bounding  box.center), scale=0.4]
\draw[thick] (5.5,1.5) -- (8.5,1.5);
\draw[thick] (5.5,.5) -- (8.5,.5);
\draw[thick] (5.5,.5) -- (5,0);
\draw[thick] (5.5,1.5) -- (5,2);
\draw[thick] (8.5,1.5) -- (9,2);
\draw[thick] (8.5,.5) -- (9,0);
\Text[x=8.5,y=-.35]{}
\end{tikzpicture}
=
\begin{tikzpicture}[baseline=(current  bounding  box.center), scale=0.4]
\draw[thick] (11.5,1.5) -- (14,1.5);
\draw[thick] (11.5,.5) -- (14,.5);
\draw[thick] (12,1) -- (11,0);
\draw[thick] (12,1) -- (11,2);
\draw[thick, fill=myorange, rounded corners=2pt] (11.5,1.5) rectangle (12.5,0.5);
\draw[thick] (12,1.3) -- (12.3,1.3) -- (12.3,1.0);
\draw[thick] (14,1) -- (15,2);
\draw[thick] (14,1) -- (15,0);
\draw[thick, fill=mygreen, rounded corners=2pt] (13.5,1.5) rectangle (14.5,.5);
\draw[thick] (13.7,1) -- (13.7,0.7) -- (14,0.7);
\Text[x=14,y=-.35]{}
\end{tikzpicture}\,.
\end{align}

It is easy to show (see Ref.~\cite{bertini2020operator}) that the unitarity of the gates causes the operators ${\mathcal T}_{1,x}$ and ${\mathcal T}_{2,x}$ to be \emph{contracting}, namely all their eigenvalues lie on or within the unit circle in the complex plane ($|\lambda|\leq 1$). Moreover, using both \eqref{eq:du1} and \eqref{eq:du2} one can immediately identify the following $x+1$ independent eigenvectors of eigenvalue $\lambda=1$
\begin{align}
\ket{\tilde 0}_x &=\frac{1}{d^{x}}
\begin{tikzpicture}[baseline=(current  bounding  box.center), scale=0.45]
\draw[thick] (0,0) parabola bend (2,-1.5) (4,0);
\draw[thick] (0.5,0) parabola bend (2,-1.25) (3.5,0);
\draw[thick] (1,0) parabola bend (2,-1) (3,0);
\node[] at (2,0) {...};
\end{tikzpicture}\,, & &\notag\\
\ket{\tilde 1}_x &=\frac{1}{d^{(x-1)}}
\begin{tikzpicture}[baseline=(current  bounding  box.center), scale=0.45]
\draw[thick] (8,0)--(8,-1.25);
\draw[thick] (12,0) -- (12,-1.25);
\draw[thick] (8.5,0) parabola bend (10,-1.25) (11.5,0);
\draw[thick] (9,0) parabola bend (10,-1) (11,0);
\path (8,-1.25) pic[rotate=0] {cross=0.1};
\path (12,-1.25) pic[rotate=0] {cross=0.1};
\node[] at (10,0) {...};
\end{tikzpicture}\,, \label{eq:rainbow_state}\\
& \ldots\notag\\
\ket{\tilde x}_x &=
\begin{tikzpicture}[baseline=(current  bounding  box.center), scale=0.45]
\draw[thick] (16,-2.5)--(16,-3.75);
\draw[thick] (17,-2.5) -- (17,-3.75);
\draw[thick] (18,-2.5) -- (18,-3.75);
\draw[thick] (21,-2.5) -- (21,-3.75);
\draw[thick] (22,-2.5) -- (22,-3.75);
\draw[thick] (23,-2.5) -- (23,-3.75);
\node at (19.5,-3) {...};
\path (16,-3.75) pic[rotate=0] {cross=0.1};
\path (17,-3.75) pic[rotate=0] {cross=0.1};
\path (18,-3.75) pic[rotate=0] {cross=0.1};
\path (21,-3.75) pic[rotate=0] {cross=0.1};
\path (22,-3.75) pic[rotate=0] {cross=0.1};
\path (23,-3.75) pic[rotate=0] {cross=0.1};
\end{tikzpicture}\,.\notag
\end{align}
for reasons made obvious by their tensor network representation these states have been named `rainbow states' in Ref.~\cite{bertini2020operator}. These vectors are linearly independent but not orthogonal. Orthonormalising them, we obtain the set $\{ \ket{y}_x\}_{y=0,\ldots,x}$ of orthonormal rainbow states (ORS)s defined as  
\begin{equation} \label{eq:ors}
    \ket{y}_x=
    \begin{cases}
     \displaystyle  \frac{d}{\sqrt{d^2-1}}(\ket{\tilde y}_x-\frac{1}{d} \ket{\widetilde{y+1}}_x), & \text{if}\ y < x \\
    \displaystyle   \ket{\tilde x}_x, & \text{if}\ y=x \\
      
    \end{cases}.
  \end{equation}
Whilst ORSs are always eigenvectors of ${\mathcal T}_{1,x}$ and ${\mathcal T}_{2,x}$ and always satisfy $|\lambda|=1$, they are not necessarily the sole eigenvectors with this property. That is, in principle, other eigenvectors with $|\lambda|=1$ may also exist. The completely chaotic subclass is the subset of dual-unitary gates for which the rainbow states are the \emph{only} eigenvectors of ${\mathcal T}_{1,x}$ and ${\mathcal T}_{2,x}$ with $|\lambda|=1$ for all values of $x$. Even though it is easy to concoct dual-unitary gates which are not completely chaotic (e.g.\ the SWAP gate), thorough numerical analyses have shown that for $d=2$ the generic case is indeed completely chaotic~\cite{bertini2020operator}. To rephrase, a randomly generated dual-unitary gate of the parametrisation in Eq.~\eqref{eq:parametrisation}  will be found to be completely chaotic with probability 1. Here we assume that the same remains true also for ${d>2}$. 

Note that completely chaotic dual-unitary circuits  have been observed to yield maximal growth of several indicators of quantum chaos and scrambling~\cite{bertini2020operator, bertini2020scrambling, claeys2020maximum} and can therefore be regarded as the ``most chaotic'' class of local quantum circuits.

\subsection{Solvable Initial States}
\label{sec:solvableIS}

As initial states for the evolution \eqref{eq:quench} we consider the family of ``solvable''  MPSs introduced in Ref.~\cite{piroli2020exact} (see also Ref.~\cite{bertini2019entanglement}). This family is composed by ``dimerised'' MPSs of the form 
\be
\!\!\!\ket{\Psi_0}=\!\!\!\!\!\!\!\! \sum_{s_1,\ldots,s_{2L}=1}^{d} \!\!\!\!\!\!{\rm tr}[A_{s_1s_2} \cdots A_{s_{2L-1}s_{2L}}] \ket{s_1}\otimes\cdots\otimes\ket{s_{2L}}\!,
\label{eq:psi0}
\ee
where $\{A_{s_1s_2}\}$ are $\chi\times \chi$ matrices --- $\chi$ is commonly referred to as ``bond dimension'' --- fulfilling the following two conditions 
\begin{itemize}
\item[(i)] Their matrix elements are specified by a $\chi d\times \chi d$ unitary matrix $W$ as follows 
\be
\braket{\alpha|A_{s r}|\beta}=\frac{1}{\sqrt d}\bra{s}\otimes\braket{\alpha|W|r}\otimes\ket{\beta},
\label{eq:Wmatrix}
\ee
where $\alpha,\beta\in\{0,\ldots,\chi-1\}$, $r,s\in\{0,\ldots,d-1\}$, and we denoted by 
\be
\{\ket{\alpha},\qquad \alpha \in\{ 0 ,\ldots, \chi-1\}\},
\ee
a basis of $\mathbb C^\chi$. 
\item[(ii)] The $\chi^2\times \chi^2$ ``state transfer matrix'' (STM)
\be
\tau = \sum_{s_1,s_{2}=1}^{d} A^*_{s_1s_2}\otimes A^{\phantom{\dag}}_{s_1s_2}
\label{eq:tau}
\ee
has unique maximal eigenvalue.  
\end{itemize} 
Note that if $\{A_{s_1s_2}\}$ fulfil (i) the maximal magnitude of the eigenvalues of $\tau$ is one and 
\be
\ket{\mcirc_\chi} = \frac{1}{\sqrt \chi}\sum_{\alpha=1}^\chi \ket{\alpha}\otimes\ket{\alpha}\,,
\ee
is a (left and right) fixed point, namely 
\be
\bra{\mcirc_\chi} \tau  = \bra{\mcirc_\chi}, \qquad \tau \ket{\mcirc_\chi} = \ket{\mcirc_\chi},
\label{eq:statefixedpoint}
\ee
where $\bra{\mcirc_\chi}=\ket{\mcirc_\chi}^\dag$. In principle there could be more unit magnitude eigenvalues of the STM, however, this requires fine tuning. Indeed, while \eqref{eq:statefixedpoint} follows directly from the unitarity of $W$, asking for additional unit-magnitude eigenvalues poses further constrains on the matrix. This means that we expect (ii) to be satisfied for generic choices of the unitary matrix $W$ (i.e., apart from a zero-measure set), which is in accord with explicit constructions of $\tau$ for $d=2$ and $\chi=1,2$~\cite{piroli2020exact}. Note that (i) and (ii) imply 
\be
\lim_{x\to\infty} \tau^x=\ket{\mcirc_\chi}\!\!\bra{\mcirc_\chi}\,.
\label{eq:infinitepowertau}
\ee

The states \eqref{eq:psi0} are conveniently represented diagrammatically as follows
\begin{align}
\!\!\!\!\!\!\!\ket{\Psi_0(W)}\!\!=\!\! \frac{1}{d^{\frac L 2}}
\begin{tikzpicture}[baseline=(current  bounding  box.center), scale=1]
\draw[very thick] (3.4,0) -- (1,0);
\draw[very thick] (-2.05,0) -- (.25,0);
\draw[very thick] (-2.05,0.15) arc (90:270:0.075);
\draw[very thick] (3.4,0.15) arc (90:-90:0.075);
\draw[very thick, dotted] (0.25,0) -- (1,0);
\foreach \i in {0,1.35}
{
\draw[thick] (-2+\i,0.5) -- (-1.5+\i,0);
\draw[thick] (-1.5+\i,0) -- (-1+\i,0.5);
\draw[ thick, fill=mygreyred, rounded corners=2pt] (-1.5-0.35+\i,0.2-0.25) rectangle (-1.5+0.35+\i,0.2+0.2);
\draw[thick] (-1.5+0.1+\i,0.15+.18)-- (-1.35+0.1+\i,0.15+.18) -- (-1.35+0.1+\i,0+.18);}
\foreach \i in {3.7-0.75,5.05-.75}
{
\draw[thick] (-2+\i,0.5) -- (-1.5+\i,0);
\draw[thick] (-1.5+\i,0) -- (-1+\i,0.5);
\draw[ thick, fill=mygreyred, rounded corners=2pt] (-1.5-0.35+\i,0.2-0.25) rectangle (-1.5+0.35+\i,0.2+0.2);
\draw[thick] (-1.5+0.1+\i,0.15+.18)-- (-1.35+0.1+\i,0.15+.18) -- (-1.35+0.1+\i,0+.18);}
\Text[x=-1.5,y=-.35]{}
\end{tikzpicture}\,,
\label{eq:psi0diagram}
\end{align}
where we reported the explicit dependence on $W$ and represented it as 
\begin{align}
\label{eq:W}
W & =
\begin{tikzpicture}[baseline=(current  bounding  box.center), scale=1]
\def\eps{0.5};
\draw[thick] (-2,0.5) -- (-1.5,0);
\draw[thick] (-1.5,0) -- (-1,0.5);
\draw[very thick] (-2,0) -- (-1,0);
\draw[ thick, fill=mygreyred, rounded corners=2pt] (-1.5-0.35,0.2-0.25) rectangle (-1.5+0.35,0.2+0.2);
\draw[thick] (-1.5+0.1,0.15+.18)-- (-1.35+0.1,0.15+.18) -- (-1.35+0.1,0+.18);
\Text[x=-1.5,y=-.35]{}
%\Text[x=-2.25,y=0]{$\alpha$}
%\Text[x=-0.75,y=0]{$\beta$}
%\Text[x=-2.25,y=0.5]{$s$}
%\Text[x=-0.75,y=0.5]{$r$}
\end{tikzpicture}\,.
\end{align}
Because of the unitarity condition (i) we have 
\begin{align}
&\qquad\begin{tikzpicture}[baseline=(current  bounding  box.center), scale=1]
\def\eps{0.5};
\draw[very thick] (-2,0) -- (.25,0);
\draw[thick, rounded corners=2pt] (-1.5,0) -- (-1,0.5) -- (-.75,0.5) -- (-1.5+1.25,0);
\draw[thick] (-1.5+1.25,0) -- (-1+1.25,0.5);
\draw[thick] (-2,0.5) -- (-1.5,0);
\draw[ thick, fill=myblue, rounded corners=2pt] (-1.5-0.35,0.2-0.25) rectangle (-1.5+0.35,0.2+0.2);
\draw[thick] (-1.75,0.17) -- (-1.75,0.02)-- (-1.6,0.02) ;
%\draw[thick] (-1.5+0.1,0.15+.18)-- (-1.35+0.1,0.15+.18) -- (-1.35+0.1,0+.18);
\Text[x=-1.5,y=-.35]{}
\draw[ thick, fill=myred, rounded corners=2pt] (-1.5-0.35+1.25,0.2-0.25) rectangle (-1.5+0.35+1.25,0.2+0.2);
\draw[thick] (-1.5+0.1+1.25,0.15+.18)-- (-1.35+0.1+1.25,0.15+.18) -- (-1.35+0.1+1.25,0+.18);
\Text[x=-1.5,y=-.35]{}
\end{tikzpicture}=\begin{tikzpicture}[baseline=(current  bounding  box.center), scale=1]
\def\eps{0.5};
\draw[very thick] (-2,0) -- (.25,0);
\draw[thick, rounded corners=2pt] (-2,0.5) -- (-1+1.25,0.5);
\Text[x=-1.5,y=-.35]{}
\end{tikzpicture}\,,\label{eq:stateuni1}\\
&\qquad\begin{tikzpicture}[baseline=(current  bounding  box.center), scale=1]
\def\eps{0.5};
\draw[very thick] (-2,0) -- (.25,0);
\draw[thick, rounded corners=2pt] (-1.5,0) -- (-1,0.5) -- (-.75,0.5) -- (-1.5+1.25,0);
\draw[thick] (-1.5+1.25,0) -- (-1+1.25,0.5);
\draw[thick] (-2,0.5) -- (-1.5,0);
\draw[ thick, fill=myred, rounded corners=2pt] (-1.5-0.35,0.2-0.25) rectangle (-1.5+0.35,0.2+0.2);
\draw[thick] (-1.5+0.1,0.15+.18)-- (-1.35+0.1,0.15+.18) -- (-1.35+0.1,0+.18);
\Text[x=-1.5,y=-.35]{}
\draw[ thick, fill=myblue, rounded corners=2pt] (-1.5-0.35+1.25,0.2-0.25) rectangle (-1.5+0.35+1.25,0.2+0.2);
%\draw[thick] (-1.5+0.1+1.25,0.15+.18)-- (-1.35+0.1+1.25,0.15+.18) -- (-1.35+0.1+1.25,0+.18);
\draw[thick] (-1.75+1.25,0.17) -- (-1.75+1.25,0.02)-- (-1.6+1.25,0.02) ;
\Text[x=-1.5,y=-.35]{}
\end{tikzpicture}=\begin{tikzpicture}[baseline=(current  bounding  box.center), scale=1]
\def\eps{0.5};
\draw[very thick] (-2,0) -- (.25,0);
\draw[thick, rounded corners=2pt] (-2,0.5) -- (-1+1.25,0.5);
\Text[x=-1.5,y=-.35]{}
\end{tikzpicture}\,,
\label{eq:stateuni2}
\end{align}
where we introduced
\be
W^\dag  =
\begin{tikzpicture}[baseline=(current  bounding  box.center), scale=1]
\def\eps{0.5};
\draw[thick] (-2,0.5) -- (-1.5,0);
\draw[thick] (-1.5,0) -- (-1,0.5);
\draw[very thick] (-2,0) -- (-1,0);
\draw[ thick, fill=mygreyblue, rounded corners=2pt] (-1.5-0.35,0.2-0.25) rectangle (-1.5+0.35,0.2+0.2);
%\draw[thick] (-1.5+0.1,0.15+.18)-- (-1.35+0.1,0.15+.18) -- (-1.35+0.1,0+.18);
\Text[x=-1.5,y=-.35]{}
%\Text[x=-2.25,y=0]{$\alpha$}
%\Text[x=-0.75,y=0]{$\beta$}
%\Text[x=-2.25,y=0.5]{$s$}
%\Text[x=-0.75,y=0.5]{$r$}
\draw[thick] (-1.75,0.17) -- (-1.75,0.02)-- (-1.6,0.02) ;
\end{tikzpicture}\,.
\ee
In this diagrammatic representation the STM is conveniently depicted as 
\be
\tau = \frac{1}{d}\,\,\begin{tikzpicture}[baseline=(current  bounding  box.center), scale=1]
\def\eps{1.};
\draw[thick, rounded corners]  (-1.5,0.35+\eps) -- (-2,-0.15+\eps) -- (-2,0.5) -- (-1.5,0);
\draw[thick, rounded corners] (-1.5,0) -- (-1,0.5) -- (-1,-0.15+\eps) -- (-1.5,0.35+\eps);
\draw[very thick] (-2,0+\eps+.35) -- (-1,0+\eps+.35);
\draw[very thick] (-2,0) -- (-1,0);
\draw[ thick, fill=myblue, rounded corners=2pt] (-1.5-0.35,0.2-0.25+\eps) rectangle (-1.5+0.35,0.2+0.2+\eps);
%\draw[thick] (-1.75,0.17+\eps) -- (-1.75,0.02+\eps)-- (-1.6,0.02+\eps) ;
\draw[ thick, fill=mygreyred, rounded corners=2pt] (-1.5-0.35,0.2-0.25) rectangle (-1.5+0.35,0.2+0.2);
\draw[thick] (-1.5+0.1,0.15+.18)-- (-1.35+0.1,0.15+.18) -- (-1.35+0.1,0+.18);
\draw[thick] (-1.5+0.1,0.15+.18+\eps)-- (-1.35+0.1,0.15+.18+\eps) -- (-1.35+0.1,0+.18+\eps);
\Text[x=-1.5,y=-.35]{}
\end{tikzpicture},
\label{eq:STM}
\ee
while its left and right fixed points are represented as 
\be
\bra{\mcirc_\chi} = \frac{1}{\sqrt \chi}\,\,\begin{tikzpicture}[baseline=(current  bounding  box.center), scale=1]
\def\eps{1.};
\draw[very thick, rounded corners] (-1.5,0) -- (-2,0) -- (-2,0.35+\eps) -- (-1.5,0.35+\eps);
\end{tikzpicture},
\qquad\qquad\ket{\mcirc_\chi} = \frac{1}{\sqrt \chi}\,\,\begin{tikzpicture}[baseline=(current  bounding  box.center), scale=1]
\def\eps{1.};
\draw[very thick, rounded corners]  (-2,0) -- (-1.5,0) -- (-1.5,0.35+\eps) -- (-2,0.35+\eps);
\end{tikzpicture}\,.
\label{eq:STMfixedpoint}
\ee
We then see that \eqref{eq:statefixedpoint} follows from \eqref{eq:stateuni1}--\eqref{eq:stateuni2} by connecting the thin lines and ``bending'' the red gate above the blue one.

Note that the simplest set $\{A_{s_1s_2}\}$ fulfilling (i) and (ii) is obtained for $\chi=1$ and 
\be
\begin{tikzpicture}[baseline=(current  bounding  box.center), scale=1]
\def\eps{0.5};
%\draw[very thick] (-.75,0) -- (.25,0);
\draw[thick, rounded corners=2pt] %(-1.5,0) -- (-1,0.5) -- 
(-.75,0.5) -- (-1.5+1.25,0);
\draw[thick] (-1.5+1.25,0) -- (-1+1.25,0.5);
\Text[x=-1.5,y=-.35]{}
\draw[ thick, fill=mygreyred, rounded corners=2pt] (-1.5-0.35+1.25,0.2-0.25) rectangle (-1.5+0.35+1.25,0.2+0.2);
\draw[thick] (-1.5+0.1+1.25,0.15+.18)-- (-1.35+0.1+1.25,0.15+.18) -- (-1.35+0.1+1.25,0+.18);
\Text[x=-1.5,y=-.35]{}
\end{tikzpicture}\,\,=\,\,\begin{tikzpicture}[baseline=(current  bounding  box.center), scale=1]
\def\eps{0.5};
\draw[thick, rounded corners=2pt] (-.75,0.5) -- (-1.5+1.,0.25) -- (-1.5+1.5,0.25) -- (-1+1.25,0.5);
\Text[x=-.5,y=-.35]{}
\end{tikzpicture}\,,
\label{eq:Bell}
\ee
which corresponds to a state \eqref{eq:psi0diagram} written as a product of Bell pairs
\begin{align}
\!\!\!\!\!\ket{\Psi_{\rm Bell}}= \frac{1}{d^{\frac L 2}}
\begin{tikzpicture}[baseline=(current  bounding  box.center), scale=1]
\draw[thick, dotted] (1.65,0.25) -- (2.1,0.25);
\foreach \i in {0,1.35}
{
\draw[thick, rounded corners=2pt] (-.75+\i,0.5) -- (-1.5+1.+\i,0.25) -- (-1.5+1.5+\i,0.25) -- (-1+1.25+\i,0.5);}
\foreach \i in {3.7-0.75,5.05-.75}
{
\draw[thick, rounded corners=2pt] (-.75+\i,0.5) -- (-1.5+1.+\i,0.25) -- (-1.5+1.5+\i,0.25) -- (-1+1.25+\i,0.5);}
\Text[x=-.5,y=-0.1]{}
\end{tikzpicture}\!.
\label{eq:psi0Bell}
\end{align}
Any solvable state with $\chi=1$ is equivalent to the Bell pair state up to a product $\mathbb L$ of local unitary matrices
\be
\mathbb L=\prod_{x \,\in \mathbb Z_L} u_x,\qquad u_x =\Pi_{2L}^{2x} (u \otimes \1^{\otimes 2L-1})\Pi_{2L}^{-2x}.
\ee
The latter can always be absorbed in the dual-unitary gates performing the time evolution because local unitaries do not change \eqref{eq:du1} and \eqref{eq:du2}.  

Another simple subclass of \eqref{eq:psi0diagram} is that obtained for $\chi=d$ and $W$ \emph{dual-unitary}. We represent them as
\be
\ket{\Psi_{\rm du}(W)}=\frac{1}{d^{\frac L 2}}\,\,
\begin{tikzpicture}[baseline=(current  bounding  box.center), scale=1]
\draw[thick, dotted] (1.75,0.75) -- (2.1,0.75);
\foreach \i in {1.35}
{
\draw[thick, rounded corners=2pt] (-.75+\i,0.5) -- (-1.5+1.+\i,0.25) -- (-1.5+1.5+\i,0.25) -- (-1+1.25+\i,0.5);}
\foreach \i in {0}
{
\draw[thick, rounded corners=2pt] %(-.75+\i,0.5) -- 
(-1.5+1.25+\i,0.25) -- (-1.5+1.5+\i,0.25) -- (-1+1.25+\i,0.5);}
\foreach \i in {3.7-0.75}
{
\draw[thick, rounded corners=2pt] (-.75+\i,0.5) -- (-1.5+1.+\i,0.25) -- (-1.5+1.5+\i,0.25) -- (-1+1.25+\i,0.5);}
\foreach \i in {5.05-.75}
{
\draw[thick, rounded corners=2pt] (-.75+\i,0.5) -- (-1.5+1.+\i,0.25) -- (-1.5+1.25+\i,0.25);}
% -- (-1+1.25+\i,0.5);}
\path (3.375,.3) pic[rotate=0,scale=1.5] {U};
\path (0.425,.3) pic[rotate=0,scale=1.5] {U};v
\Text[x=.5,y=-0.1]{}
\draw[thick] (-.25,0.4) arc (90:270:0.075);
\draw[thick] (4.05,0.4) arc (90:-90:0.075);
\end{tikzpicture},
\label{eq:duDef}
\ee
where we used $W$ of the shape \eqref{eq:U} to stress that it is dual-unitary. We remark that, due to the dual-unitarity of $W$, for these states the STM is equal to a projector
\be
\tau=\ket{\mcirc_d}\!\!\bra{\mcirc_d}\,.
\label{eq:DUdefSTM}
\ee
Moreover, note that the class \eqref{eq:duDef} can be obtained from the Bell-pair state \eqref{eq:psi0Bell} by applying half a time-step of dual-unitary evolution and shifting the lattice by one. For this reason we will refer to these states as dual-unitary deformations of Bell-pair states.

\section{Results and Discussion}
\label{sec:results}

Before embarking on the survey of our results we note that --- as proved in Ref.~\cite{piroli2020exact} and reviewed in Sec.~\ref{sec:rhoBm} --- in dual-unitary circuits evolving from the solvable states \eqref{eq:psi0}, the reduced density matrix $\rho_B(t)$ approaches the infinite temperature state at time $t=|B|/2$. This is a consequence of the purely Markovian dynamics of subsystems in this setting~\cite{piroli2020exact}. Indeed, the rest of the system acts as a Markovian bath attached to the boundary of any given finite block of qudits. This causes each qubit to thermalise as soon as it is reached by the information about the boundary (which travels at the maximal speed ${v_{\rm max}=1}$). As a consequence, all operator entanglement entropies \eqref{eq:Renyigen} vanish for $2t\geq|B|$ and we can restrict to $2t<|B|$. In the latter regime we find two main results (see Tables~\ref{tab:SBell} and \ref{tab:SCC} for a summary).

\begin{table}[t]
\begin{tabular}{|| l || l ||}
\hline
{\rm Regime}  &  {\rm Entanglement}\\ 
\hline
$0<2t<\ell-x$ & $\min(2 t,x) \log d^2$ \\
\hline
 & {\rm Gate Dependent}\\
$\ell-x\leq 2t <\ell$ & {\rm SWAP gate}: Eq.~\eqref{eq:SSWAP} \\
 & {\rm Completely Chaotic gates}: Eq.~\eqref{eq:Schaotic}\\
\hline
\end{tabular}
\caption{Exact behaviour of the entanglement of $\rho^\beta_B(t)$ with $\beta\in\mathbb R^+$ after a quench from a Bell pair state. Here $\ell = |B| = |A\bar A|$ and $x=|A|$.}
\label{tab:SBell}
\end{table}

\begin{table}[b]
\begin{tabular}{|| l || l ||}
\hline
{\rm Regime}  &  {\rm Entanglement}\\ 
\hline
$0<2t<\ell-x$ & Eq.~\eqref{eq:Sunigenstate} \\
\hline
 & {\rm Gate Dependent}\\
$\ell-x\leq 2t <\ell$ & {\rm SWAP gate}: Eq.~\eqref{eq:SSWAPgen} \\
 & {\rm Completely Chaotic gates}: Eq.~\eqref{eq:Schaogen}\\
\hline
\end{tabular}
\caption{Asymptotic behaviour of the entanglement of $\rho^m_B(t)$ with $m\in\mathbb N$ after a quench from a generic solvable state. Here $\ell = |B| = |A\bar A|$ and $x=|A|$.}
\label{tab:SCC}
\end{table}

\begin{figure*}
\centering
\includegraphics[width=0.75\textwidth]{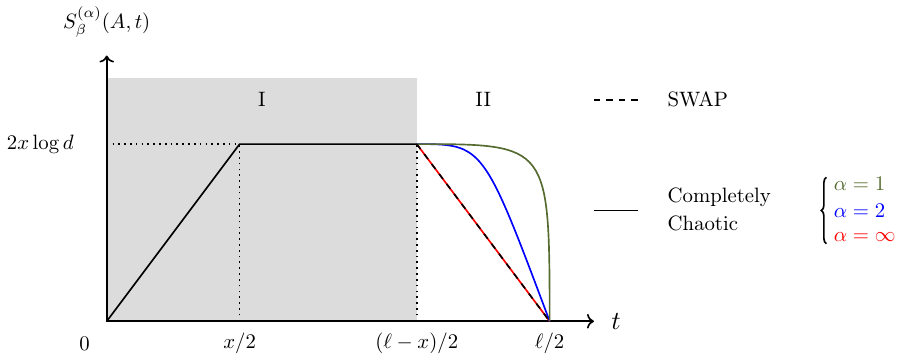}
\caption{Operator R\'enyi entropies of the $\beta$-th power of the reduced density matrix (cf.~\eqref{eq:Renyigen}) after a quench from the Bell state \eqref{eq:psi0Bell}. The result is independent of $\beta$. Moreover, for ${t \leq (\ell-x)/2}$ the evolution is \emph{universal}, i.e., it is independent of the specific dual-unitary gate generating the dynamics and of the R\'enyi index $\alpha$. Instead, for ${(\ell-x)/2\leq t \leq \ell/2}$ the behaviour depends on the specific dual-unitary gate generating the dynamics and on the R\'enyi index.}
    \label{fig:graph}
\end{figure*}

\subsection{Exact results for Bell pairs}
 Our first main result is an exact expression for $S^{(n)}_{m}(A,t)$ when evolving from the Bell-pair state $\ket{\Psi_{\rm Bell}}$, see Fig.~\ref{fig:graph}. Specifically, we identify two physically different regimes:
\begin{itemize} 
\centering
\item[(I)]    $0<2t<\ell-x$;
\item[(II)] $\ell-x\leq 2t <\ell$;
\end{itemize}
where, to ease the notation, we set  
\be
\ell\equiv |B|,\qquad\qquad x\equiv |A|\,,
\label{eq:ellx}
\ee
and, without loss of generality, we assumed $x\leq\ell/2$. From now on we always use this shorthand notation unless we explicitly state otherwise. 

In the first regime we find a universal growth of entanglement followed by saturation (cf.\ Secs.~\ref{sec:n=2earlytimes}, \ref{sec:n=2midtimes} and~\ref{sec:highern})
 \be
S^{(n)}_{m}(A,t)\bigr |_{\rm Bell} = \min(2 t,x) \log d^2,
\label{eq:Suni}
 \ee
 for any dual-unitary local gate $U$, R\'enyi index $n$, and power $m$ of the reduced density matrix. 
 
In the second regime we find that the operator entanglement decays but its dynamics depends on the specific dual-unitary gate considered and on the R\'enyi index $n$, while it remains completely independent of $m$. In particular, when evolving with the SWAP gate we find  
 \be
S^{(n)}_{m}(A, t)\bigr |_{\rm SWAP, Bell} = (\ell-2 t) \log d^2,
\label{eq:SSWAP}
 \ee
so that the barrier is symmetric around $t=\ell/2$: when the most symmetric bipartition becomes completely entangled the operator entanglement starts to decay. On the other hand, for gates in the completely chaotic class we can access the entanglement entropies in the scaling limit 
\be
\limsc\!\!:\quad\ell,t\to\infty,\qquad \ell-2t\equiv k \leq  x  \,\,\, {\rm fixed}.
\label{eq:scaling1}
\ee
In this limit we find (cf.\ Secs.~\ref{sec:n=2chaos} and \ref{sec:highern}) 
 \begin{align}
&\limsc S^{(n)}_{m}(A, t)\bigr |_{\rm cc, Bell}\notag\\
&\qquad\qquad=\!\frac{\log[ d^{-2n k} \!\!+d^{2(1-n)x}(1-d^{-2 k}))]}{1-n}\,.
\label{eq:Schaotic}
 \end{align}
Eq.~\eqref{eq:Schaotic} implies that all R\'enyi entropies display a longer plateau for completely chaotic gates than for the SWAP gate. This supports the idea --- put forward in Ref.~\cite{wang2019barrier} --- that the extension of the plateaux of the operator R\'enyi entropies can be used as an indicator of quantum chaos. Moreover, from Eq.~\eqref{eq:Schaotic} we also see that the time when a given operator R\'enyi entropy starts leaving the plateau depends on the R\'enyi index. For large $k$ and $x$ we can estimate this time by finding when $d^{-2n k}$ dominates in the argument of the logarithm in \eqref{eq:Schaotic}. Namely
\be
 d^{-2n k} \geq d^{2(1-n)x} \quad \Rightarrow \quad t \geq t^*(n)=\frac{\ell -x}{2} + \frac{x}{2n}\,.
\ee
From this formula we immediately see that the min entropy (${n=\infty}$) behaves always like the SWAP gate, while the deviation increases when decreasing $n$. This signals that the ``chaotic'' behaviour is due to the sub-leading eigenvalues of the superoperatorial reduced density matrix ${\rm P}_{\!A,\beta}(t)$ (cf.\ Eq.~\eqref{eq:rhosuper}). A similar behaviour has also been observed when considering the entanglement of a local operator in completely chaotic dual-unitary circuits~\cite{bertini2020operator}.  

To find the cases of maximal deviation from the SWAP behaviour we note that Eq.~\eqref{eq:Schaotic} (as well as \eqref{eq:Suni} and \eqref{eq:SSWAP}) can be directly analytically continued by replacing $(n,m)$ with $(\alpha,\beta)\in\mathcal D$ (cf.~\eqref{eq:D})~\cite{Note1}.\footnotetext[1]{Indeed, Carlson's Theorem~\cite{rubel1956necessary} guarantees that the only analytic continuation of \eqref{eq:Suni}, \eqref{eq:SSWAP}, and \eqref{eq:Schaotic} to $\mathcal D $
which is bounded at infinity (and hence compatible with \eqref{eq:Renyigen}).} In particular, this means that we can compute   
\be
\!\!\!\!\lim_{\alpha\to1^+}\!\!\limsc\, S^{(\alpha)}_{\beta}\!(A,t)\bigr |_{\rm cc, Bell} = x \log d^2 + \frac{(k-x)}{d^{2k}} \log d^2\!.
\label{eq:SvNchaos}
\ee
This expression shows that the von Neumann entropy remains exponentially close (in $\ell-2t$) to the plateau $x\log d^2$ (see Fig.~\ref{fig:graph}). In fact \eqref{eq:SvNchaos} becomes discontinuous in the scaling limit $x,k\to\infty$ with $k/x$ fixed, more precisely
\be
\lim_{\substack{x,k\to\infty\\ k/x=\zeta}}\lim_{\alpha\to1^+}\!\!\limsc\, \frac{S^{(\alpha)}_{\beta}\!(A,t)}{x} = \begin{cases}
 0 & \zeta=0\\
 \log d^2 & 0<\zeta \leq 1
\end{cases}.
\ee 

Eqs.~\eqref{eq:Suni} and \eqref{eq:Schaotic} (but not \eqref{eq:SSWAP}) are immediately generalised to the case when the initial state is $\ket{\Psi_{{\rm du}}(W)}$, i.e.\ a dual-unitary deformation of the Bell-pair state (cf.~\eqref{eq:duDef}). For these initial states the formulae continue to hold with the replacement
\be
t\mapsto t+\frac{1}{2}.
\label{eq:replacement}
\ee
This can be understood recalling that these states are nothing but the Bell-pair state after half a time step of dual-unitary evolution (cf.\ the discussion after Eq.~\eqref{eq:DUdefSTM}). In the case of \eqref{eq:SSWAP} this argument can only be applied to the special case of $\ket{\Psi_{{\rm du}}(\rm SWAP)}$, i.e. when the Bell-pair state is deformed with the SWAP gate. In this case we have 
 \be
S^{(n)}_{m}(A, t)\bigr |_{\rm SWAP, SWAP} = (\ell-2 t) \log d^2-\log d^2.
\label{eq:SSWAPdu}
 \ee
Note that, even though $\ket{\Psi_{{\rm du}}(\rm SWAP)}$ has higher entanglement than $\ket{\Psi_{\rm Bell}}$, at any fixed time in the ``disentangling'' regime (II) \eqref{eq:SSWAPdu} is \emph{smaller} than \eqref{eq:SSWAP}. The difference between the two is precisely the operator entanglement of
\be
\!\!\!\!\rho_B(0)\bigr |_{\rm SWAP}={\rm tr}_{\bar B}[\ket{\Psi_{{\rm du}}(\rm SWAP)}\!\!\bra{\Psi_{{\rm du}}(\rm SWAP)}]\,.
\ee
This can be explained noting that when we evolve from $\ket{\Psi_{{\rm du}}(\rm SWAP)}$ we reach the maximal possible operator entanglement before and hence the disentangling phase starts earlier. Note that this also happens in the completely chaotic case for $t\geq t^*(n)$ where the difference reads as 
\be
\frac{n}{n-1}\log d^2\,.
\ee
Our exact results \eqref{eq:Suni}, \eqref{eq:SSWAP}, and \eqref{eq:Schaotic}  for a specific class of lattice models can be compared with the findings of Refs.~\cite{dubail2017entanglement, wang2019barrier}, which studied the dynamics of \eqref{eq:Renyigen} after a quench from a low entangled initial state in CFT (both rational and holographic) and in random unitary circuits with infinite-dimensional local Hilbert space. The qualitative behaviour observed in these systems in the regime (I) agrees with our exact result \eqref{eq:Suni}. One can make this statement more precise by defining the rescaled quantities 
\be
\mathcal S^{(n)}_{m}(z) \equiv \frac{S^{(n)}_{m}(A,z x/v_{{\rm E},n})}{x s_n},\qquad z=\frac{v_{{\rm E}, n} t}{x},
\label{eq:Snrescaled}
\ee  
where $s_n$ is the entropy density of the stationary state and $v_{E,n}$ is the slope of the the $n$-th state R\'enyi entropy divided by $s_n$ (for us ${s_n=\log d}$ and $v_{E,n}=1$)~\cite{Note2}. \footnotetext[2]{In general one has also to rescale all lengths by a factor $(2-{v_{\rm B}}/{v_{\rm E,n}})$, where $v_{\rm B}$ is the velocity characterising the spreading of operators~\cite{wang2019barrier}.} At the leading order, the rescaled quantities \eqref{eq:Snrescaled} computed in the aforementioned systems show exactly the behaviour one finds from \eqref{eq:Suni}. This gives further evidence in support of the universal behaviour of the operator entanglement in the regime (I)~\cite{Note3}.\footnotetext[3]{Note that, however, in general integrable models the scaling form is expected to be different as one can see by employing the quasi-particle picture for the entanglement spreading~\cite{calabrese2005evolution}.} In the regime (II), instead, the different systems considered in Ref.~\cite{dubail2017entanglement, wang2019barrier} give different results. In particular the result for rational CFT, which is expected to describe integrable-like behaviour, agrees with our non-interacting result~\eqref{eq:SSWAP}, while that for holographic CFTs (only available for $\alpha=1$ and $\beta=1/2$) agrees with \eqref{eq:SvNchaos}. Interestingly, however, the result for random unitary circuits with $d\to\infty$ is rather different from ours. Indeed, even if it differs from the non-interacting result~\eqref{eq:SSWAP}, it does not show any dependence on the R\'enyi index. Note that for \eqref{eq:Schaotic} this is not the case even in the $d\to\infty$ limit. Note that $S_m^{(2)}(A,t)$ computed for random unitary circuits agrees with \eqref{eq:Schaotic} in the scaling limit. This, however, seems just a coincidence.

\subsection{Exact asymptotic results for generic solvable states}
 
Our second main result is to show that \eqref{eq:Suni}, \eqref{eq:SSWAP} and \eqref{eq:Schaotic} describe the leading order behaviour of the operator R\'enyi entropies after a quench from \emph{generic} solvable initial states~\eqref{eq:psi0}. In particular, in Secs.~\ref{sec:purity} and \ref{sec:highern} we find 
\be
\!\!\!\!\lim_{\ell\to\infty}S^{(n)}_{m}(A, t) = \begin{cases}
\displaystyle 2t \log d^2 + \frac{\log \eta_{x-2t,n}}{1-n} & 2t \leq x \\
x \log d^2 & 2t > x\\
\end{cases}.
\label{eq:Sunigenstate}
\ee
Here we introduced the following tensor network depending on the initial state only  
\be
\eta_{x,m} =\frac{1}{\chi^m d^{2mx}}\begin{tikzpicture}[baseline=(current  bounding  box.center), scale=0.5]

\foreach \i in {1,...,4}
{
\draw[ thick] (2*\i+2.5-1,6.65) arc (0:180:0.15);
\draw[ thick] (2*\i+2-1.5,6.65) arc (0:-180:-0.15);
\draw[ thick] (2*\i+2-1.5,-1.65) arc (0:180:-0.15);
\draw[ thick] (2*\i+2-0.5,-1.65) arc (0:-180:0.15);
}

\foreach \j in {0,2,6}
{\foreach \i in {1,...,2}{
\path (1+2*\i,\j) pic[scale=1] {state};
\path (1+2*\i,\j-1) pic[scale=1] {state_dag};
}}

\foreach \j in {0,2,6}
{\draw[thick] (2,-0.2+\j)--(2,-0.8+\j);
\draw[thick] (10,-0.2+\j)--(10,-0.8+\j);}

%\draw[thick] (0.5,0.3)--(0.5,0.7);

\foreach \j in {-2,0,2,4,6}{
\foreach \i in {1,...,2}{
\draw[thick] (0.5+2*\i,0.3+\j)--(0.5+2*\i,0.7+\j);
\draw[thick] (1.5+2*\i,0.3+\j)--(1.5+2*\i,0.7+\j);
}}

\foreach \j in {0,2,6}
{\foreach \i in {3,...,4}{
\path (1+2*\i,\j-1) pic[scale=1] {state_dag2};
\path (1+2*\i,\j) pic[scale=1] {state2};
}}

%\draw[thick] (0.5,0.3)--(0.5,0.7);

\foreach \j in {-2,0,2,4,6}{
\foreach \i in {3,...,4}{
\draw[thick] (0.5+2*\i,0.3+\j)--(0.5+2*\i,0.7+\j);
\draw[thick] (1.5+2*\i,0.3+\j)--(1.5+2*\i,0.7+\j);
}}

\draw [thick,decorate,decoration={brace}]
(10.5,6)--(10.5,-1);
\node[scale=1] at (11.35,2.5) {$m$};

\draw [thick,decorate,decoration={brace}]
(5.75,-1.85)--(2.25,-1.85);
\draw [thick,decorate,decoration={brace}]
(9.75,-1.85)--(6.25,-1.85);
\node[scale=1] at (4,-2.5) {$x$};
\node[scale=1] at (8,-2.5) {$x$};

\node[scale=1.25] at (6,3.5) {$\ldots$};
\end{tikzpicture}.
\label{eq:eta}
\ee
Note that \eqref{eq:infinitepowertau} implies  
 \be
 \lim_{x\to\infty}\eta_{x,m} = {\chi^{2-2m}},
 \ee
 which is the result one finds for any $x$ for the Bell-pair state \eqref{eq:psi0Bell} and its dual-unitary deformations. 
 
Since Eq.~\eqref{eq:Sunigenstate} is valid in the $\ell\to\infty$ limit, it should be compared with the regime (I) result~\eqref{eq:Suni}. We see that Eq.~\eqref{eq:Sunigenstate} coincides with  \eqref{eq:Suni} for $2t\geq x$ and, moreover, for $1 \ll 2t \ll x$ we have
 \be
 \lim_{\ell\to\infty}S^{(n)}_{m}(A,t) \simeq 2t \log d^2 + \log \chi^2,
 \label{eq:Sunigenasy}
 \ee
where $\simeq$ denotes the leading contributions in $t$. This equation agrees with \eqref{eq:Suni} and its modification for initial states $\ket{\Psi_{\rm du}(W)}$ (for which $\chi=d$).

To probe the asymptotic behaviour in the regime (II) we consider $t,\ell,x\gg1$ such that  $1 \ll \ell-2t \leq x$. In this regime, when evolving with the SWAP gate we find (cf. Sec~\ref{sec:purity} and \ref{sec:highern}) 
\be
S^{(n)}_{m}(A,t)|_{\rm SWAP} \simeq \frac{1}{1-n} \log {\rm tr}(\tilde {\mathcal B}^n),
\label{eq:SWAPgenstate}
\ee
where $\simeq$ now denotes the leading order in $\ell-2t$ and we defined the following tensor that depends only on the initial-state matrix $W$ (cf.~\eqref{eq:psi0diagram})
\be
\!\!\!\!\!\tilde {\mathcal B}\! =\!\frac{1}{d^{2(\!\ell-\!2t)}}\!\!
\begin{tikzpicture}[baseline=(current  bounding  box.center), scale=0.5]

\foreach \j in {0}
{\foreach \i in {1,...,3}{
\path (1+2*\i,\j-1) pic[scale=1] {state};
\path (1+2*\i,\j) pic[scale=1] {state_dag};
}}

%\draw[thick] (0.5,0.3)--(0.5,0.7);

\foreach \j in {0}
{\foreach \i in {4,...,6}{
\path (1+2*\i,\j) pic[scale=1] {state_dag2};
\path (1+2*\i,\j-1) pic[scale=1] {state2};
}}

\draw [thick,decorate,decoration={brace}]
(8,-1.85)--(2.25,-1.85);
\draw [thick,decorate,decoration={brace}]
(14,-1.85)--(8.25,-1.85);
\node[scale=1] at (5.5,-2.5) {$\ell-2t$};
\node[scale=1] at (11.5,-2.5) {$\ell-2t$};

\foreach \j in {0}{
\foreach \i in {1,...,6}{
\draw[thick] (0.5+2*\i,0.3+\j-1)--(0.5+2*\i,0.7+\j-1);
\draw[thick, rounded corners] (1.5+2*\i,0.7+\j-1)--(1.9+2*\i,0.7+\j-1) --(1.9+2*\i,1.75+\j-1);
\draw[thick, rounded corners] (1.5+2*\i,0.3+\j-1)--(1.9+2*\i,0.3+\j-1) --(1.9+2*\i,-.75+\j-1);
}}

\draw[thick] (2,0.2) arc (90:270:0.15);
\draw[thick] (14,0.2) arc (90:-90:0.15);

\draw[thick] (2,-0.9) arc (90:270:0.15);
\draw[thick] (14,-0.9) arc (90:-90:0.15);

\Text[x=3,y=1.75]{}

\end{tikzpicture}\!.
\label{eq:def_B_tilde} 
\ee
Moreover, in Appendix~\ref{app:SWAP} we argue that for generic $W$ (cf.~\eqref{eq:psi0diagram}) Eq.~\eqref{eq:SWAPgenstate} can be reduced to
\be
S^{(n)}_{m}(A, t)|_{\rm SWAP} \simeq (\ell-2t)\log d^2-\log \chi^2. 
\label{eq:SSWAPgen}
\ee
Instead, in the completely chaotic subclass we find the following asymptotic result (cf.\ Secs.~\ref{sec:n=2chaos}--\ref{sec:highern}) 
\be
\!\!\!\!\limsc S^{(n)}_{A, m}(t)|_{\rm cc}\simeq\frac{\log[\chi^{2n} d^{-2n k}+ d^{2(1-n)x}]}{1-n},\,\,\, n\geq2\,,
\label{eq:Schaogen}
\ee
where $k= \ell-2t\gg1$. 

At the leading order the results \eqref{eq:Sunigenasy}, \eqref{eq:SSWAPgen}, and \eqref{eq:Schaogen} agree with the corresponding formulae obtained for the Bell-pair initial state (i.e.\ Eqs.~\eqref{eq:Suni}, \eqref{eq:SSWAP}, and \eqref{eq:Schaotic} respectively). Moreover, the first sub-leading corrections in \eqref{eq:Sunigenasy}, \eqref{eq:SSWAPgen}, and \eqref{eq:Schaogen} can be accounted for by performing the shift  
\be
t\mapsto t+\frac{\log \chi}{2\log d}
\label{eq:genericshift}
\ee 
in the Bell-pair state results. This shows that the asymptotic dynamics of operator entanglement from the initial states \eqref{eq:psi0} is largely initial-state independent. For a fixed local gate $U$ and different initial states the operator entanglement describes always the same curve: the only effect of the initial state is to shift the curve to the left to account for the initial entanglement.

\subsection{Approximability}

Our results have immediate implications also concerning the approximability of the reduced density matrix by an MPO. Indeed, since Eq.~\eqref{eq:Sunigenstate} proves that the operator R\'enyi entropies of $\rho_B(t)$ grow linearly in time for \emph{any} choice of dual-unitary gates and solvable initial state, it implies that the state $\rho_B(t)$ \emph{is never} efficiently approximable by an MPO in our setting (see the discussion around Eq.~\eqref{eq:ineqRal1} in Appendix~\ref{app:approximability}).

\section{Derivation}
\label{sec:derivation}

Here we present the main steps of our derivation. We begin in Sec.~\ref{sec:rhoBm} by obtaining a simplified expression for the thermodynamic limit of $\rho_B^m(t)$ when evolving from solvable initial states. Then, in Sec.~\ref{sec:RhoA}, we use the latter to write a simplified thermodynamic-limit expression for the super-operatorial reduced density matrix ${\rm P}_{\!A,m}(t)$ (cf.~\eqref{eq:Pbeta}), represented schematically by
\begin{equation} {\rm P}_{\!A,m}(t) =
    \begin{tikzpicture}[baseline=(current  bounding  box.center), scale=0.5]

\foreach \i in {2,3,4,5}
{
\draw[thick,black] (\i,2.9)..controls (0.5+\i,6) and (0.5+\i,-7) .. (\i,-3.9);
}

\foreach \i in {0,1}
{\draw[thick] (\i,2.8)--(\i,3.5);
\draw[thick] (\i,-3.8)--(\i,-4.5);
}

\draw[thick, fill=blue!20, rounded corners=2pt] (0,0) rectangle (5,3);
\draw[thick, fill=red!20, rounded corners=2pt] (0,-4) rectangle (5,-1);

\node[scale=1] at (2.5,1.5) {$\ket{\rho_B^m(t)}$};
\node[scale=1] at (2.5,-2.5) {$\bra{\rho_B^m(t)}$};

\draw [thick,decorate,decoration={brace}]
(-0.2,4)--(1.4,4);
\draw [thick,decorate,decoration={brace}]
(1.6,4)--(5.2,4);

\node[scale=1] at (0.6,4.75) {$x$};
\node[scale=1] at (3.4,4.75) {$\ell-x$};

\end{tikzpicture} \,,
\end{equation}
(we remind the reader that $\ell=|B|$ and $x=|A|$, cf.~\eqref{eq:ellx}). Finally, we move on to the calculation of 
\be
{\rm tr}({\rm P}^n_{\!A,m}(t)) = e^{(1-n) S^{(n)}_{m}(A,t)}\,.
\label{eq:npurity}
\ee
In particular, in Sec.~\ref{sec:purity}, we illustrate the main ideas by considering the simplest case $n=2$, which corresponds to the purity of ${\rm P}_{\!A,m}(t)$, and in Sec.~\ref{sec:highern} we show how to generalise the derivation to $n>2$. 

\subsection{Thermodynamic limit of $\rho_B^m(t)$  for solvable initial states}
\label{sec:rhoBm}
We begin by considering $\rho_B(t)$ for solvable initial states and review the simplification derived in Ref.~\cite{piroli2020exact}.  Adopting the diagrammatic representation introduced in Sec.~\ref{sec:setting} we can represent the reduced density matrix as 
{\be
\!\!\!\rho_{B}(t) =
\begin{tikzpicture}[baseline=(current  bounding  box.center), scale=0.45]
%\draw[thick, fill=black, rounded corners=2pt] (-8.75,2.5) rectangle (.6,2.25);
%\draw[thick, fill=black, rounded corners=2pt] (-8.75,.25) rectangle (.6,0);

\foreach \i in {-2,...,2}
{
\path (1+2*\i-5,2.2) pic[rotate=0] {state};
\path (1+2*\i-5,.3) pic[rotate=0] {state_dag};
}

\draw[thick] (-9,2.15) arc (90:270:0.075);
\draw[thick] (1,2.15) arc (90:-90:0.075);

\draw[thick] (-9,.5) arc (90:270:0.075);
\draw[thick] (1,0.5) arc (90:-90:0.075);

\foreach \i in {1,...,2}{
\draw[thick, rounded corners=1pt]  (\i-9.5,-6) -- (\i-9.5,-6-0.2*\i) -- (-0.2*\i-10,-6-0.2*\i) -- (-0.2*\i-10,8.5+0.2*\i) -- (\i-9.5,8.5+0.2*\i) -- (\i-9.5,8.5) ;
}
\foreach \i in {9,...,10}{
\draw[thick, rounded corners=1pt]  (\i-9.5,-6) -- (\i-9.5,-8.25+0.2*\i) -- (-0.2*\i+4,-8.25+0.2*\i) -- (-0.2*\i+4,10.75-0.2*\i) -- (\i-9.5,10.75-0.2*\i) -- (\i-9.5,8.5) ;
}
\foreach \kk in {0}{
\def\eps{13*\kk};

\foreach \ii[evaluate=\ii as \i using \ii-0.5] in {1,...,5}
{
\draw[thick] (-.5-2*\i,-1+\eps) -- (0.5-2*\i,\eps);
\draw[thick] (-0.5-2*\i,\eps) -- (0.5-2*\i,-1+\eps);
\draw[thick, fill=myblue, rounded corners=2pt] (-0.25-2*\i,-0.25+\eps) rectangle (.25-2*\i,-0.75+\eps);
\draw[thick] (-2*\i,-0.35+\eps) -- (.15-2*\i,-.35+\eps) -- (.15-2*\i,-0.5+\eps);
}
\foreach \ii[evaluate=\ii as \i using \ii-0.5] in {1,...,5}
{
\draw[thick] (.5-2*\i,-6+\eps) -- (1-2*\i,-5.5+\eps);
\draw[thick] (1.5-2*\i,-6+\eps) -- (1-2*\i,-5.5+\eps);
}
\foreach \jj[evaluate=\jj as \j using 2*(ceil(\jj/2)-\jj/2)] in {0,...,3}
\foreach \i in {1,...,5}
{
\draw[thick] (.5-2*\i-1*\j+1,-2-1*\jj+\eps) -- (1-2*\i-1*\j+1,-1.5-\jj+\eps);
\draw[thick] (1-2*\i-1*\j+1,-1.5-1*\jj+\eps) -- (1.5-2*\i-1*\j+1,-2-\jj+\eps);
}
\foreach \jj[evaluate=\jj as \j using 2*(ceil(\jj/2)-\jj/2)] in {0,...,4}
\foreach \i in {1,...,5}
{
\draw[thick] (.5-2*\i-1*\j+1,-1-1*\jj+\eps) -- (1-2*\i-1*\j+1,-1.5-\jj+\eps);
\draw[thick] (1-2*\i-1*\j+1,-1.5-1*\jj+\eps) -- (1.5-2*\i-1*\j+1,-1-\jj+\eps);
\draw[thick, fill=myblue, rounded corners=2pt] (0.75-2*\i-1*\j+1,-1.75-\jj+\eps) rectangle (1.25-2*\i-1*\j+1,-1.25-\jj+\eps);
\draw[thick] (1-2*\i-1*\j+1,-1.35-1*\jj+\eps) -- (1.15-2*\i-1*\j+1,-1.35-1*\jj+\eps) -- (1.15-2*\i-1*\j+1,-1.5-1*\jj+\eps);
}
}
\foreach \kk in {0}{
\def\eps{13*\kk+8.5};
\foreach \i in {0,...,4}
{
\draw[thick] (-.5-2*\i,-1+\eps) -- (0.5-2*\i,\eps);
\draw[thick] (-0.5-2*\i,\eps) -- (0.5-2*\i,-1+\eps);
\draw[thick, fill=myred, rounded corners=2pt] (-0.25-2*\i,-0.25+\eps) rectangle (.25-2*\i,-0.75+\eps);
\draw[thick] (-2*\i,-0.35+\eps) -- (.15-2*\i,-.35+\eps) -- (.15-2*\i,-0.5+\eps);
}
\foreach \i in {1,...,5}
{
\draw[thick] (.5-2*\i,-6+\eps) -- (1-2*\i,-5.5+\eps);
\draw[thick] (1.5-2*\i,-6+\eps) -- (1-2*\i,-5.5+\eps);
}
\foreach \jj[evaluate=\jj as \j using -2*(ceil(\jj/2)-\jj/2)] in {0,...,3}
\foreach \i in {1,...,5}
{
\draw[thick] (.5-2*\i-1*\j,-2-1*\jj+\eps) -- (1-2*\i-1*\j,-1.5-\jj+\eps);
\draw[thick] (1-2*\i-1*\j,-1.5-1*\jj+\eps) -- (1.5-2*\i-1*\j,-2-\jj+\eps);
}
\foreach \jj[evaluate=\jj as \j using -2*(ceil(\jj/2)-\jj/2)] in {0,...,4}
\foreach \i in {1,...,5}
{
\draw[thick] (.5-2*\i-1*\j,-1-1*\jj+\eps) -- (1-2*\i-1*\j,-1.5-\jj+\eps);
\draw[thick] (1-2*\i-1*\j,-1.5-1*\jj+\eps) -- (1.5-2*\i-1*\j,-1-\jj+\eps);
\draw[thick, fill=myred, rounded corners=2pt] (0.75-2*\i-1*\j,-1.75-\jj+\eps) rectangle (1.25-2*\i-1*\j,-1.25-\jj+\eps);
\draw[thick] (1-2*\i-1*\j,-1.35-1*\jj+\eps) -- (1.15-2*\i-1*\j,-1.35-1*\jj+\eps) -- (1.15-2*\i-1*\j,-1.5-1*\jj+\eps);
}
}
\foreach \i in{1.5+4.5,2.5+4.5,3.5+4.5}{
\draw[thick] (0.5,2*\i-0.5-3.5) arc (45:-90:0.17);
\draw[thick] (-10+0.5+0,2*\i-0.5-3.5) arc (90:270:0.15);
}
\foreach \i in{0.5+4.5,1.5+4.5,2.5+4.5}
{
\draw[ thick] (0.5,1+2*\i-0.5-3.5) arc (-45:90:0.17);
\draw[ thick] (-10+0.5,1+2*\i-0.5-3.5) arc (270:90:0.15);
}
\foreach \i in{1.5-3,2.5-3,3.5-3}{
\draw[thick] (0.5,2*\i-0.5-3.5) arc (45:-90:0.17);
\draw[thick] (-10+0.5+0,2*\i-0.5-3.5) arc (90:270:0.15);
}
\foreach \i in{0.5-3,1.5-3,2.5-3}
{
\draw[ thick] (0.5,1+2*\i-0.5-3.5) arc (-45:90:0.17);
\draw[ thick] (-10+0.5,1+2*\i-0.5-3.5) arc (270:90:0.15);
}
\draw [thick,decorate,decoration={brace}]
(-6.75,8.75)--(-1.5,8.75) ;
\node[scale=1] at (-4.125,9.5) {$\ell$};
\label{eq:rhoB}
\end{tikzpicture}\,.
\ee
Note that the curved lines on left and right edges (at the same height) are connected because we choose periodic boundary conditions, moreover, for our choice of units, a region of length $\ell$ contains $2\ell$ qubits ($\ell=3$ and $t=3$ in the above diagram). 
\bw
Exploiting now the unitarity relations \eqref{eq:du1} we have 
\be
\rho_{B}(t) = {\rm tr}_{\bar B}[\ket{\Psi_t}\!\bra{\Psi_t}] =\frac{1}{d^{L}}\begin{tikzpicture}[baseline=(current  bounding  box.center), scale=0.5]
\foreach \i in {1,...,6}
{\path (2*\i,0.5) pic[rotate=0] {Ur};}
\foreach \i in {1,...,5}{
\path (2*\i+1,1.5) pic[rotate=0] {Ur};}
\foreach \i in {2,...,5}
{\path (2*\i,2.5) pic[rotate=0] {Ur};}
\foreach \i in {2,...,4}
{\path (2*\i+1,3.5) pic[rotate=0] {Ur};}

\foreach \i in {-2,...,8}
{
\path (1+2*\i,-0.3) pic[rotate=0] {state};
\path (1+2*\i,-2.7) pic[rotate=0] {state_dag};
}

\foreach \j in {3} 
{
\foreach \i in {1,...,6}
{\path (2*\i,-0.5-\j) pic[rotate=0] {Ur_dag};}
\foreach \i in {1,...,5}{
\path (2*\i+1,-1.5-\j) pic[rotate=0] {Ur_dag};}
\foreach \i in {2,...,5}
{\path (2*\i,-2.5-\j) pic[rotate=0] {Ur_dag};}
\foreach \i in {2,...,4}
{\path (2*\i+1,-3.5-\j) pic[rotate=0] {Ur_dag};}}

\path (-4,-.5) pic[rotate=0] {curl_3};
\path (-4,-2.5) pic[rotate=0] {curl_4};
\path (18,-.5) pic[rotate=0] {curl_5};
\path (18,-2.5) pic[rotate=0] {curl_6};

\draw[thick,color=gray] (0.5,0)--(0.3,0)--(0.3,-3)--(0.5,-3);

\foreach \i in {0,-2} 
{
\draw[thick,color=gray] (-0.5+\i,0)--(-0.3+\i,0)--(-0.3+\i,-3)--(-0.5+\i,-3);
\draw[thick,color=gray] (-1.5+\i,0)--(-1.7+\i,0)--(-1.7+\i,-3)--(-1.5+\i,-3);

\draw[thick,color=gray] (14.5-\i,0)--(14.3-\i,0)--(14.3-\i,-3)--(14.5-\i,-3);
\draw[thick,color=gray] (15.5-\i,0)--(15.7-\i,0)--(15.7-\i,-3)--(15.5-\i,-3);
}

\foreach \i in {0,...,2}
{\draw[thick,color=gray] (1.5+\i,1+\i)--(0.15-0.15*\i,1+\i)--(0.15-0.15*\i,-4-\i)--(1.5+\i,-4-\i);
}

\draw[thick,color=gray] (13.5,0)--(13.7,0)--(13.7,-3)--(13.5,-3);

\foreach \i in {0,...,2}
{\draw[thick,color=gray] (12.5-\i,1+\i)--(13.85+0.15*\i,1+\i)--(13.85+0.15*\i,-4-\i)--(12.5-\i,-4-\i);
}

\draw [thick,decorate,decoration={brace}]
(13.5,-7.25)--(0.5,-7.25) ;
\draw [dashed,orange] (0.5,-3)--(0.5,-7.25) ;
\draw [dashed,orange] (13.5,-3)--(13.5,-7.25) ;
\node[scale=1] at (7,-8) {$\ell+2t$};

\draw [thick,decorate,decoration={brace}]
(18,4)--(18,0) ;
\draw [dashed,orange] (9.5,4)--(18,4) ;
\node[scale=1] at (18.75,2) {$t$};

\draw [thick,decorate,decoration={brace}]
(4.5,4.25)--(9.5,4.25) ;
\node[scale=1] at (7,5) {$\ell$};

\end{tikzpicture},
\ee
where we took $L>t$ ($\ell=3$ and $t=2$ in the above diagram). The outer ladder of contracted initial states can be rewritten as a power of the STM (cf.~\eqref{eq:STM}). Namely 
\be
\begin{tikzpicture}[baseline=(current  bounding  box.center), scale=0.75]
\foreach \j in {0}
{\foreach \i in {0,2,6}{
\path (1+.65*\i,\j-1) pic[scale=0.8, scale=1.5] {state};
\path (1+.65*\i,\j) pic[scale=0.8, scale=1.5] {state_dag};
}}

\node at (3.75,-.5) {...};

\foreach \j in {0}{
\foreach \i in {0,2,6}{
\draw[thick] (0.6+.65*\i,0.3+\j-1-0.05)--(0.6+.65*\i,0.7+\j-1+0.05);
\draw[thick] (1.5+.65*\i-0.1,0.3+\j-1-0.05)--(1.5+.65*\i-0.1,0.7+\j-1+0.05);
}}
\draw [thick,decorate,decoration={brace}]
(6,-1.35)--(0,-1.35) ;
\node[scale=1] at (3,-2) {$L-\ell-2t$};
\node at (3.75,1) {};
\end{tikzpicture} = \tau^{L-\ell-2t}\,.
\ee
This, together with \eqref{eq:infinitepowertau}, leads to the following simplification in the thermodynamic limit 
\be
\rho_{B, \rm th}(t) = \lim_{L\to \infty}{\rm tr}_{\bar B}[\ket{\Psi_t}\!\bra{\Psi_t}] =\frac{1}{\chi d^{\ell+2t}}\begin{tikzpicture}[baseline=(current  bounding  box.center), scale=0.5]

\draw[ thick] (0,-0.5)--(-0.8,-0.5)--(-1,-0.7)--(-1,-2.3)--(-0.8,-2.5)--(0,-2.5);
\draw[ thick] (14,-0.5)--(14.8,-0.5)--(15,-0.7)--(15,-2.3)--(14.8,-2.5)--(14,-2.5);

\foreach \i in {1,...,6}
{\path (2*\i,0.5) pic[rotate=0] {Ur};}
\foreach \i in {1,...,5}{
\path (2*\i+1,1.5) pic[rotate=0] {Ur};}
\foreach \i in {2,...,5}
{\path (2*\i,2.5) pic[rotate=0] {Ur};}
\foreach \i in {2,...,4}
{\path (2*\i+1,3.5) pic[rotate=0] {Ur};}

\foreach \i in {0,...,6}
{
\path (1+2*\i,-0.3) pic[rotate=0] {state};
\path (1+2*\i,-2.7) pic[rotate=0] {state_dag};
}

\foreach \j in {3} 
{
\foreach \i in {1,...,6}
{\path (2*\i,-0.5-\j) pic[rotate=0] {Ur_dag};}
\foreach \i in {1,...,5}{
\path (2*\i+1,-1.5-\j) pic[rotate=0] {Ur_dag};}
\foreach \i in {2,...,5}
{\path (2*\i,-2.5-\j) pic[rotate=0] {Ur_dag};}
\foreach \i in {2,...,4}
{\path (2*\i+1,-3.5-\j) pic[rotate=0] {Ur_dag};}}

\draw[thick,color=gray] (0.5,0)--(0.3,0)--(0.3,-3)--(0.5,-3);

\foreach \i in {0,...,2}
{\draw[thick,color=gray] (1.5+\i,1+\i)--(0.15-0.15*\i,1+\i)--(0.15-0.15*\i,-4-\i)--(1.5+\i,-4-\i);
}

\draw[thick,color=gray] (13.5,0)--(13.7,0)--(13.7,-3)--(13.5,-3);

\foreach \i in {0,...,2}
{\draw[thick,color=gray] (12.5-\i,1+\i)--(13.85+0.15*\i,1+\i)--(13.85+0.15*\i,-4-\i)--(12.5-\i,-4-\i);
}

\draw [thick,decorate,decoration={brace}]
(13.5,-7.25)--(0.5,-7.25) ;
\draw [dashed,orange] (0.5,-3)--(0.5,-7.25) ;
\draw [dashed,orange] (13.5,-3)--(13.5,-7.25) ;
\node[scale=1] at (7,-8) {$\ell+2t$};
\draw [thick,decorate,decoration={brace}]
(15,4)--(15,0) ;
\draw [dashed,orange] (9.5,4)--(15,4) ;
\node[scale=1] at (15.5,2) {$t$};

\draw [thick,decorate,decoration={brace}] (4.5,4.25)--(9.5,4.25) ;
\node[scale=1] at (7,5) {$\ell$};
\end{tikzpicture}\,,
\ee
where we introduced the subscript ``$\rm th$'' to stress that we took the thermodynamic limit. 

We then repeatedly use solvability  \eqref{eq:stateuni1}--\eqref{eq:stateuni2} and dual-unitarity \eqref{eq:du2} to contract the network further.
\be
%\rho_{B, \rm th}(t)=\frac{1}{\chi d^{\ell+2t}}
\begin{tikzpicture}[baseline=(current  bounding  box.center), scale=0.5]

\foreach \i in {1,2}
{\path (2*\i,0.5) pic[rotate=0] {Ur};}
\foreach \i in {1}{
\path (2*\i+1,1.5) pic[rotate=0] {Ur};}
\foreach \i in {0,1}
{
\path (1+2*\i,-0.3) pic[rotate=0] {state};
\path (1+2*\i,-2.7) pic[rotate=0] {state_dag};
}
\foreach \j in {3} 
{
\foreach \i in {1,2}
{\path (2*\i,-0.5-\j) pic[rotate=0] {Ur_dag};}
\foreach \i in {1}{
\path (2*\i+1,-1.5-\j) pic[rotate=0] {Ur_dag};}
}
\draw[thick] (0.5,0)--(-0.3,0)--(-0.3,-3)--(0.5,-3);
\draw[thick] (0,-0.5)--(0,-2.5);

\foreach \i in {0}
{\draw[thick] (1.5+\i,1+\i)--(-.6-0.3*\i,1+\i)--(-.6-0.3*\i,-4-\i)--(1.5+\i,-4-\i);
}
\draw[thick,->] (5.5,-1.5)--(7,-1.5);

\foreach \k in {9} 
{
\foreach \i in {1,2}
{\path (2*\i+\k,0.5) pic[rotate=0] {Ur};}
\foreach \i in {1}{
\path (2*\i+1+\k,1.5) pic[rotate=0] {Ur};}
\foreach \i in {1}
{
\path (1+2*\i+\k,-0.3) pic[rotate=0] {state};
\path (1+2*\i+\k,-2.7) pic[rotate=0] {state_dag};
}
\foreach \j in {3} 
{
\foreach \i in {1,2}
{\path (2*\i+\k,-0.5-\j) pic[rotate=0] {Ur_dag};}
\foreach \i in {1}{
\path (2*\i+1+\k,-1.5-\j) pic[rotate=0] {Ur_dag};}
}
\draw[thick] (0.5+\k,0)--(-0.3+\k,0)--(-0.3+\k,-3)--(0.5+\k,-3);

\draw[thick] (0.5+\k,0)--(1.5+\k,0);
\draw[thick] (0.5+\k,-3)--(1.5+\k,-3);

\draw[thick] (2+\k,-0.5)--(0+\k,-0.5)--(0+\k,-2.5)--(2+\k,-2.5);

\foreach \i in {0}
{\draw[thick] (1.5+\i+\k,1+\i)--(-.6-0.3*\i+\k,1+\i)--(-.6-0.3*\i+\k,-4-\i)--(1.5+\i+\k,-4-\i);
}
\draw[thick,->] (5.5+\k,-1.5)--(7+\k,-1.5);
}

\foreach \k in {18} 
{
\foreach \i in {2}
{\path (2*\i+\k,0.5) pic[rotate=0] {Ur};}
\foreach \i in {1}{
\path (2*\i+1+\k,1.5) pic[rotate=0] {Ur};}
\foreach \i in {1}
{
\path (1+2*\i+\k,-0.3) pic[rotate=0] {state};
\path (1+2*\i+\k,-2.7) pic[rotate=0] {state_dag};
}
\foreach \j in {3} 
{
\foreach \i in {2}
{\path (2*\i+\k,-0.5-\j) pic[rotate=0] {Ur_dag};}
\foreach \i in {1}{
\path (2*\i+1+\k,-1.5-\j) pic[rotate=0] {Ur_dag};}
}
\draw[thick] (0.5+\k,0)--(-0.3+\k,0)--(-0.3+\k,-3)--(0.5+\k,-3);

\draw[thick] (0.5+\k,0)--(1.5+\k,0)--(1.8+\k,0.3)--(2.2+\k,0.3)--(2.5+\k,0);
\draw[thick] (0.5+\k,-3)--(1.5+\k,-3)--(1.8+\k,-3.3)--(2.2+\k,-3.3)--(2.5+\k,-3);

\draw[thick] (1.5+\k,1)--(1.8+\k,0.7)--(2.2+\k,0.7)--(2.5+\k,1);
\draw[thick] (1.5+\k,-4)--(1.8+\k,-3.7)--(2.2+\k,-3.7)--(2.5+\k,-4);

\draw[thick] (2+\k,-0.5)--(0+\k,-0.5)--(0+\k,-2.5)--(2+\k,-2.5);

\foreach \i in {0}
{\draw[thick] (1.5+\i+\k,1+\i)--(-.6-0.3*\i+\k,1+\i)--(-.6-0.3*\i+\k,-4-\i)--(1.5+\i+\k,-4-\i);
}
}

\node[scale=1] at (6.25,-0.75) {$(i)$};
\node[scale=1] at (15.25,-0.75) {$(ii)$};

\end{tikzpicture}\,.
\label{eq:rhot}
\ee
\ew
The result depends on the time elapsed. In particular, for $2t \leq \ell$, we obtain 
\be
\!\!\!\!\!\!\rho_{B, \rm th}(t)=\frac{1}{\chi d^{\ell+2t}}
\begin{tikzpicture}[baseline=(current  bounding  box.center), scale=0.5]
\foreach \i in {2,...,5}
{\path (2*\i,0.5) pic[rotate=0] {Ur};}
\foreach \i in {1,...,5}{
\path (2*\i+1,1.5) pic[rotate=0] {Ur};}

\foreach \i in {2,...,4}
{
\path (1+2*\i,-0.3) pic[rotate=0] {state};
\path (1+2*\i,-2.7) pic[rotate=0] {state_dag};
}

\foreach \j in {3} 
{
\foreach \i in {2,...,5}
{\path (2*\i,-0.5-\j) pic[rotate=0] {Ur_dag};}
\foreach \i in {1,...,5}{
\path (2*\i+1,-1.5-\j) pic[rotate=0] {Ur_dag};}}

\draw[thick] (3.5,0)--(3.5,-3);
\draw[thick] (2.5,1)--(2.5,-4);
\draw[thick] (10.5,0)--(10.5,-3);
\draw[thick] (11.5,1)--(11.5,-4);
\draw[thick] (4,-0.5)--(4,-2.5);
\draw[thick] (10,-0.5)--(10,-2.5);

\draw [thick,decorate,decoration={brace}]
(9.75,-5.5)--(4.25,-5.5) ;
\draw [dashed,orange] (9.75,-5.5)--(9.75,-2.5) ;
\draw [dashed,orange] (4.25,-5.5)--(4.25,-2.5) ;
\node[scale=1] at (7,-6.25) {$\ell-2t$};

\draw [thick,decorate,decoration={brace}]
(12,2)--(12,0) ;
\node[scale=1] at (12.5,1) {$t$};

\draw [thick,decorate,decoration={brace}]
(2.5,2.25)--(11.5,2.25) ;
\node[scale=1] at (7,3) {$\ell$};

\end{tikzpicture}
\label{eq:rhotherm}\!\!\!\!\!\!,
\end{equation}
where, for example, we took $\ell=5$ and $t=1$. On the other hand, for $2t>\ell$, we find
\begin{equation}
\!\!\!\!\rho_{B, \rm th}(t) = \frac{1}{d^{2\ell}}\, \begin{tikzpicture}[baseline=(current  bounding  box.center), scale=0.8]
\foreach \i in {0,...,5}{
\draw[thick] (-2+0.5*\i,0.5) -- (-2+0.5*\i,-0.5);}
\Text[x=-1.75,y=-0.65]{}
\end{tikzpicture} \,= \frac{1}{d^{2\ell}} \mathbbm{1}_{d}^{\otimes 2 \ell},
\label{eq:rhotherm}
\end{equation}
where $\1_d$ denotes the identity matrix in $\mathbb C^{d}$. Note that the reduced density matrix becomes equal to the identity, i.e.\ the subsystem thermalises to the infinite temperature state.

The form \eqref{eq:rhot}--\eqref{eq:rhotherm} of the reduced density matrix allows us to easily evaluate its $m$-th power. In particular in the non-trivial case $2t\leq \ell$ we have
\be
\!\!\!\rho^m_{B, \rm th}(t) = \frac{1}{d^{4tm}} \begin{tikzpicture}[baseline=(current  bounding  box.center), scale=0.5]

\foreach \i in {2,...,5}
{\path (2*\i,0.5) pic[rotate=0] {Ur};}
\foreach \i in {1,...,5}{
\path (2*\i+1,1.5) pic[rotate=0] {Ur};}

\draw[thick, fill=mygrey1,  rounded corners=3pt] (4,-0.5) rectangle (10,-2.5);

\foreach \i in {0,...,5}
{\draw[thick](4.5+\i,-0.5)--(4.5+\i,0);
\draw[thick](4.5+\i,-2.5)--(4.5+\i,-3);
}

\foreach \j in {3} 
{
\foreach \i in {2,...,5}
{\path (2*\i,-0.5-\j) pic[rotate=0] {Ur_dag};}
\foreach \i in {1,...,5}{
\path (2*\i+1,-1.5-\j) pic[rotate=0] {Ur_dag};}}

\draw[thick] (3.5,0)--(3.5,-3);
\draw[thick] (2.5,1)--(2.5,-4);
\draw[thick] (10.5,0)--(10.5,-3);
\draw[thick] (11.5,1)--(11.5,-4);

%\draw [thick,decorate,decoration={brace}]
%(9.75,-5.5)--(4.25,-5.5) ;
%\draw [dashed,orange] (9.75,-5.5)--(9.75,-2.5) ;
%\draw [dashed,orange] (4.25,-5.5)--(4.25,-2.5) ;
%\node[scale=1] at (7,-6.25) {$\ell-2t$};

%\draw [thick,decorate,decoration={brace}]
%(12,2)--(12,0) ;
%\node[scale=1] at (12.5,1) {$t$};

%\draw [thick,decorate,decoration={brace}]
%(2.5,2.25)--(11.5,2.25) ;
%\node[scale=1] at (7,3) {$\ell$};

\end{tikzpicture}
\label{eq:rhomt}
\ee
where we introduced 
\be
\!\!\!\!\!\!\!\begin{tikzpicture}[baseline=(current  bounding  box.center), scale=0.4]
\foreach \k in {11}
{
\draw[thick,fill=mygrey1,rounded corners=2pt] (0+\k,0.5) rectangle (6+\k,4.5);

\foreach \i in {0,...,5}
{
\draw[thick,color=black] (0.5+\i+\k,0.5)--(0.5+\i+\k,0);
\draw[thick,color=black] (0.5+\i+\k,4.5)--(0.5+\i+\k,5);
}
}
\end{tikzpicture}=\frac{1}{\chi^m d^{m(\ell-2t)}}\begin{tikzpicture}[baseline=(current  bounding  box.center), scale=0.4]
\foreach \j in {0,2,6}
{\foreach \i in {0,...,2}{
\path (1+2*\i,\j) pic[scale=0.8] {state};
\path (1+2*\i,\j-1) pic[scale=0.8] {state_dag};
}}

\foreach \j in {0,2,6}
{\draw[thick] (0,-0.2+\j)--(0,-0.8+\j);
\draw[thick] (6,-0.2+\j)--(6,-0.8+\j);}

\draw[thick] (0.5,0.3)--(0.5,0.7);

\foreach \j in {-2,0,2,4,6}{
\foreach \i in {0,...,2}{
\draw[thick] (0.5+2*\i,0.3+\j)--(0.5+2*\i,0.7+\j);
\draw[thick] (1.5+2*\i,0.3+\j)--(1.5+2*\i,0.7+\j);
}}

\draw [thick,decorate,decoration={brace}]
(6.5,6)--(6.5,-1);
\node[scale=1] at (7.35,2.5) {$m$};

\node[scale=1.25] at (4,3.5) {$\ldots$};
\end{tikzpicture}.
\label{eq:mlayers}
\ee
In particular, note that for the Bell-pair state \eqref{eq:psi0Bell} and its dual-unitary deformations \eqref{eq:duDef} we have 
\be
\!\!\!\!\!\!\!\begin{tikzpicture}[baseline=(current  bounding  box.center), scale=0.4]
\foreach \k in {11}
{
\draw[thick,fill=mygrey1,rounded corners=2pt] (0+\k,0.5) rectangle (6+\k,4.5);

\foreach \i in {0,...,5}
{
\draw[thick,color=black] (0.5+\i+\k,0.5)--(0.5+\i+\k,0);
\draw[thick,color=black] (0.5+\i+\k,4.5)--(0.5+\i+\k,5);
}
}
\end{tikzpicture}=\frac{1}{\chi^{2m-1} d^{(\ell-2t)}}\begin{tikzpicture}[baseline=(current  bounding  box.center), scale=0.4]
\foreach \j in {0}
{\foreach \i in {0,...,2}{
\path (1+2*\i,\j) pic[scale=0.8] {state};
\path (1+2*\i,\j-1) pic[scale=0.8] {state_dag};
}}

\foreach \j in {0}
{\draw[thick] (0,-0.2+\j)--(0,-0.8+\j);
\draw[thick] (6,-0.2+\j)--(6,-0.8+\j);}

\draw[thick] (0.5,0.3)--(0.5,0.7);

\foreach \j in {-2,0}{
\foreach \i in {0,...,2}{
\draw[thick] (0.5+2*\i,0.3+\j)--(0.5+2*\i,0.7+\j);
\draw[thick] (1.5+2*\i,0.3+\j)--(1.5+2*\i,0.7+\j);
}}
\end{tikzpicture},
\label{eq:mlayersBell}
\ee
namely 
\be
\!\!\!\!\rho^m_{B, \rm th}(t)|_{\rm Bell} = \frac{1}{({\chi^2 d^{4t}})^{m-1}} \rho_{B, \rm th}(t)|_{\rm Bell}\,,\quad 2t\leq \ell\,.
\ee
This means that $\chi^2 d^{4 t}\rho_{B, \rm th}(t)|_{\rm Bell}$ is a projector, i.e., all non-zero eigenvalues of $\rho_{B, \rm th}(t)|_{\rm Bell}$ are equal to $\chi^{-2} d^{-4 t}$ (since ${\rm tr}(\rho_{B, \rm th}(t))=1$ their number must be $\chi^2 d^{4 t}$). We remark that for more general states \eqref{eq:psi0diagram} the property \eqref{eq:mlayersBell} holds in the infinite $\ell$ limit. This is an immediate consequence of \eqref{eq:infinitepowertau}.

\subsection{Thermodynamic limit of ${\rm P}_{\!A,m}(t)$  for solvable initial states}
\label{sec:RhoA}

The denominator in the definition of ${\rm P}_{\!A,m}(t)$ (cf. \eqref{eq:rhosuper}) is directly evaluated using the results of the previous subsection
\be
\!\!\!\!\!\!\!{\rm tr}(\rho^{2m}_{B, \rm th}(t)) \!=\! 
\begin{cases}
\displaystyle \frac{\eta'_{\ell-2t, 2m}}{d^{4t(2m-1)}}  & 2t \leq \ell\\
\displaystyle \frac{1}{d^{2\ell(2m-1)}} & 2t > \ell\\
\end{cases}\!.
\label{eq:trhoB2m}
\ee
Here we used the short-hand notation 
\be
\eta'_{x,m} =\frac{1}{\chi^m d^{mx}}\begin{tikzpicture}[baseline=(current  bounding  box.center), scale=0.5]

\foreach \i in {1,...,4}
{
\draw[ thick] (2*\i+2.5-1,6.65) arc (0:180:0.15);
\draw[ thick] (2*\i+2-1.5,6.65) arc (0:-180:-0.15);
\draw[ thick] (2*\i+2-1.5,-1.65) arc (0:180:-0.15);
\draw[ thick] (2*\i+2-0.5,-1.65) arc (0:-180:0.15);
}

\foreach \j in {0,2,6}
{\foreach \i in {1,...,2}{
\path (1+2*\i,\j) pic[scale=1] {state};
\path (1+2*\i,\j-1) pic[scale=1] {state_dag};
}}

\foreach \j in {0,2,6}
{\draw[thick] (2,-0.2+\j)--(2,-0.8+\j);
\draw[thick] (10,-0.2+\j)--(10,-0.8+\j);}

%\draw[thick] (0.5,0.3)--(0.5,0.7);

\foreach \j in {-2,0,2,4,6}{
\foreach \i in {1,...,2}{
\draw[thick] (0.5+2*\i,0.3+\j)--(0.5+2*\i,0.7+\j);
\draw[thick] (1.5+2*\i,0.3+\j)--(1.5+2*\i,0.7+\j);
}}

\foreach \j in {0,2,6}
{\foreach \i in {3,...,4}{
\path (1+2*\i,\j-1) pic[scale=1] {state_dag};
\path (1+2*\i,\j) pic[scale=1] {state};
}}

%\draw[thick] (0.5,0.3)--(0.5,0.7);

\foreach \j in {-2,0,2,4,6}{
\foreach \i in {3,...,4}{
\draw[thick] (0.5+2*\i,0.3+\j)--(0.5+2*\i,0.7+\j);
\draw[thick] (1.5+2*\i,0.3+\j)--(1.5+2*\i,0.7+\j);
}}

\draw [thick,decorate,decoration={brace}]
(10.5,6)--(10.5,-1);
\node[scale=1] at (11.35,2.5) {$m$};

\draw [thick,decorate,decoration={brace}]
(9.75,-1.85)--(2.25,-1.85);
\node[scale=1] at (6,-2.5) {$x$};

\node[scale=1.25] at (6,3.5) {$\ldots$};
\end{tikzpicture},
\label{eq:eta'}
\ee
for the trace of the operator in Eq.~\eqref{eq:mlayers} (curved lines on top and bottom edges at corresponding positions are connected because of the trace). Note that \eqref{eq:infinitepowertau} implies  
 \be
 \lim_{x\to\infty}\eta'_{x,m} = {\chi^{2-2m}}.
 \ee
We also remark that \eqref{eq:trhoB2m} are nothing but 
\be
e^{(1-2m) S^{(2m)}_B(t)},
\ee
where $S^{(m)}_B(t)$ are the R\'enyi entropies of $\rho_{B, \rm th}(t)$ computed in Ref.~\cite{piroli2020exact}. 

To express the numerator of ${\rm P}_{\!A,m}(t)$ we map $\rho^m_{B, \rm th}(t)$ onto the corresponding state $\ket{\rho^m_{B, \rm th}(t)}$, as discussed in Sec.~\ref{section:osee_intro}. Diagrammatically, this amounts to folding up the upper and lower parts of the diagrams Eq.~\eqref{eq:rhomt} and Eq.~\eqref{eq:rhotherm}, bending the lower legs on top of the upper ones. In particular, for $2t\leq\ell$ we find
\be
\ket{\rho^m_{B}(t)}= \frac{1}{d^{4t(m-1)}} 
\begin{tikzpicture}[baseline=(current  bounding  box.center), scale=0.5]

\foreach \i in {0,...,2}
{\draw[thick] (9.5+2*\i,0) -- (10.5+2*\i,1);
\draw[thick] (9.5+2*\i,1) -- (10.5+2*\i,0);
\draw[thick, fill=myorange, rounded corners=2pt] (9.75+2*\i,0.75) rectangle (10.25+2*\i,0.25);
\draw[thick] (10+2*\i,0.65) -- (10+.15+2*\i,0.65) -- (10.15+2*\i,0.5);
}

\foreach \i in {-1,...,2}
{\draw[thick] (10.5+2*\i,1) -- (11.5+2*\i,2);
\draw[thick] (10.5+2*\i,2) -- (11.5+2*\i,1);
\draw[thick, fill=myorange, rounded corners=2pt] (10.75+2*\i,1.75) rectangle (11.25+2*\i,1.25);
\draw[thick] (11+2*\i,1.65) -- (11.15+2*\i,1.65) -- (11.15+2*\i,1.5);}

\draw[thick] (10.5,-0.25) -- (13.5,-0.25);

\draw[thick] (11.5,0) -- (11.5,-0.25);
\draw[thick] (12.5,0) -- (12.5,-0.25);

\draw[thick, fill=black, rounded corners=2pt] (10.25,0) rectangle (13.75,-0.25);

\draw[thick, fill=white, rounded corners=2pt] (9.5,0) circle (.15);
\draw[thick, fill=white, rounded corners=2pt] (8.5,1) circle (.15);
\draw[thick, fill=white, rounded corners=2pt] (14.5,0) circle (.15);
\draw[thick, fill=white, rounded corners=2pt] (15.5,1) circle (.15);
%\draw[thick, fill=white, rounded corners=2pt] (10.5,-.5) circle (.15);
%\draw[thick, fill=white, rounded corners=2pt,] (13.5,-.5) circle (.15);
\end{tikzpicture}\, ,
\ee
where the folded orange gate is defined in Eq.~\eqref{eq:doublegates1}. We also introduced a folded initial state
\be \label{eq:define_folded}
\begin{tikzpicture}[baseline=(current  bounding  box.center), scale=0.5]

\draw[thick] (0,0) parabola (0.5,3);
\draw[thick,fill=gray] (0.5,2.5)--(-0.416,2.5) parabola bend (0,0) (0.5,2.5);

\draw[thick,fill=mygrey1] (0.5,3) parabola[bend at end] (0,0) -- (4,0) parabola (4.5,3) -- (0.5,3) ;

\foreach \j in {0,...,8}
{\draw[thick] (0.5+0.5*\j,3)--(0.5+0.5*\j,3.5);}

\foreach \j in {0,...,1}
{\draw[thick] (-.4166+0.5*\j,2.5)--(-.4166+0.5*\j,3);}

\node[scale=1.5] at (5.5,1.5) {$=$};

\draw[thick, fill=black, rounded corners=2pt] (7,1.5-0.125) rectangle (11,1.5+0.125);

\foreach \j in {0,...,8}
{\draw[thick] (7+0.5*\j,1.5)--(7+0.5*\j,2+0.125);}

\draw [thick,decorate,decoration={brace}]
(4,-.25)--(0,-0.25);
\node[scale=1] at (2,-0.75) {$\ell-2t$};
\end{tikzpicture} \, ,
\ee
and we indicated a folded wire as in Eq.~\eqref{eq:foldedwire}. Note that the latter is a normalised state of $\mathbb C^{d^2}$, namely $\braket{\mcirc|\mcirc}=1$. Instead, for $2t>\ell$ we find 
\be
\ket{\rho^m_{B}(t)}= \frac{1}{d^{2\ell (m-1)}} \begin{tikzpicture}[baseline=(current  bounding  box.center), scale=0.8]
\foreach \i in {0,...,5}{
\draw[very thick] (-0.15+0.75*\i,0.25) -- (-0.15+0.75*\i,-0.251);
\draw[thick, fill=white] (-.15+0.75*\i,-0.25) circle (0.1cm); }
\Text[x=-0.15,y=-0.65]{}
\end{tikzpicture}\,.
\ee
Next, we construct the super-operator $\ket{\rho^m_{B}(t)} \bra{\rho^m_{B}(t)}$, consider the bipartition $B=A\bar A$, and trace out $\bar A$. Dividing by \eqref{eq:trhoB2m}, we finally obtain 
\be
\!\!\!\!\!{\rm P}_{\!A,m}(t)= \frac{1}{\eta'_{\ell-2t,2m}}
\begin{tikzpicture}[baseline=(current  bounding  box.center), scale=0.5]

\foreach \i in {0,...,2}
{\path (2*\i,0) pic {Ufolded};
}
\foreach \i in {0,...,3}
{\path (2*\i-1,1) pic {Ufolded};
}

\draw[thick] (0.5,-0.25) -- (3.5,-0.25);

\foreach \i in {0,...,1}
{
\draw[thick] (0.5+2*\i,0)--(0.5+2*\i,-0.25) ;
\draw[thick] (1.5+2*\i,0) -- (1.5+2*\i,-0.25);}

\draw[thick, fill=white] (-.5,0) circle (.15);
\draw[thick, fill=white] (-1.5,1) circle (.15);
\draw[thick, fill=white] (4.5,0) circle (.15);
\draw[thick, fill=white] (5.5,1) circle (.15);

\draw[thick] (0.5,-1.25) -- (3.5,-1.25);

\foreach \i in {0,...,1}
{
\draw[thick] (0.5+2*\i,-1.5)--(0.5+2*\i,-1.25) ;
\draw[thick] (1.5+2*\i,-1.5) -- (1.5+2*\i,-1.25);
}

\foreach \i in {0,...,2}
{\path (2*\i,-2.5) pic {U_dagfolded};
}
\foreach \i in {0,...,3}
{\path (2*\i-1,-3.5) pic {U_dagfolded};
}

\draw[thick, fill=white] (-.5,-1.5) circle (.15);
\draw[thick, fill=white] (-1.5,-2.5) circle (.15);
\draw[thick, fill=white] (4.5,-1.5) circle (.15);
\draw[thick, fill=white] (5.5,-2.5) circle (.15);
%%%%

\draw[thick, fill=black, rounded corners=2pt] (.25,0) rectangle (3.75,-0.25);

\draw[thick, fill=black, rounded corners=2pt] (.25,-1.25) rectangle (3.75,-1.5);

\foreach \i in {-1,0,...,4}
{\draw[thick] (4.5-\i,2) -- (4.5-\i,2.2+0.2*\i) -- (6.25+0.25*\i,2.2+0.2*\i) -- (6.25+0.25*\i,0);
}

\foreach \i in {-1,0,...,4}
{\draw[thick] (4.5-\i,-3.5) -- (4.5-\i,-3.7-0.2*\i) -- (6.25+0.25*\i,-3.7-0.2*\i) -- (6.25+0.25*\i,0);
}

\draw [thick,decorate,decoration={brace}]
(-0.25,-3.75) -- (-1.75,-3.75);
%\draw [thick,decorate,decoration={brace}]
%(5.5,-4.75) -- (0.25,-4.75);

\node[scale=1] at (-1,-4.25) {$x$};
%\node[scale=1] at (2.875,-5.5) {$\ell-x$};

\end{tikzpicture}\,,
\label{eq:P2tlB}
\ee
for $2t\leq \ell$ ($\ell=4$ and $t=1$ in the diagram) and 
\be
{\rm P}_{\!A,m}(t)= \begin{tikzpicture}[baseline=(current  bounding  box.center), scale=0.8]
\foreach \i in {0,...,1}{
\draw[very thick] (-0.15+0.75*\i,0.25) -- (-0.15+0.75*\i,-0.251);
\draw[very thick] (-0.15+0.75*\i,0.25-.85) -- (-0.15+0.75*\i,-0.251-.85);
\draw[thick, fill=white] (-.15+0.75*\i,-0.25) circle (0.1cm);
\draw[thick, fill=white] (-.15+0.75*\i,-0.25-.35) circle (0.1cm); }
\Text[x=-0.15,y=-1.25]{}
\end{tikzpicture}\,,
\label{eq:PBl2t}
\ee
for $2t>\ell$. In \eqref{eq:P2tlB} we denoted by $x$ the size of $A$ and we made use of the hermitian conjugate folded gate defined in Eq.~\eqref{eq:doublegates2}.

\subsection{Evaluation of \eqref{eq:npurity} for $n=2$}
\label{sec:purity}

We begin by observing that the calculation is trivial for $2t\geq \ell$, due to the simple projector form that ${\rm P}_{\!A,m}(t)$ takes in this case. Indeed, inserting \eqref{eq:PBl2t} in  \eqref{eq:npurity} we immediately find 
\be
\textrm{tr}({\rm P}_{\!A,m}^2(t))=1\,.
\ee
Therefore, from now on we restrict to $2t\leq \ell$. In this regime we construct a tensor network representation for the purity of ${\rm P}_{\!A,m}(t)$ by contracting two copies of \eqref{eq:P2tlB}, i.e.
\bw 
\be
\textrm{tr}({\rm P}_{\!A,m}^2(t))=
\begin{tikzpicture}[baseline=(current  bounding  box.center), scale=0.65]
\foreach \i in {0,...,2}
{\path (2*\i,0) pic[scale = 1.3] {Ufolded};
}
\foreach \i in {0,...,3}
{\path (2*\i-1,1) pic[scale = 1.3] {Ufolded};
}

\draw[thick] (0.5,0) -- (3.5,0);
%\draw[thick] (0.5,0) -- (.5,-.5);
%\draw[thick] (3.5,0) -- (3.5,-.5);

%\draw[thick, fill=white] (.5,-.5) circle (.125);
%\draw[thick, fill=white] (3.5,-.5) circle (.125);
\draw[thick, fill=white] (-.5,0) circle (.125);
\draw[thick, fill=white] (-1.5,1) circle (.125);
\draw[thick, fill=white] (4.5,0) circle (.125);
\draw[thick, fill=white] (5.5,1) circle (.125);

\draw[thick] (0.5,-1.5) -- (3.5,-1.5);
%\draw[thick] (0.5,-1.5) -- (.5,-1);
%\draw[thick] (3.5,-1.5) -- (3.5,-1);

\foreach \i in {0,...,2}
{\path (2*\i,-2.5) pic[scale = 1.3] {U_dagfolded};
}
\foreach \i in {0,...,3}
{\path (2*\i-1,-3.5) pic[scale = 1.3] {U_dagfolded};
}

%\draw[thick, fill=white] (.5,-1) circle (.125);
%\draw[thick, fill=white] (3.5,-1) circle (.125);
\draw[thick, fill=white] (-.5,-1.5) circle (.125);
\draw[thick, fill=white] (-1.5,-2.5) circle (.125);
\draw[thick, fill=white] (4.5,-1.5) circle (.125);
\draw[thick, fill=white] (5.5,-2.5) circle (.125);

%%%%

\foreach \i in {0,...,2}
{\path (8+2*\i,1) pic[rotate=180, scale = 1.3] {U_dagfolded};
}
\foreach \i in {0,...,3}
{\path (8+2*\i-1,2) pic[rotate=180, scale = 1.3] {U_dagfolded};
}

\draw[thick] (8.5,0) -- (11.5,0);
%\draw[thick] (8.5,0) -- (8.5,-.5);
%\draw[thick] (11.5,0) -- (11.5,-.5);

%\foreach \i in {0,...,1}
%{
%\draw[thick, fill=black] (8.5+2*\i,0) circle (0.1);
%\draw[thick, fill=white] (9.5+2*\i,0) circle (0.1);
%}

%\draw[thick, fill=white] (8.5,-.5) circle (.125);
%\draw[thick, fill=white] (11.5,-.5) circle (.125);
\draw[thick, fill=white] (8-.5,0) circle (.125);
\draw[thick, fill=white] (8-1.5,1) circle (.125);
\draw[thick, fill=white] (12.5,0) circle (.125);
\draw[thick, fill=white] (13.5,1) circle (.125);

\draw[thick] (8.5,-1.5) -- (11.5,-1.5);
%\draw[thick] (8.5,-1.5) -- (8.5,-1);
%\draw[thick] (11.5,-1.5) -- (11.5,-1);

\foreach \i in {0,...,2}
{\path (8+2*\i,-1.5) pic[rotate=180, scale = 1.3] {Ufolded};
}
\foreach \i in {0,...,3}
{\path (8+2*\i-1,-2.5) pic[rotate=180, scale = 1.3] {Ufolded};
}

%\draw[thick, fill=white] (8.5,-1) circle (.125);
%\draw[thick, fill=white] (11.5,-1) circle (.125);
\draw[thick, fill=white] (8-.5,-1.5) circle (.125);
\draw[thick, fill=white] (8-1.5,-2.5) circle (.125);
\draw[thick, fill=white] (12.5,-1.5) circle (.125);
\draw[thick, fill=white] (13.5,-2.5) circle (.125);

\foreach \i in {-1,0,...,4}
{\draw[thick] (4.5-\i,2) -- (4.5-\i,2.2+0.2*\i) -- (7.5+\i,2.2+0.2*\i) -- (7.5+\i,2);
}

\foreach \i in {-1,0,...,4}
{\draw[thick] (4.5-\i,-3.5) -- (4.5-\i,-3.7-0.2*\i) -- (7.5+\i,-3.7-0.2*\i) -- (7.5+\i,-3.5);
}

\draw[thick] (-1.5,2) -- (-1.9,2) -- (-1.9,-3.5) -- (-1.5,-3.5);
\draw[thick] (13.5,2) -- (13.9,2) -- (13.9,-3.5) -- (13.5,-3.5);
\draw[thick] (-.5,2) -- (-.5,2.2) -- (-2.1,2.2) -- (-2.1,-3.7) -- (-.5,-3.7) -- (-.5,-3.5);
\draw[thick] (12.5,2) -- (12.5,2.2) -- (14.1,2.2) -- (14.1,-3.7) -- (12.5,-3.7) -- (12.5,-3.5);

\draw[thick, fill=black, rounded corners=2pt] (.25,0) rectangle (3.75,-0.25);
\draw[thick, fill=black, rounded corners=2pt] (.25,-1.25) rectangle (3.75,-1.5);
\draw[thick, fill=black, rounded corners=2pt] (.25+8,0) rectangle (3.75+8,-0.25);
\draw[thick, fill=black, rounded corners=2pt] (.25+8,-1.25) rectangle (3.75+8,-1.5);

\end{tikzpicture} \, .   
\ee
\ew
To ease the notation we absorbed a factor of $\eta^{\prime\,1/2}_{\ell-2t,2m}$ in each one of the four copies of the initial state. 

\begin{figure*}[t]
    \centering
\includegraphics[width=0.65\textwidth]{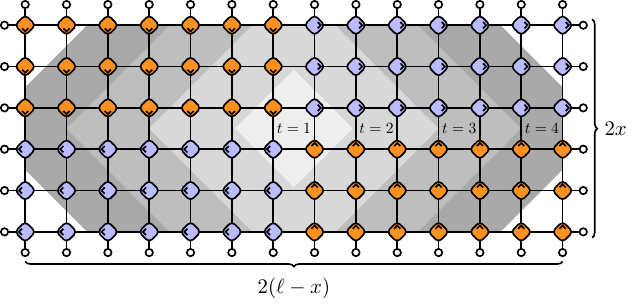}
    \caption{Schematic for the network representing $\textrm{tr}({\rm P}_{\!A,m}^2(t))$ for $\ell=10$ and $x=3$. The concentric shaded regions indicate which regions are present at a given time $t$. The network evolves from a diamond ($t=1$) to hexagon ($t=2,3$), then finally to an octagon ($t=4$). Length $\ell-2t$ of the folded initial state is inserted along the edges of each shape. }
    \label{fig:schematic}
\end{figure*}

This object can be contracted further by appealing to unitarity. The precise shape of the network this yields depends on the time elapsed since the quench. Figure \ref{fig:schematic} summarises the possible outcomes for a given choice of $\ell$ and $x$. In particular, 
\begin{itemize}
    \item At early times $t<x/2$ we obtain a diamond (e.g. $t=1$);
    \item At intermediate times $x/2 <t <(\ell-x)/2$ we obtain a hexagon (e.g. $t=2,3$);
    \item At late times $(\ell-x)/2 < t < \ell/2$ we obtain an octagon (e.g. $t=4$).
\end{itemize}
The edges of each network will be occupied by a folded initial state \eqref{eq:define_folded} of length $\ell-2t$. Comparison of the schematic to Eqs. \eqref{eq:earlytimeTN}, \eqref{eq:midtimeTN} and \eqref{eq:chaoticTN} makes its interpretation at different times clear.  

We now evaluate this network in these different time regimes.

\subsubsection{$0 \leq t < x/2$}
\label{sec:n=2earlytimes}

At early times, the schematic in Figure~\ref{fig:schematic} takes the form of a small central diamond with ladders of initial states protruding from each vertex. Namely
\bw
\be
 \begin{tikzpicture}[baseline=(current  bounding  box.center), scale=0.6]

\path (0,0) pic[rotate=225, scale=1.2] {Urfolded};
\path (1.4,0) pic[rotate=-45, scale=1.2] {Ur_dagfolded};
\path (1.4,-1.4) pic[rotate=45, scale=1.2] {Urfolded};
\path (0,-1.4) pic[rotate=135, scale=1.2] {Ur_dagfolded};

\draw[line width=4,line cap=round] (0.3,1)--(-1,-0.3);
\draw[line width=4,line cap=round] (1.1,1)--(2.4,-0.3);
\draw[line width=4,line cap=round] (1.1,-2.4) -- (2.4,-1.1);
\draw[line width=4,line cap=round] (0.3,-2.4)--(-1,-1.1);

\draw[thick] (1.1,1.1)--(1.1,1)--(0.3,1)--(0.3,1.1);
\draw[thick] (1.1,-2.5)--(1.1,-2.4)--(0.3,-2.4)--(0.3,-2.5);
\draw[thick] (-1.1,-.3)--(-1,-.3)--(-1,-1.1)--(-1.1,-1.1);
\draw[thick] (2.5,-.3)--(2.4,-.3)--(2.4,-1.1)--(2.5,-1.1);

%\draw[thick, fill=black] (0.3,1.5) circle (0.1);
%\draw[thick, fill=black] (1.1,1.5) circle (0.1);
%\draw[thick, fill=white] (0.3,2) circle (0.1);
%\draw[thick, fill=white] (1.1,2) circle (0.1);

\path (0.7,1.8) pic[rotate=90,scale=0.65] {state_pair};
\path (0.7,-3.2) pic[rotate=-90,scale=0.65] {state_pair};

\foreach \i in {0,...,4}
{\path (3.2+1.2*\i,-0.7) pic[rotate=180,scale=0.65] {state_pair};
\path (-1.8-1.2*\i,-0.7) pic[rotate=0,scale=0.65] {state_pair};
}

\draw [thick,decorate,decoration={brace}]
(2.8,-1.4)--(1.4,-2.8);

\node[scale=1] at (2.8,-2.2) {$2t$};

\draw [thick,decorate,decoration={brace}]
(2.8,-1.4)--(1.4,-2.8);

\draw [thick,decorate,decoration={brace}]
(-1.25,-1.5)--(-7.25,-1.5);
\node[scale=1] at (-4.25,-2) {$\ell-x-2t$};

\draw [thick,decorate,decoration={brace}]
(1.5,2.5)--(1.5,1);
\node[scale=1] at (2.75,1.75) {$x-2t$};

\end{tikzpicture},
\label{eq:earlytimeTN}
\ee
\ew
where, for example, we took $\ell=10$, $x=3$, and $t=1$. Let us start by considering the case of the Bell-pair state. In this case the ladders of initial states simplify leading to  
\be
d^{-8t}\, \begin{tikzpicture}[baseline=(current  bounding  box.center), scale=0.6]
\path (-4,-6) pic[rotate=225, scale=1.2] {Urfolded};
\path (1.4-4,0-6) pic[rotate=-45, scale=1.2] {Ur_dagfolded};
\path (1.4-4,-1.4-6) pic[rotate=45, scale=1.2] {Urfolded};
\path (0-4,-1.4-6) pic[rotate=135, scale=1.2] {Ur_dagfolded};
\draw[thick] (0-4,0.7-6)--(1.4-4,0.7-6);
\draw[thick] (-.7-4,0-6) -- (-0.7-4,-1.4-6);
\draw[thick] (0-4,-2.1-6)--(1.4-4,-2.1-6);
\draw[thick] (2.1-4,0-6) -- (2.1-4,-1.4-6);
\end{tikzpicture}
= d^{-8t}\, \begin{tikzpicture}[baseline=(current  bounding  box.center), scale=0.6]
\def\eps{9.5};
\draw[thick, rounded corners=1pt] (\eps-4,-5.3)--(\eps-2.6,-5.3)--(\eps-2.6,-5.6)--(\eps-2.2,-6)--(\eps-1.9,-6)--(\eps-1.9,-7.4)--(\eps-2.2,-7.4)--(\eps-2.6,-7.8)--(\eps-2.6,-8.1)--(\eps-4,-8.1)--(\eps-4,-7.8)--(\eps-4.4,-7.4)--(\eps-4.7,-7.4)--(\eps-4.7,-6)--(\eps-4.4,-6)--(\eps-4,-5.6)--(\eps-4,-5.3);

\draw[thick, rounded corners=1pt] (\eps-3.6,-6)--(\eps-3,-6)--(\eps-2.6,-6.4)--(\eps-2.6,-7)--(\eps-3,-7.4)--(\eps-3.6,-7.4)--(\eps-4,-7)--(\eps-4,-6.4)--(\eps-3.6,-6);
\end{tikzpicture}\,.
\label{eq:loopstlx/2}
\ee
For generic $t$, the r.h.s.\ of \eqref{eq:loopstlx/2} contains $2t$ loops, every one contributing a factor of $d^2$. The $d^{-8t}$ prefactor comes from the normalisation of the Bell pairs. Combining these observations, we deduce
\begin{equation}
\textrm{tr}({\rm P}_{\!A,m}^2(t))|_{\rm Bell} = d^{-4t}.
\end{equation}
For dual-unitary deformations of the Bell-pair state (cf.~\eqref{eq:duDef}) the exact same reasoning goes through with $t$ replaced by $t+1/2$ so the final result reads as 
\begin{equation}
\textrm{tr}({\rm P}_{\!A,m}^2(t))|_{\rm du} = d^{-2} d^{-4t}.
\end{equation}
Finally, this calculation can be extended to more general states \eqref{eq:psi0diagram} in the limit of infinite $\ell$. Indeed, in this case one can use \eqref{eq:mlayersBell} to find
\bw
\begin{align}
\lim_{\ell\to\infty}\textrm{tr}({\rm P}_{\!A,m}^2(t)) &= \frac{1}{d^{4x}}
\begin{tikzpicture}[baseline=(current  bounding  box.center), scale=0.6]
\draw[thick] (-2,-0.7)--(-2,-1.8);
\draw[thick] (-1.25,-0.5)--(-1.25,-2.5);
\draw[thick] (4.5,-0.67)--(4.5,-1.8);
\draw[thick] (3.75,-0.5)--(3.75,-2.5);
\draw[thick] (1,4.1)--(1.5,4.1);
\draw[thick] (1,4.8)--(1.5,4.8);
\draw[thick] (1,4.1-1.5)--(1.5,4.1-1.5);
\draw[thick] (1,4.8-1.5)--(1.5,4.8-1.5);
\draw[thick] (1,4.1-9.9)--(1.5,4.1-9.9);
\draw[thick] (1,4.8-9.9)--(1.5,4.8-9.9);
\draw[thick] (1,4.1-10.7)--(1.5,4.1-10.7);
\draw[thick] (1,3.4-10.7)--(1.5,3.4-10.7);
\draw[thick] (0.5,1.25)--(2.5,1.25);
\draw[thick] (0.5,-3.75)--(2.5,-3.75);
\draw[thick] (0,0)--(2.5,0)--(2.5,-2.5)--(0,-2.5)--(0,0);
\path (0,0) pic[rotate=225, scale=1.2] {Urfolded};
\path (2.5,0) pic[rotate=-45, scale=1.2] {Ur_dagfolded};
\path (2.5,-2.5) pic[rotate=45, scale=1.2] {Urfolded};
\path (0,-2.5) pic[rotate=135, scale=1.2] {Ur_dagfolded};
\def\eps{1.1};
\path (2.55+\eps,-1.27-\eps) pic[rotate=45, scale=1.2] {statefolded};
\path (2.55-1.25+\eps,-1.27-1.25-\eps) pic[rotate=45, scale=1.2] {statefolded};
\def\eps{2.41};
\path (2.55-\eps,-1.27+\eps) pic[rotate=225, scale=1.2] {statefolded};
\path (2.55-1.28-\eps,-1.27-1.28+\eps) pic[rotate=225, scale=1.2] {statefolded};
\def\eps{0.15};
\def\xeps{2.4};
\path (2.55-\eps,-1.27+\xeps) pic[rotate=135, scale=1.2] {state_dagfolded};
\path (2.55+1.25-\eps,-1.27-1.25+\xeps) pic[rotate=135, scale=1.2] {state_dagfolded};
\def\eps{3.7};
\def\xeps{-1.1};
\path (2.55-\eps,-1.27+\xeps) pic[rotate=315, scale=1.2] {state_dagfolded};
\path (2.55+1.25-\eps,-1.27-1.25+\xeps) pic[rotate=315, scale=1.2] {state_dagfolded};
\def\xeps{-0.9};
\def\yeps{4.45};
\path (2.55+\xeps,\yeps) pic[rotate=90, scale=1.2] {state_dagfolded};
\path (2.55+\xeps,-1.5+\yeps) pic[rotate=90, scale=1.2] {state_dagfolded};
\def\xeps{-1.65};
\def\yeps{4.45};
\path (2.55+\xeps,\yeps) pic[rotate=270, scale=1.2] {statefolded};
\path (2.55+\xeps,-1.5+\yeps) pic[rotate=270, scale=1.2] {statefolded};
\def\xeps{-0.9};
\def\yeps{-5.45};
\path (2.55+\xeps,\yeps) pic[rotate=90, scale=1.2] {statefolded};
\path (2.55+\xeps,-1.5+\yeps) pic[rotate=90, scale=1.2] {statefolded};
\def\xeps{-1.67};
\def\yeps{-5.45};
\path (2.55+\xeps,\yeps) pic[rotate=270, scale=1.2] {state_dagfolded};
\path (2.55+\xeps,-1.5+\yeps) pic[rotate=270, scale=1.2] {state_dagfolded};
\path (0.7,5.4) pic[rotate=0] {cross_chi=0.2};
\path (1.85,5.4) pic[rotate=0] {cross_chi=0.2};
\path (0.68,-7.9) pic[rotate=0] {cross_chi=0.2};
\path (1.85,-7.9) pic[rotate=0] {cross_chi=0.2};
\end{tikzpicture} = \frac{\eta_{x-2t,2}}{d^{4t}}\,.
\label{eq:genstateearly}
\end{align}
Where we defined folded $W$
\be \label{eq:statefolded1}
\begin{tikzpicture}[baseline=(current  bounding  box.center), scale=0.75]
\path (0,0) pic[rotate=0,scale=1.5] {statefoldedlegs};
\path (3.5,0.3) pic[rotate=0,scale=1.5] {state_daglegs};
\path (3,0) pic[rotate=0,scale=1.5] {statelegs};
\node[scale=1] at (1.5,0.1) {$=$};
\draw[thick] (2*0.075,2*0.07) -- (2*0.15,2*0.07) -- (2*0.15,-2*0.005);
\draw[thick] (3+2*0.075,2*0.07) -- (3+2*0.15,2*0.07) -- (3+2*0.15,-2*0.005);
\draw[thick] (3.5+2*0.075,0.3-2*0.07) -- (3.5+2*0.15,0.3-2*0.07) -- (3.5+2*0.15,0.3+2*0.005);
\end{tikzpicture} \, ,
\ee
and its Hermitian conjugate $W^\dag$,
\be
 \label{eq:statefolded2}
\begin{tikzpicture}[baseline=(current  bounding  box.center), scale=0.75]
\path (0,0) pic[rotate=0,scale=1.5] {state_dagfoldedlegs};
\path (3.5,0.3) pic[rotate=0,scale=1.5] {statelegs};
\path (3,0) pic[rotate=0,scale=1.5] {state_daglegs};
\node[scale=1] at (1.5,0.1) {$=$};
\draw[thick] (-2*0.075,-2*0.07) -- (-2*0.15,-2*0.07) -- (-2*0.15,2*0.005);
\draw[thick] (3-2*0.075,-2*0.07) -- (3-2*0.15,-2*0.07) -- (3-2*0.15,2*0.005);
\draw[thick] (3.5-2*0.075,0.3+2*0.07) -- (3.5-2*0.15,0.3+2*0.07) -- (3.5-2*0.15,0.3-2*0.005);
\end{tikzpicture} \, .
\ee
In the last step of \eqref{eq:genstateearly}, we used the solvability conditions \eqref{eq:stateuni1}--\eqref{eq:stateuni2}. Namely, we used that the matrix $W$ is unitary. 
 
\subsubsection{$x/2 \leq t < (\ell-x)/2$} 
\label{sec:n=2midtimes}

In this case, which is realised for example for $t=2$ or $t=3$ in Fig.~\ref{fig:schematic},  our network takes the form of a hexagon with just 2 protruding ladders of initial states. Namely
\begin{equation}
\begin{tikzpicture}[baseline=(current  bounding  box.center), scale=0.5]

\foreach \i in {0,...,2}
{\path (1.4*\i,0) pic[rotate=225] {Urfolded};
\path (4.2+1.4*\i,0) pic[rotate=-45] {Ur_dagfolded};
\path (4.2+1.4*\i,-7) pic[rotate=45] {Urfolded};
\path (1.4*\i,-7) pic[rotate=135] {Ur_dagfolded};}
\foreach \i in {0,...,3}
{\path (-1.4+1.4*\i,-1.4) pic[rotate=225] {Urfolded};
\path (4.2+1.4*\i,-1.4) pic[rotate=-45] {Ur_dagfolded};
\path (4.2+1.4*\i,-5.6) pic[rotate=45] {Urfolded};
\path (-1.4+1.4*\i,-5.6) pic[rotate=135] {Ur_dagfolded};}
\foreach \i in {0,...,4}
{\path (-2.8+1.4*\i,-2.8) pic[rotate=225] {Urfolded};
\path (4.2+1.4*\i,-2.8) pic[rotate=-45] {Ur_dagfolded};
\path (4.2+1.4*\i,-4.2) pic[rotate=45] {Urfolded};
\path (-2.8+1.4*\i,-4.2) pic[rotate=135] {Ur_dagfolded};
}

\draw[line width=4,line cap=round] (-3.8,-3.1)--(-0.7,0);
\draw[line width=4,line cap=round] (-3.8,-3.9)--(-0.7,-7);
\draw[line width=4,line cap=round] (10.8,-3.1)--(7.7,0);
\draw[line width=4,line cap=round] (10.8,-3.9)--(7.7,-7);
\draw[thick] (-3.9,-3.1)--(-3.8,-3.1)--(-3.8,-3.9)--(-3.9,-3.9);
\draw[thick] (10.9,-3.1)--(10.8,-3.1)--(10.8,-3.9)--(10.9,-3.9);

%\draw[thick] (-0.7,0)--(-0.7,0.7);
%\draw[thick] (-0.7,-7)--(-0.7,-7.7);
%draw[thick] (7.7,0)--(7.7,0.7);
%draw[thick] (7.7,-7)--(7.7,-7.7);

\path (11.6,-3.5) pic[rotate=180,scale=0.5] {state_pair};
\path (-4.6,-3.5) pic[rotate=0,scale=0.5] {state_pair};

\foreach \i in {0,...,5}
{\path (1.4*\i,0.7) pic[rotate=0] {arrowhead=0.1};
\path (1.4*\i,-7.7) pic[rotate=180] {arrowhead=0.1};
}

%\path (-5.25,-3.1) pic[rotate=90,scale=0.1] {arrowhead};
%\path (-5.25,-3.9) pic[rotate=90,scale=0.1] {arrowhead};
%\path (12.25,-3.1) pic[rotate=270,scale=0.1] {arrowhead};
%\path (12.25,-3.9) pic[rotate=270,scale=0.1] {arrowhead};

%\path (-0.7,0.7) pic[rotate=0] {cross=0.1};
%\path (7.7,0.7) pic[rotate=0] {cross=0.1};
%\path (-0.7,-7.7) pic[rotate=180] {cross=0.1};
%\path (7.7,-7.7) pic[rotate=180] {cross=0.1};

\draw [thick,decorate,decoration={brace}]
(8.3,0.2)--(11.1,-2.6);

\node[scale=1] at (10.3,-.7) {$x$};

\end{tikzpicture}\,,
\label{eq:midtimeTN}
\end{equation}
\ew
where, for example, we took $\ell=10$, $x=3$, and $t=3$. Once again, let us start considering Bell pairs, in this case the tensor network reduces to 
\be
\!\!\!\!\!\frac{1}{d^{4x}}\,\begin{tikzpicture}[baseline=(current  bounding  box.center), scale=0.5]
\path (0,-12) pic[scale=0.5] {int_times};
\end{tikzpicture}, \label{eq:mid_timesBell}
\ee
where the factor $d^{-4x}$ comes from the Bell pairs' normalisation. Using dual-unitarity the network is decomposed in a collection of open lines and closed loops,
\be
\begin{tikzpicture}[baseline=(current  bounding  box.center), scale=0.4]
\path (0,-24) pic[scale=0.4] {coloured_lines};
\end{tikzpicture}\,. \label{eq:mid_timesloops}
\ee
In Eq.~\eqref{eq:mid_timesloops} we have $x$ red central loops, each contributing a factor of $d^2$. Hence, the red loops give a factor of $d^{2x}$. Recalling that the state $\mcirc$ is normalised, the black lines contribute with a factor 1 and can be discarded. Combining these factors we arrive at
\begin{equation}
\textrm{tr}({\rm P}_{\!A,m}^2(t))|_{\rm Bell} = d^{-2x}.
\end{equation}
The same reasoning applies for dual-unitary deformations of the Bell-pair state leading to the same result. The case of more general solvable states can again be treated in the limit of infinite $\ell$, where we are left with the tensor network
\bw
\be
\!\!\!\!\!d^{-4x}\,\begin{tikzpicture}[baseline=(current  bounding  box.center), scale=0.5]

\foreach \i in {0,...,2}
{\path (1.4*\i,0) pic[rotate=225] {Urfolded};
\path (4.2+1.4*\i,0) pic[rotate=-45] {Ur_dagfolded};
\path (4.2+1.4*\i,-7) pic[rotate=45] {Urfolded};
\path (1.4*\i,-7) pic[rotate=135] {Ur_dagfolded};}
\foreach \i in {0,...,3}
{\path (-1.4+1.4*\i,-1.4) pic[rotate=225] {Urfolded};
\path (4.2+1.4*\i,-1.4) pic[rotate=-45] {Ur_dagfolded};
\path (4.2+1.4*\i,-5.6) pic[rotate=45] {Urfolded};
\path (-1.4+1.4*\i,-5.6) pic[rotate=135] {Ur_dagfolded};}
\foreach \i in {0,...,4}
{\path (-2.8+1.4*\i,-2.8) pic[rotate=225] {Urfolded};
\path (4.2+1.4*\i,-2.8) pic[rotate=-45] {Ur_dagfolded};
\path (4.2+1.4*\i,-4.2) pic[rotate=45] {Urfolded};
\path (-2.8+1.4*\i,-4.2) pic[rotate=135] {Ur_dagfolded};
}

\draw[thick] (-1.4,-0.7)--(-0.7,0);
\draw[thick] (-2.1,-1.4)--(-2.8,-2.1);
%\draw[thick] (-3.5,-2.8)--(-3.5,-4.2);
\draw[thick] (8.4,-0.7)--(7.7,0);
\draw[thick] (9.1,-1.4)--(9.8,-2.1);
%\draw[thick] (10.5,-2.8)--(10.5,-4.2);
\draw[thick] (9.8,-4.9)--(9.1,-5.6);
\draw[thick] (8.4,-6.3)--(7.7,-7);
\draw[thick] (-2.8,-4.9)--(-2.1,-5.6);
\draw[thick] (-1.4,-6.3)--(-0.7,-7);
\draw[thick] (-3.5,-2.8)--(-5,-2.8);
\draw[thick] (-3.5,-4.2)--(-5,-4.2);
\draw[thick] (10.5,-2.8)--(12.2,-2.8);
\draw[thick] (10.5,-4.2)--(12.2,-4.2);

\draw[thick] (-5.3,-2.5)--(-5.3,-4.5);
\draw[thick,rounded corners=2pt] (-6.1,-2.3)--(-6.5,-2.3)--(-6.5,-4.7)--(-6.1,-4.7);

\draw[thick,rounded corners=2pt] (13.1,-2.25)--(13.6,-2.25)--(13.6,-4.7)--(13.2,-4.7);

%\def\eps{-2}
%\draw[thick] (-4.55+\eps,-2.5)--(-4.55+\eps,-4.5);
%\draw[thick] (-5.3+\eps,-2.5)--(-5.3+\eps,-4.5);

\def\eps{16.8}
\draw[thick] (-4.55+\eps,-2.5)--(-4.55+\eps,-4.5);
\def\eps{18.9}
%\draw[thick] (-4.55+\eps,-2.5)--(-4.55+\eps,-4.5);
%\draw[thick] (-5.3+\eps,-2.5)--(-5.3+\eps,-4.5);

\foreach \i in {0,...,5}
{\path (1.4*\i,0.7) pic[rotate=0] {arrowhead=0.1};
\path (1.4*\i,-7.7) pic[rotate=180] {arrowhead=0.1};
}

\path (-2.6,-1.6) pic[rotate=225,scale=1.25] {statefolded};
\path (-1.2,-0.2) pic[rotate=225,scale=1.25] {statefolded};

\def\epsx{10.9}
\def\epsy{-5.2}
\path (\epsx-2.6,\epsy-1.6) pic[rotate=45,scale=1.25] {statefolded};
\path (\epsx-1.2,\epsy-0.2) pic[rotate=45,scale=1.25] {statefolded};

\def\epsx{8.25}
\def\epsy{-0.15}
\path (\epsx+0,\epsy+0) pic[rotate=135,scale=1.25] {state_dagfolded};
\path (\epsx+1.4,\epsy-1.4) pic[rotate=135,scale=1.25] {state_dagfolded};

\def\epsx{-2.6}
\def\epsy{-5.4}
\path (\epsx+0,\epsy+0) pic[rotate=315,scale=1.25] {state_dagfolded};
\path (\epsx+1.4,\epsy-1.4) pic[rotate=315,scale=1.25] {state_dagfolded};

\path (-0.5,.85) pic[rotate=0] {cross=0.2};
\path (-0.5,-7.85) pic[rotate=0] {cross=0.2};
\path (7.62,-7.85) pic[rotate=0] {cross=0.2};
\path (7.6,0.85) pic[rotate=0] {cross=0.2};

%\draw[thick] (-0.5,.85) arc (-45:90:0.2);
%\draw[thick] (-0.5-3.1,.85-3.1) arc (-45:-180:0.2);
%\draw[thick] (-0.5-3.1,.85-5.7) arc (45:180:0.2);
%\draw[thick] (-0.5,-7.85) arc (45:-90:0.2);
%\draw[thick] (7.62,-7.85) arc (135:270:0.2);
%\draw[thick] (10.75,-4.7) arc (135:0:0.2);
%\draw[thick] (10.65,-2.2) arc (225:350:0.2);
%\draw[thick] (7.6,0.85) arc (225:90:0.2);

\path (11.93,-2.5) pic[rotate=180,scale=1.25] {state_dagfolded};
%\path (13.93,-2.5) pic[rotate=180,scale=1.25] {state_dagfolded};
\path (-4.9,-2.55) pic[rotate=180,scale=1.25] {statefolded};
%\path (-6.9,-2.55) pic[rotate=180,scale=1.25] {statefolded};

\path (12,-4.45) pic[rotate=0,scale=1.25] {statefolded};
%\path (14,-4.45) pic[rotate=0,scale=1.25] {statefolded};
\path (-4.9,-4.45) pic[rotate=0,scale=1.25] {state_dagfolded};
%\path (-6.9,-4.45) pic[rotate=0,scale=1.25] {state_dagfolded};

%\draw[thick] (-8.1,-2.3) arc (90:225:0.2);
%\draw[thick] (-8.1,-4.7) arc (-90:-225:0.2);
%\draw[thick] (15.15,-2.25) arc (90:-45:0.2);
%\draw[thick] (15.15,-4.7) arc (-90:45:0.2);

\end{tikzpicture}
\label{eq:mid_timesgeneral}
\ee
Using the unitarity of $W$ once again, we can contract the network to find 
\be
\lim_{\ell\to\infty}\textrm{tr}({\rm P}_{\!A,m}^2(t)) = d^{-2x}.
\ee

\subsubsection{$(\ell-x)/2 \leq t < \ell/2$} 
\label{sec:n=2chaos}

At later times the tensor network takes the form of an octagon, see e.g. Figure~\ref{fig:schematic} for $t=4$. Namely we have 

\be
\textrm{tr}({\rm P}_{\!A,m}^2(t))
\begin{tikzpicture}[baseline=(current  bounding  box.center), scale=0.5]

\foreach \i in {0,...,4}
{\path (2.8+1.4*\i,0) pic[rotate=225] {Urfolded};
\path (9.8+1.4*\i,0) pic[rotate=-45] {Ur_dagfolded};
\path (2.8+1.4*\i,-7) pic[rotate=135] {Ur_dagfolded};
\path (9.8+1.4*\i,-7) pic[rotate=45] {Urfolded};}

\foreach \i in {0,...,5}
{\path (1.4+1.4*\i,-1.4) pic[rotate=225] {Urfolded};
\path (9.8+1.4*\i,-1.4) pic[rotate=-45] {Ur_dagfolded};
\path (1.4+1.4*\i,-5.6) pic[rotate=135] {Ur_dagfolded};
\path (9.8+1.4*\i,-5.6) pic[rotate=45] {Urfolded};}

\foreach \i in {0,...,6}
{\path (1.4*\i,-2.8) pic[rotate=225] {Urfolded};
\path (9.8+1.4*\i,-2.8) pic[rotate=-45] {Ur_dagfolded};
\path (1.4*\i,-4.2) pic[rotate=135] {Ur_dagfolded};
\path (9.8+1.4*\i,-4.2) pic[rotate=45] {Urfolded};}

\foreach \i in {2,3}
{\path (18.9,-1.4*\i) pic[rotate = 270] {arrowhead=0.1};
}
\foreach \i in {2,3}
{\path (-0.7,-1.4*\i) pic[rotate = 90] {arrowhead=0.1};
}
\foreach \i in {2,...,11}
{\path (1.4*\i,0.7) pic[rotate =0] {arrowhead=0.1};
}
\foreach \i in {2,...,11}
{\path (1.4*\i,-7.7) pic[rotate =180] {arrowhead=0.1};
}

\draw [thick,decorate,decoration={brace}]
(19.7,0.2) -- (19.7,-7.2);
\draw [thick,decorate,decoration={brace}]
(18.2,-8.5) -- (0,-8.5);

\draw [thick,decorate,decoration={brace}]
(-0.2,-1.8) -- (1.8,0.2);

\draw [thick,decorate,decoration={brace}]
(2.8,1) -- (15.4,1);
\node[scale=1] at (9.1,1.75) {$2(2t-x)$};

\node[scale=1] at (9.1,-9.4) {$2(\ell-x)$};
\node[scale=1] at (20.5,-3.5) {$2x$};
\node[scale=1] at (-0.1-0.2,-0.4+0.2) {$\ell-2t$};

\draw[line width=4,line cap=round] (2.1,0)--(0,-2.1);
\draw[line width=4,line cap=round] (2.1,-7)--(0,-4.9);
\draw[line width=4,line cap=round] (16.1,0)--(18.2,-2.1);
\draw[line width=4,line cap=round] (16.1,-7)--(18.2,-4.9);

\end{tikzpicture}\,,
\label{eq:chaoticTN}
\ee

\ew
with $\ell=10$, $x=3$, and $t=4$. Assuming dual-unitarity alone, this network does not contract completely. Rather, we find that the result of the contraction depends on our particular choice of dual-unitary gate. To see this, here we explicitly contract the network in the two extreme cases of the SWAP gate --- which implements non-interacting dynamics --- and completely-chaotic dual-unitary circuits --- which, as discussed in Sec.~\ref{section:dual_unitary}, can be considered the ``most chaotic'' local quantum circuits.

Specialising to the SWAP gate (cf.~\eqref{eq:SWAP}), and considering the Bell-pair initial state, the network \eqref{eq:chaoticTN} specialises to 
\begin{equation}
\frac{1}{d^{4(\ell-2t)}}\begin{tikzpicture}[baseline=(current  bounding  box.center), scale=0.35]
\path (0,-14) pic[scale=0.35] {cage};
\end{tikzpicture} \,.
\label{eq:swap}
\end{equation}
Once again black lines make no contribution due the normalisation of $\ket{\mcirc}$. We also have $2(\ell-2t)$ red loops, each contributing a factor of $d$, and a prefactor of $d^{-4(\ell-2t)}$ from the initial state. Therefore
\begin{equation}
\textrm{tr}({\rm P}_{\!A,m}^2(t))|_{\rm SWAP, Bell} = d^{-2(\ell-2t)}.   
\label{eq:chaoticSWAPBell}
\end{equation}
For generic solvable states \eqref{eq:psi0diagram} one has 
\begin{equation}
\textrm{tr}({\rm P}_{\!A,m}^2(t))|_{\rm SWAP} = \textrm{tr}(\mathcal B^2),   
\label{eq:chaoticSWAP}
\end{equation}
where $\mathcal B$ is defined by

\be \label{eq:def_B}
\mathcal{B}=
\begin{tikzpicture}[baseline=(current  bounding  box.center), scale=0.5]

\draw [thick,decorate,decoration={brace}]
(3,0)--(3,-5);
\node[scale=1] at (4.5,-2.5) {$\ell-2t$};

\draw[line width=8,line cap=round] (0,0)--(0,-5);
\draw[line width=8,line cap=round] (1.5,0)--(1.5,-5);

\foreach \i in {0,2}
{ \filldraw[thick, fill=black, draw=white] (1.3,-1.05-\i) rectangle (2.8,-0.95-\i);
  \filldraw[thick, fill=black, draw=white] (0.2,-1.05-\i) rectangle (-1.5,-0.95-\i);
  }
 \draw[thick] (0,-2)--(1.5,-2);
  \draw[thick] (0,-4)--(1.5,-4);

\end{tikzpicture} \, .
\ee
Eq.~\eqref{eq:chaoticSWAP} simplifies in the limit of large $\ell -2t$. Indeed, using \eqref{eq:infinitepowertau} we find
\begin{equation} 
\label{eq:Btildefolded}
\mathcal{B}\simeq \tilde{\cal B}=\frac{\chi^2}{d^{2(\ell-2t)}}
    \begin{tikzpicture}[baseline=(current  bounding  box.center), scale=0.5]

\foreach \i in {0,2.4,4.8}
{
\draw[thick,rounded corners=1pt] (0,-0.5-\i)--(1.5,-0.5-\i);
\draw[thick,rounded corners=1pt] (0,0.5-\i)--(0.5,0.5-\i)--(0.5,1-\i)--(-1,1-\i);
\draw[thick,rounded corners=1pt] (1.5,0.5-\i)--(1,0.5-\i)--(1,1-\i)--(2.5,1-\i);
}

\foreach \i in {0,-2.4,-4.8}
{
\path (0,\i) pic[rotate=-90,scale=1.5] {state_dagfolded};
\path (1.5,\i) pic[rotate=90,scale=1.5] {statefolded};
}

\path (-0.3,1.5) pic[rotate=0] {arrowhead=0.2};
\path (1.8,1.5) pic[rotate=0] {arrowhead=0.2};
\path (-0.3,1.5-7.8) pic[rotate=180] {arrowhead=0.2};
\path (1.8,1.5-7.8) pic[rotate=180] {arrowhead=0.2};

\draw [thick,decorate,decoration={brace}]
(2.75,1.5)--(2.75,-6.3);

\node[scale=1] at (4,-2.4) {$\ell-2t$};
\end{tikzpicture},
\end{equation}
where $\tilde{\cal B}$ is nothing but Eq.~\eqref{eq:def_B_tilde} in the folded representation (cf.~\eqref{eq:statefolded1}--\eqref{eq:statefolded2}). The trace in \eqref{eq:chaoticSWAP} can be expressed in terms of the following transfer matrix 
\begin{equation}
\label{eq:Mmap2}
\mathcal{M}_2=
    \begin{tikzpicture}[baseline=(current  bounding  box.center), scale=0.5]

\draw[thick,rounded corners=1pt] (4.5,0.5)--(4,0.5)--(4,1)--(5.5,1)--(5.5,2)--(-1,2)--(-1,1);

\foreach \i in {0}
{
\draw[thick,rounded corners=1pt] (1.5+\i,0.5)--(1+\i,0.5)--(1+\i,1)--(3.5+\i,1)--(3.5+\i,0.5)--(3+\i,0.5);
}

\draw[thick,rounded corners=1pt] (0,-0.5)--(1.5,-0.5);
\draw[thick,rounded corners=1pt] (3,-0.5)--(4.5,-0.5);
%\draw[thick,rounded corners=1pt] (6,-0.5)--(7.5,-0.5);
\draw[thick,rounded corners=1pt] (0,0.5)--(0.5,0.5)--(0.5,1)--(-1,1);

\foreach \i in {0,3}
{
\path (\i,0) pic[rotate=-90,scale=1.5] {state_dagfolded};
\path (\i+1.5,0) pic[rotate=90,scale=1.5] {statefolded};
}

\end{tikzpicture}, 
\end{equation}
and,  as discussed in Appendix~\ref{app:SWAP}, it yields
\be
\textrm{tr}({\rm P}_{\!A,m}^2(t))|_{\rm SWAP} \simeq \chi^2 d^{-2(\ell-2t)}\,,
\ee
for almost all initial-state matrices $W$.

Going back to generic dual-unitary gates we see that the diagram \eqref{eq:chaoticTN} is explicitly written in terms of the transfer matrices ${\mathcal T}_{1,x}$ and ${\mathcal T}_{2,x}$ introduced in Sec.~\ref{section:dual_unitary}. In particular, we have 
\begin{align}
\textrm{tr}({\rm P}_{\!A,m}^2(t)) & = \|{\mathcal T}^{2t-x}_{1,x}\ket{\phi_{1}}_x\|^2 \notag\\
     &= \|{\mathcal T}^{2t-(\ell-x)}_{2, \ell-x}\ket{\phi_{2}}_{\ell-x}\|^2,\label{eq:TMPn2}
\end{align}
where we defined 
\begin{equation} \label{eq:define_T} \ket{\phi_{1}}_x=
\begin{tikzpicture}[baseline=(current  bounding  box.center), scale=0.5]

\foreach \i in {0,...,3}
\draw[thick] (1.4+1.4*\i,-1.3)--(1.4+1.4*\i,-2.5);

\foreach \i in {0}
{\path (2.8+1.4*\i,0) pic[rotate=45] {Ur_dagfolded};
\path (4.2+1.4*\i,0) pic[rotate=135] {Urfolded};
}
\foreach \i in {0,...,1}
{\path (1.4+1.4*\i,-1.4) pic[rotate=45] {Ur_dagfolded};
\path (4.2+1.4*\i,-1.4) pic[rotate=135] {Urfolded};
}

\foreach \i in {2,...,3}
{\path (1.4*\i,0.7) pic[rotate=0] {arrowhead=0.2};
}

\draw[dashed,thick,orange] (3.5,1.3)--(3.5,-2.4);
\draw[dashed,thick,orange] (0,1.3)--(0,-1.3);

\draw [thick,decorate,decoration={brace}]
(0,1.3)--(3.4,1.3);
\draw [thick,decorate,decoration={brace}]
(3.6,1.3)--(5,1.3);
\draw [thick,decorate,decoration={brace}]
(5.6,1)--(8.1,-1.5);

\node[scale=1] at (1.75,2) {$x$};
\node[scale=1] at (4.4,2) {$x-(\ell-2t)$};
\node[scale=1] at (8,0.5) {$\ell-2t$};

\draw[thick] (1.4,-0.7)--(2.1,0);
\draw[thick] (5.6,-0.7)--(4.9,0);
\draw[thick] (0.2,-1.5)--(0.2,-2.5);
\draw[thick] (6.8,-1.5)--(6.8,-2.5);

\draw[line width=4,line cap=round] (1.9,0)--(0.2,-1.7);
\draw[line width=4,line cap=round] (5.1,0)--(6.8,-1.7);

\end{tikzpicture},
\end{equation}
and 
\begin{equation}
\ket{\phi_{2}}_x=
    \begin{tikzpicture}[baseline=(current  bounding  box.center), scale=0.5]

\foreach \i in {0,...,3}
\draw[thick] (1.4+1.4*\i,-1.3)--(1.4+1.4*\i,-2.5);

\foreach \i in {0}
{\path (2.8+1.4*\i,0) pic[rotate=225] {Urfolded};
\path (4.2+1.4*\i,0) pic[rotate=-45] {Ur_dagfolded};
}
\foreach \i in {0,...,1}
{\path (1.4+1.4*\i,-1.4) pic[rotate=225] {Urfolded};
\path (4.2+1.4*\i,-1.4) pic[rotate=-45] {Ur_dagfolded};
}

\foreach \i in {2,...,3}
{\path (1.4*\i,0.7) pic[rotate=0] {arrowhead=0.2};
}

\draw[dashed,thick,orange] (3.5,1.3)--(3.5,-2.4);
\draw[dashed,thick,orange] (0,1.3)--(0,-1.3);

\draw [thick,decorate,decoration={brace}]
(0,1.3)--(3.4,1.3);
\draw [thick,decorate,decoration={brace}]
(3.6,1.3)--(5,1.3);
\draw [thick,decorate,decoration={brace}]
(5.6,1)--(8.1,-1.5);

\node[scale=1] at (1.75,2) {$\ell-x$};
\node[scale=1] at (4.4,2) {$2t-x$};
\node[scale=1] at (8,0.5) {$\ell-2t$};

\draw[thick] (1.4,-0.7)--(2.1,0);
\draw[thick] (5.6,-0.7)--(4.9,0);
\draw[thick] (0.2,-1.5)--(0.2,-2.5);
\draw[thick] (6.8,-1.5)--(6.8,-2.5);

\draw[line width=4,line cap=round] (1.9,0)--(0.2,-1.7);
\draw[line width=4,line cap=round] (5.1,0)--(6.8,-1.7);

\end{tikzpicture}\,.
\end{equation}

The first line of \eqref{eq:TMPn2} suggests that $\textrm{tr}({\rm P}_{\!A,m}^2(t))$ becomes accessible in the scaling limit \eqref{eq:scaling1}, namely $\ell,t\to\infty$ with $\ell-2t\equiv k \leq  x$ fixed. Indeed, in this limit the finite, contracting, matrix ${\mathcal T}_{1,x}$ is taken to an infinite power. This means that it can be replaced by a projector on the eigenspaces corresponding to eigenvectors with unit magnitude. Analogously, the second line of  \eqref{eq:TMPn2} accounts for the fact that that one can keep fixed $\ell-x$ instead of $x$. This is expected as the entanglement is symmetric under the exchange of $A$ and $\bar A$.

The evaluation of the limit \eqref{eq:scaling1} becomes particularly simple for completely chaotic dual-unitary gates (cf. Sec.~\ref{section:dual_unitary}). Indeed, for gates in this class the only unit-magnitude eigenvalue of ${\mathcal T}_{1,x}$ is one and the corresponding invariant subspace is spanned by the ORS $\{\ket{y}_x\}_{i=0}^x$ (cf. Eq.~\eqref{eq:ors}). Therefore we have 
\begin{equation}
\limsc {\mathcal T}_{1,x}^{2t-x} = \sum_{y=0}^{x} \ket{{y}}_{\!xx}\!\!{\bra{{y}}}.
\end{equation}
Making this replacement, we find 
\begin{equation} \label{eq:projectors}
\limsc \textrm{tr}({\rm P}_{\!A,m}^2(t))=\sum_{y=0}^{x} |{}_x\!\!\braket{y|\phi_{1}}_x|^2.
\end{equation}
For the Bell-pair state~\eqref{eq:psi0Bell} the overlaps ${}_x\!\!\braket{y|\phi_{1}}_x$ can be evaluated exactly. The calculation is shown explicitly in  Appendix~\ref{app:overlaps} while the result reads as 
\begin{equation} \label{eq:ors_cont}
   {}_x\!\!\braket{y|\phi_{1}}_x = 
    \begin{cases}
      d^{-x-y-1} \sqrt{d^2-1}, & \text{if}\ y< k\\
      0, & \text{if}\ k\leq y<x\\
      d^{-2k} , & \text{if}\ y=x\\
    \end{cases}\,.
  \end{equation}
Plugging this result in Eq.~\eqref{eq:projectors}, we arrive at
\begin{equation}
\limsc \textrm{tr}({\rm P}_{\!A,m}^2(t))\bigr |_{\rm cc, Bell} = d^{-4 k} + d^{-2x}(1-d^{-2k}).
\label{eq:completelychaoticn2Bell}
\end{equation}
For dual-unitary deformations of the Bell-pair state (cf.~\eqref{eq:duDef}) we obtain the same result with the replacement \eqref{eq:replacement}, while for generic initial states the overlap can be evaluated explicitly only at the leading order in $k$. In this case, as shown in Appendix~\ref{app:overlaps}, we find 
\be \label{eq:ors_contgen}
   {}_x\!\!\braket{y|\phi_{1}}_x \simeq 
    \begin{cases}
     d^{-x-y-1} \sqrt{d^2-1} ,& \text{if}\ y\ll k\\
      \chi^2 d^{-2k } , & \text{if}\ y=x\\
    \end{cases}.
\ee
Plugging back into \eqref{eq:projectors} we find the following leading order result for large $k$
\be
\limsc \textrm{tr}({\rm P}_{\!A,m}^2(t))\bigr |_{\rm cc} \simeq \chi^4 d^{-4 k} + d^{-2x},
\label{eq:completelychaoticn2gen}
\ee
which agrees with \eqref{eq:completelychaoticn2Bell} when we make the replacement \eqref{eq:genericshift}.

\subsection{Evaluation of \eqref{eq:npurity} for $n>2$} 
\label{sec:highern}

Now we generalise our results to arbitrary $n \in \mathbb{Z}^+$. Once again, the case $2 t\geq\ell$ is immediate. Indeed, using Eq.~\eqref{eq:PBl2t} we find 
\be
\textrm{tr}({\rm P}_{\!A,m}^n(t))=1\,.
\ee
From now on we consider $2 t<\ell$. We begin noting that all the networks we evaluated for $n=2$ consist of identical left and right and halves, $\mathcal H$, related by a 180\degree \, rotation. $\mathcal{H}$ can be thought of as a $d^{2x}\times d^{2x}$ matrix (cf. Eqs~\eqref{eq:earlytimeTN}, \eqref{eq:midtimeTN}, and \eqref{eq:chaoticTN}). For example, in the regime $(\ell-x) \leq 2 t < \ell$ we have 
\be
[\mathcal H]_{a_1a_2a_3}^{b_1b_2b_3}=\!\!\!\!\begin{tikzpicture}[baseline=(current  bounding  box.center), scale=0.5]

\foreach \i in {0,...,4}
{\path (2.8+1.4*\i,0) pic[rotate=225] {Urfolded};
\path (2.8+1.4*\i,-7) pic[rotate=135] {Ur_dagfolded};
}

\foreach \i in {0,...,5}
{\path (1.4+1.4*\i,-1.4) pic[rotate=225] {Urfolded};
\path (1.4+1.4*\i,-5.6) pic[rotate=135] {Ur_dagfolded};
}

\foreach \i in {0,...,6}
{\path (1.4*\i,-2.8) pic[rotate=225] {Urfolded};
\path (1.4*\i,-4.2) pic[rotate=135] {Ur_dagfolded};}

\foreach \i in {2,3}
{\path (-0.7,-1.4*\i) pic[rotate = 90] {arrowhead=0.1};
}
\foreach \i in {2,...,6}
{\path (1.4*\i,0.7) pic[rotate =0] {arrowhead=0.1};
}
\foreach \i in {2,...,6}
{\path (1.4*\i,-7.7) pic[rotate =180] {arrowhead=0.1};
}

\draw [thick,decorate,decoration={brace}]
(-0.2,-1.8) -- (1.8,0.2);

\draw [thick,decorate,decoration={brace}]
(2.8,1) -- (9.1,1);
\node[scale=1] at (5.95,1.75) {$2t-x$};

\node[scale=1] at (-0.1-0.2,-0.4+0.2) {$\ell-2t$};

\draw[line width=4,line cap=round] (2.1,0)--(0,-2.1);
\draw[line width=4,line cap=round] (2.1,-7)--(0,-4.9);

\foreach \i in {1,...,3}
{\node[scale=1] at (9.5,-8.4+1.4*\i) {$a_\i$};
}

\foreach \i in {1,...,3}
{\node[scale=1] at (9.5,1.4-1.4*\i) {$b_\i$};
}

\end{tikzpicture},
\ee
with $a_i,b_i\in\{0,\ldots 3\}$. In this language, higher traces of ${\rm P}_{\!A,m}(t)$ are expressed as 
\begin{equation}
   \textrm{tr}({\rm P}_{\!A,m}^n(t)) = \textrm{tr}(\mathcal H^n). 
   \label{eq:PH}
\end{equation}
The matrix $\cal H$ can be simplified by repeating the steps presented in the previous subsection. In particular, for the Bell-pair state and its unitary deformations we have 
\begin{align}
{\mathcal H} &= \frac{1}{d^{4t}} \mathcal Q\otimes\1_{d^2}^{\otimes 2 t} & & 2t < x\,, \label{eq:Hearlytimes}\\
{\mathcal H} &= \frac{1}{d^{2x}}\1_{d^2}^{\otimes x} & & x/2\leq 2t < (\ell-x)\,,
\label{eq:Hmidtimes}
\end{align}
where we introduced $\mathcal Q$ 
\begin{equation}
\mathcal{Q}=
    \begin{tikzpicture}[baseline=(current  bounding  box.center), scale=0.5]

\foreach \i in {0,-2.4}
\foreach \j in {0,-2.65}
{
{
\draw[thick] (\j+1.7,0.4+\i)--(\j+3,0.4+\i);
\draw[thick] (\j+1.7,-0.4+\i)--(\j+3,-0.4+\i);
}}

\foreach \i in {0,-2.4}
{
\path (0,\i) pic[rotate=90,scale=1.5] {state_dagfolded};
\path (2,\i) pic[rotate=-90,scale=1.5] {statefolded};
}

\draw[thick, rounded corners =2pt] (0.3,-3.5)--(0.3,-4.25)--(1.7,-4.25)--(1.7,-3.5);

\path (0.3,1.5) pic[rotate=0] {cross_chi=0.2};
\path (1.7,1.5) pic[rotate=0] {cross_chi=0.2};

\draw [thick,decorate,decoration={brace}]
(3.5,1.5)--(3.5,-4.3);

\node[scale=1] at (5,-1.4) {$x-2t$};

\end{tikzpicture}\,.
\label{eq:Qdiagram}
\end{equation}
Eqs.~\eqref{eq:Hearlytimes}--\eqref{eq:Hmidtimes} immediately lead to  
\be
\!\!\!\textrm{tr}({\rm P}_{\!A,m}^n(t))  \!\!=\! 
 \begin{cases}
d^{4t (1-n)}  \eta_{x-2t,n} & 2t < x \\
d^{2x (1-n)} & x\leq 2t < (\ell-x)
 \end{cases}\,,
 \label{eq:trnmidearly}
\ee
where we used $\textrm{tr}(\mathcal Q^n)=\eta_{x-2t,n}$ (cf.\ Eq.~\eqref{eq:eta}). For generic initial states \eqref{eq:psi0} the results \eqref{eq:Hearlytimes}, \eqref{eq:Hmidtimes}, and \eqref{eq:trnmidearly} are recovered in the limit of infinite $\ell$. Note that for the Bell-pair state and its dual-unitary deformations we have
\be
\eta_{x-2t,n} =\chi^{2(1-n)},
\ee
while for generic initial states this result holds only in the limit of infinite $x$.

Once again, in the region ${(\ell-x) < 2t < \ell}$ dual-unitarity is not enough to fully contract the tensor networks and the result depends on the specific (families of) gates considered. Considering SWAP gates we have 
\be
{\mathcal H}\bigr |_{\rm SWAP} = \mathcal B \otimes \mathcal P^{\otimes x-\ell+2t}\,, \quad (\ell-x) < 2t < \ell
\label{eq:HchaosSWAP}
\ee
where $\mathcal B$ is defined in Eq.~\eqref{eq:def_B}, and 
\be
\mathcal P=
\begin{tikzpicture}[baseline=(current  bounding  box.center), scale=0.5]
\draw[thick] (16.5,-2.5) -- (16.5,-3.25);
\draw[thick] (16.5,-4.5) -- (16.5,-3.75);

\path (16.5,-3.25) pic[rotate = 180] {arrowhead=0.1};
\path (16.5,-3.75) pic[rotate = 0] {arrowhead=0.1};
\end{tikzpicture}\,.
\ee
The form \eqref{eq:HchaosSWAP} leads to 
\be
\!\!\!\textrm{tr}({\rm P}_{\!A,m}^n(t))\bigr |_{\rm SWAP} = \textrm{tr}(\mathcal B^n)\,.
\ee
In particular, for the Bell-pair state we have $\mathcal B=\1/d^{2(\ell-2t)}$, which gives 
\be
%\lim_{\substack{\ell,t,x\to\infty \\ (\ell-x) < 2t < \ell }}
\textrm{tr}({\rm P}_{\!A,m}^n(t))\bigr |_{\rm SWAP, Bell} = d^{2(\ell-2t)(1-n)}.
\ee
Instead, for generic solvable initial states we find (see Appendix~\ref{app:SWAP})
\be
\textrm{tr}({\rm P}_{\!A,m}^n(t))|_{\rm SWAP} \simeq \chi^2 d^{2(\ell-2t)(1-n)}\,,
\ee
for large $\ell-2t$. 

For gates in the completely chaotic family, we can compute $\mathcal H$ in the scaling limit \eqref{eq:scaling1}. The result reads as 
\be
\limsc {\mathcal H}\bigr |_{\rm cc} = \sum_{y=0}^x {}_x\!\braket{y|\psi_{1}}_x \mathcal R_{x,y}
\label{eq:scalinglimitH}
\ee
where we introduced 
\begin{align}
\!\!\!\! \mathcal R_{x,y}\! &= \frac{d^{1+y-x}}{\sqrt{d^2-1}} \mathcal P^{\otimes y}\!\otimes\!(\1_{d^2}\!-\! \mathcal P)\!\otimes\!\! \1_{d^2}^{\otimes (x-y-1)}\, & & \!\!\!\!\!y<x,\\
\!\!\!\! \mathcal R_{x,x}\! &=  \mathcal P^{\otimes x}\, & & \!\!\!\!\!y=x.
\end{align}
In particular, for the Bell-pair state the expression \eqref{eq:scalinglimitH} can be simplified using  \eqref{eq:ors_cont}. The end result reads as  
\begin{align} 
\!\!\!\limsc {\mathcal H}\bigr |_{\rm cc, Bell} =& \frac{1}{d^{2k}}\mathcal P^{\otimes x}  \label{eq:def_p}\\
&+ \frac{1}{d^{2x}} (\1_{d^2}^{\otimes k} - \mathcal P ^{\otimes k}) \otimes \1_{d^2}^{\otimes (x-k)}\!\!. 
\end{align}  
Plugging into \eqref{eq:PH} we then find 
\be
\!\!\!\limsc \textrm{tr}({\rm P}_{\!A,m}^n(t))\bigr |_{\rm cc, Bell}\!\! = d^{-2n k} \!\!+d^{2(1-n)x}(1-d^{-2 k}).
\label{eq:trPnchaosBell}
\ee
Once again, for dual-unitary deformations of the Bell-pair state (cf.~\eqref{eq:duDef}) we obtain the same result with the replacement \eqref{eq:replacement}, while for generic initial states we can access the leading order in $k=\ell-2t$. In particular, using \eqref{eq:ors_contgen} we find 
\begin{align}
\!\!\!\limsc \textrm{tr}({\rm P}_{\!A,m}^n(t))\bigr |_{\rm cc}\simeq&  \,\chi^{2n} d^{-2n k}+ d^{2(1-n)x},\,\,\, n\geq2\,,
\end{align}
which agrees with \eqref{eq:trPnchaosBell} with the replacement \eqref{eq:genericshift}. 

\section{Conclusions}
\label{sec:conclusions}

In this paper we have computed analytically the ``entanglement barriers'' described by the operator entanglement entropies of (positive powers of) the reduced density matrix after quantum quenches in dual-unitary circuits. To the best of our knowledge, this is the first analytical characterisation of these quantities in a concrete (microscopic) lattice system. 

Using simple and intuitive diagrammatic calculations we have provided exact expressions for the barriers when the initial state is a collection of Bell pairs and showed that, up to a shift in time, they describe the asymptotic shape of the barrier obtained for general (solvable) initial MPSs with higher entanglement. We have identified a universal linear growth phase, which is independent of the specific dual-unitary gate governing the dynamics, of the R\'enyi index, and of the power of the reduced density matrix. We have also identified a decay phase which is sensitive to the choice of gate and to the R\'enyi {index,} while remaining independent of the power of the reduced density matrix. We have provided an exact characterisation of the latter for two prototypical classes of dual-unitary gates: the noninteracting family, represented by the SWAP gate, and the completely chaotic family~\cite{bertini2020operator}. These represent the two extreme points of the ergodicity spectrum; the former implements integrable (or better free) behaviour whereas the latter is `structureless' and generic (i.e.\ quantum chaotic). Whilst the SWAP circuits give entanglement dynamics as in rational CFTs~\cite{dubail2017entanglement}, the completely chaotic family give results in quantitative agreement with holographic CFTs~\cite{wang2019barrier}, showing a longer barrier followed by a sharp drop for the von Neumann operator entanglement entropy. Moreover we find that for each fixed R\'enyi index the barrier is longer for the completely chaotic case. This confirms in a concrete class of lattice systems the proposal of Ref.~\cite{wang2019barrier}, i.e., that the length of the entanglement barrier provides a good indicator of quantum chaos. Interestingly, the completely chaotic family gives a nontrivial R\'enyi spectrum such that the barriers for non-interacting and completely chaotic circuits converge in the limit of infinite replicas. Finally, we have argued that the linear growth in time of all R\'enyi entropies of ${\rho_B(t)}$ suggests that the reduced density matrix \emph{cannot} be efficiently approximated by an MPO throughout its evolution in time~\cite{schuch2008entropy} (see also \cite{guth2020efficient}). In the paper we considered the case of clean (or homogeneous) systems but the generalisation to the case of  inhomogeneous dual-unitary circuits is straightforward.

An immediate question concerns the stability of our results when considering more general (non solvable) initial states and/or perturbations to dual-unitarity in the time evolution. Based on our results for generic solvable MPSs and on the analysis of state entanglement dynamics carried out in Ref.~\cite{bertini2019entanglement}, we expect our results to describe the asymptotic shape of the entanglement barriers after a quench from any initial state in dual-unitary circuits. In the case of of chaotic dual-unitary circuits this is very intuitive, because the asymptotic dynamics in a chaotic system should not be affected by the initial condition. Our results for the SWAP dynamics from general solvable MPSs, however, suggest that the initial-state dependence effectively drops out also in the integrable case if one does not fine-tune the initial condition. The case of perturbations of the dual-unitary condition is more difficult to assess because our results heavily rely on it. However, following Ref.~\cite{kos2021correlations}, it is reasonable to expect qualitative stability of our results, at least in the chaotic case. Indeed, generic perturbations are not expected to modify the chaotic character of the dynamics. This is confirmed by the recent findings of Ref.~\cite{kos2021correlations} concerning dynamical correlations at equilibrium, and should also apply to the entanglement dynamics given its inherent universality. A more thorough analysis of this question, however, is left to future work. 

Another interesting direction for future research concerns the study of (operator) entanglement dynamics in dual-unitary lattices of higher dimension. Indeed, since the dual-unitarity condition is directly generalised to higher dimension~\cite{bertini2019exact}, it offers the rare opportunity of an exact microscopic characterisation of the entanglement growth in this more realistic context.

\acknowledgments
We thank Katja Klobas, Toma{\v z} Prosen, Tianci Zhou, and Pavel Kos for useful comments on the manuscript and especially Lorenzo Piroli for key clarifications about approximability. This work has been supported by the Royal Society through the University Research Fellowship No.\ 201101.

\appendix 

\onecolumngrid

\section{Operator Space Entanglement and Approximability}
\label{app:approximability}

One of the central achievements of tensor network theory has been the establishment of precise connections between the behaviour of the R\'enyi entropies for the bipartitions of a given state $\ket{\Psi}$ and the approximability of that state by means of MPSs~\cite{verstraete2006matrix, schuch2008entropy} (see also~\cite{cirac2020matrix}). In particular, Ref.~\cite{verstraete2006matrix} proved that a state $\ket{\Psi}$ defined on a lattice of $N$ sites can be approximated by a carefully chosen MPS with bond dimension $\chi$, $\ket{\Psi_{\rm MPS}}$, in the following sense
\be
|\braket{\Psi|\Psi_{\rm MPS}}| \geq 1- \delta_{\chi ,N},
\label{eq:bound}
\ee
where 
\be
\delta_{\chi ,N} \leq N \left[\frac{\chi}{1-\alpha}\right]^{\frac{\alpha-1}{\alpha}} \exp\left[{\frac{\alpha-1}{\alpha} \bar S^{(\alpha)}}\right],\quad \alpha\in[0,1[,
\label{eq:deltasmaller}
\ee
and $\bar S^{(\alpha)}$ is the $\alpha$-th R\'enyi entropy $S^{(\alpha)}(A)$ maximised over all possible bipartitions $A\bar A= N$ of $\ket{\Psi}$. The bound \eqref{eq:bound} ensures that if a R\'enyi entropy with $\alpha\in[0,1[$ scales at most logarithmically with $N$ the state can be efficiently approximated by an MPS with polynomially growing bond dimension. Conversely, Ref.~\cite{schuch2008entropy} proved that 
\be
\log \chi  \geq  S^{(\alpha)}(A)+ \frac{\alpha}{\alpha-1} |\log(1-\delta_{\chi ,N})|,
\label{eq:deltalarger}
\ee
where $\alpha>1$. This means that if a R\'enyi entropy with $\alpha>1$ scales faster than logarithmically with $N$ then, to have a good MPS approximation, the bond dimension has to grow exponentially with $N$, making the MPS description redundant. The relations \eqref{eq:deltasmaller}--\eqref{eq:deltalarger} provide a quantitative way to connect the behaviour of R\'enyi entropies with approximability.  

In this appendix we want to understand whether the operator R\'enyi entropies give information on the approximability of $\rho_B(t)$ in expectation values of local operators. Namely on whether of not $\rho_B(t)$ can be approximated by a MPO in the expectation value of an arbitrary (bounded) observable $\mathcal O$. 

Let us begin by asking whether a bound like \eqref{eq:bound} for the operator state $\ket{\rho_B(t)}$ (cf.~\eqref{eq:operatortostate}) is sufficient to ensure this kind of approximability. To proceed let us introduce Shatten's $p$-norms~\cite{bhatia1997matrix}
\be
\label{eq:shattennorms}
\|A\|_p = ({\rm tr}((A^\dag A)^{p/2}))^{1/p}\,,\qquad p=1,2,\ldots\,,
\ee
where $\|A\|_\infty$ is the largest singular value of $A$ (for Hermitian operators is the absolute value of the eigenvalue with largest magnitude). Now, undoing the operator-to-state mapping \eqref{eq:operatortostate} we can write the bound \eqref{eq:bound} for the state $\ket{\rho_B(t)}$ as follows 
\be
\| \rho_B(t) -  \rho_{\rm MPO} \|_2 \leq 2 \delta_{\chi ,2\ell},
\label{eq:bound1}
\ee 
where $\ell=|B|$ and delta fulfils \eqref{eq:deltasmaller} and \eqref{eq:deltalarger} with $N=2\ell$. In this language our question becomes whether \eqref{eq:bound1} can be used to bound 
\be
\sup_{{\|{\cal O}\|_\infty = 1}} |{\rm tr}(\rho_B(t) \mathcal O)- {\rm tr}(\rho_{\rm MPO} \mathcal O)|\,,
\label{eq:goal}
\ee
from above or below. We begin by noting that the lower bound is easily obtained considering ${\cal O} = (\rho_B(t)-\rho_{\rm MPO})/\|\rho_B(t)-\rho_{\rm MPO}\|_\infty$ which gives 
\be
\sup_{{\|{\cal O}\|_\infty = 1}} |{\rm tr}(\rho_B(t) \mathcal O)- {\rm tr}(\rho_{\rm MPO} \mathcal O)|\geq  \frac{\| \rho_B(t) -  \rho_{\rm MPO} \|_2 ^2}{\|\rho_B(t)-\rho_{\rm MPO}\|_\infty}\geq \frac{1}{2} {\| \rho_B(t) -  \rho_{\rm MPO} \|_2 ^2}\,.
\label{eq:ineqRal1}
\ee
Using \eqref{eq:deltalarger} this implies that, if an operatorial R\'enyi entropy of $\rho_B(t)$ with $\alpha>1$ scales faster than logarithmically with $\ell$, \eqref{eq:goal} can be made arbitrary small only increasing exponentially the bond dimension. Namely, $\rho_B(t)$ is not efficiently approximable by an MPO. 

Unfortunately, instead, it is generically \emph{not} possible to use \eqref{eq:bound1} to bound \eqref{eq:goal} from above in the limit of infinite $B$. For instance, one might think to connect \eqref{eq:goal} to \eqref{eq:bound1} using the Cauchy-Schwartz inequality 
\be
|{\rm tr}(\rho_B(t) \mathcal O)- {\rm tr}(\rho_{\rm MPO} \mathcal O)|  \leq \|\mathcal O \|_2  \| \rho_B(t) -  \rho_{\rm MPO} \|_2 
\ee
but this bound becomes meaningless in the limit of infinite $\ell$ because $\|\mathcal O \|_2$ is generically unbounded (for example the identity operator gives $\|\1 \|_2=d^{2\ell}$). A more useful bound is the following~\cite{bhatia1997matrix} 
\be
\!\!\!\! |{\rm tr}(\rho_B(t) \mathcal O)- {\rm tr}(\rho_{\rm MPO} \mathcal O)|  \leq \|\mathcal O \|_\infty  \| \rho_B(t) -  \rho_{\rm MPO} \|_1.
\label{eq:usefulbound}
\ee
Since $\|\mathcal O \|_\infty$ is bounded in $\ell$ by assumption in \eqref{eq:goal}, the bound \eqref{eq:usefulbound} remains meaningful also in the limit of infinite $B$. In this limit, however, Eq.~\eqref{eq:bound1} does not give relevant information because 1- and 2-norms are not equivalent for general operators (take for example $A=\1_d^{\otimes \ell} /d^\ell$; in the limit $\ell\to\infty$ this operator has vanishing 2-norm while its 1-norm is one). Importantly, however, all $p$-norms are equivalent when considering pure states. In particular, one can easily show
\be
\|\ket{\Psi}\bra{\Psi}-\ket{\Phi}\bra{\Phi}\|_p = 2^{\frac{1-p}{p}} \|\ket{\Psi}\bra{\Psi}-\ket{\Phi}\bra{\Phi}\|_1\,, \qquad p=1,\ldots,
\ee
by using that $\ket{\Psi}\bra{\Psi}-\ket{\Phi}\bra{\Phi}$ lives in a two-dimensional subspace. 

This suggests a simple way to circumvent the problem by considering the \emph{square root} of the density operator. Indeed, expectation values can be written as  
\be
{\rm tr}(\rho_B(t) \mathcal O) = {\rm tr}( \ket{\sqrt{\rho_{B}(t)}}\!\bra{\sqrt{\rho_{B}(t)}}\cdot \mathcal O\otimes\1)
\label{eq:sqrtrhoEV}
\ee
where $\ket{\sqrt{\rho_{B}(t)}}$ is the state corresponding to ${\sqrt{\rho_{B}(t)}}$  under the mapping \eqref{eq:operatortostate}. Note that $\ket{\sqrt{\rho_{B}(t)}}$ is a \emph{purification}~\cite{nielsen2011quantum} of $\rho_{B}(t)$ in the following sense 
\be
\sum_{r_j\in\{0,\ldots,d-1\}}  \braket{d\boldsymbol s'+\boldsymbol r|\sqrt{\rho_{B}(t)}} \braket{\sqrt{\rho_{B}(t)}|d\boldsymbol s+\boldsymbol r} = \braket{\boldsymbol s'|\rho_B(t)|\boldsymbol s} .
\ee
Eq.~\eqref{eq:sqrtrhoEV} means that if we can find an MPS $\ket{\Psi_{\rm MPS}}$ such that  
\be
\| \ket{\sqrt{\rho_{B}(t)}}\!\bra{\sqrt{\rho_{B}(t)}} -   \ket{\Psi_{\rm MPS}}\!\bra{\Psi_{\rm MPS}}\|_2 \leq \delta,
\label{eq:squarerootbound}
\ee
then the bound \eqref{eq:usefulbound} gives 
\be
| {\rm tr}( \ket{\sqrt{\rho_{B}(t)}}\!\bra{\sqrt{\rho_{B}(t)}} \cdot \mathcal O\otimes\1)
- {\rm tr}( \ket{\Psi_{\rm MPS}}\!\bra{\Psi_{\rm MPS}} \cdot \mathcal O\otimes\1)|  \leq \sqrt 2 \|\mathcal O \|_\infty  \delta.
\ee
To find a bound like \eqref{eq:squarerootbound} we use \eqref{eq:bound} for $\ket{\rho_{B}(t)}$, which implies 
\be
\| \ket{\sqrt{\rho_{B}(t)}}\!\bra{\sqrt{\rho_{B}(t)}} -   \ket{\Psi_{\rm MPS}}\!\bra{\Psi_{\rm MPS}}\|_2 \leq 2 \delta_{1/2, \chi ,2\ell},  
\ee
where 
\be
\delta_{1/2, \chi ,2\ell} \leq N \left[\frac{\chi }{1-\alpha}\right]^{\frac{\alpha-1}{\alpha}} \exp\left[{\frac{\alpha-1}{\alpha} \bar S^{(\alpha)}_{1/2}}\right],\qquad \alpha\in[0,1]\,,
\ee
where $\bar S_{1/2}^{(\alpha)}$ is the $\alpha$-th R\'enyi entropy $S^{(\alpha)}(A, \sqrt{\rho_{A \bar A}(t)})$ (cf.~\eqref{eq:Renyisquareroot}) maximised over all bipartitions $A\bar A=N$ of $\ket{\sqrt{\rho_{B}(t)}}$. 
This means that if $S^{(\alpha)}(A, \sqrt{\rho_{A \bar A}(t)})$ scales at most logarithmically with $\ell$, $\rho_{B}(t)$ can be efficiently approximated by an MPO with polynomially growing bond dimension. 

Note that the cases discussed here do not exhaust all possibilities, in principle one can have all $S^{(\alpha)}(A, {\rho_{A \bar A}(t)})$ with $\alpha>1$ scaling at most logarithmically in $\ell$ and all $S^{(\alpha)}(A, \sqrt{\rho_{A \bar A}(t)})$ with $\alpha\in[0,1]$ scaling faster than logarithmically so that no conclusion can be drawn (for more details on the approximability problem for mixed states see Refs.~\cite{delascuevas2013, delascuevasfundamental2016, guth2020efficient}).

\section{Evaluation of Eq. \eqref{eq:SWAPgenstate} for generic unitary $W$}
\label{app:SWAP}

Our task is to evaluate $\textrm{Tr}(\tilde {\mathcal B}^n)$ for $\tilde {\mathcal B}$ of the form of Eq.~\eqref{eq:def_B_tilde}, reproduced here for convenience 
\begin{equation} 
\tilde {\mathcal B}=\frac{\chi^2}{d^{2(\ell-2t)}}
    \begin{tikzpicture}[baseline=(current  bounding  box.center), scale=0.5]

\foreach \i in {0,2.4,4.8}
{
\draw[thick,rounded corners=1pt] (0,-0.5-\i)--(1.5,-0.5-\i);
\draw[thick,rounded corners=1pt] (0,0.5-\i)--(0.5,0.5-\i)--(0.5,1-\i)--(-1,1-\i);
\draw[thick,rounded corners=1pt] (1.5,0.5-\i)--(1,0.5-\i)--(1,1-\i)--(2.5,1-\i);
}

\foreach \i in {0,-2.4,-4.8}
{
\path (0,\i) pic[rotate=-90,scale=1.5] {state_dagfolded};
\path (1.5,\i) pic[rotate=90,scale=1.5] {statefolded};
}

\path (-0.3,1.5) pic[rotate=0] {arrowhead=0.2};
\path (1.8,1.5) pic[rotate=0] {arrowhead=0.2};
\path (-0.3,1.5-7.8) pic[rotate=180] {arrowhead=0.2};
\path (1.8,1.5-7.8) pic[rotate=180] {arrowhead=0.2};

\draw [thick,decorate,decoration={brace}]
(2.75,1.5)--(2.75,-6.3);

\node[scale=1] at (4,-2.4) {$\ell-2t$};
\end{tikzpicture}.
\end{equation}
In particular, we will derive Eq. \eqref{eq:SWAPgenstate}, corresponding to the entanglement entropy $S^{(n)}_{m}(A,t)|_{\rm SWAP}$ in the late time regime, $\ell-x<2t<\ell$, and large $k\equiv \ell-2t\gg1$. 

Begin by noting that $\textrm{tr}(\tilde{\cal B}^n)$ can be written as the matrix element
\begin{equation}
    \textrm{tr}(\tilde{\cal B}^n)=\frac{\chi^{2n}}{d^{2n(\ell-2t)}} \braket{\Omega|\mathcal{M}_n^{\ell-2t}|\Omega} \,
\end{equation}
where we defined the tensor

\begin{equation}
\label{eq:Mmap}
\mathcal{M}_n=
    \begin{tikzpicture}[baseline=(current  bounding  box.center), scale=0.5]

\draw[thick,rounded corners=1pt] (7.5,0.5)--(7,0.5)--(7,1)--(8.5,1)--(8.5,2)--(-1,2)--(-1,1);

\foreach \i in {0,3}
{
\draw[thick,rounded corners=1pt] (1.5+\i,0.5)--(1+\i,0.5)--(1+\i,1)--(3.5+\i,1)--(3.5+\i,0.5)--(3+\i,0.5);
}

\draw[thick,rounded corners=1pt] (0,-0.5)--(1.5,-0.5);
\draw[thick,rounded corners=1pt] (3,-0.5)--(4.5,-0.5);
\draw[thick,rounded corners=1pt] (6,-0.5)--(7.5,-0.5);
\draw[thick,rounded corners=1pt] (0,0.5)--(0.5,0.5)--(0.5,1)--(-1,1);

\foreach \i in {0,3,6}
{
\path (\i,0) pic[rotate=-90,scale=1.5] {state_dagfolded};
\path (\i+1.5,0) pic[rotate=90,scale=1.5] {statefolded};
}

\draw [thick,decorate,decoration={brace}]
(8,-1.75)--(-0.5,-1.75);
\node[scale=1] at (3.75,-2.5) {$n$};

\end{tikzpicture}
\end{equation}
and the state 

\begin{equation}
    \ket{\Omega}=
    \begin{tikzpicture}[baseline=(current  bounding  box.center), scale=0.5]
\foreach \i in {0,...,5}
{
\draw[thick] (\i+-0.3,0)--(\i+-0.3,-2);
\path (\i+-0.3,-2) pic[rotate=0] {arrowhead=0.2};
}
\draw [thick,decorate,decoration={brace}]
(5.5,-2.25)--(-0.5,-2.25);
\node[scale=1] at (2.5,-3) {$2n$};
\end{tikzpicture} \, .
\end{equation}

Recalling the solvability condition \eqref{eq:stateuni1}--\eqref{eq:stateuni2} and the definition of the folded gates \eqref{eq:statefolded1}--\eqref{eq:statefolded2}, it is simple to see that
\begin{align}
&\qquad\begin{tikzpicture}[baseline=(current  bounding  box.center), scale=1]
\def\eps{0.5};
\draw[very thick] (-2,0) -- (.25,0);
\draw[thick, rounded corners=2pt] (-1.5,0) -- (-1,0.5) -- (-.75,0.5) -- (-1.5+1.25,0);
\draw[thick] (-1.5+1.25,0) -- (-1+1.25,0.5);
\draw[thick] (-2,0.5) -- (-1.5,0);
\draw[ thick, fill=mygreen, rounded corners=2pt] (-1.5-0.35,0.2-0.25) rectangle (-1.5+0.35,0.2+0.2);
\draw[thick] (-1.75,0.17) -- (-1.75,0.02)-- (-1.6,0.02) ;
%\draw[thick] (-1.5+0.1,0.15+.18)-- (-1.35+0.1,0.15+.18) -- (-1.35+0.1,0+.18);
\Text[x=-1.5,y=-.35]{}
\draw[ thick, fill=myorange, rounded corners=2pt] (-1.5-0.35+1.25,0.2-0.25) rectangle (-1.5+0.35+1.25,0.2+0.2);
\draw[thick] (-1.5+0.1+1.25,0.15+.18)-- (-1.35+0.1+1.25,0.15+.18) -- (-1.35+0.1+1.25,0+.18);
\Text[x=-1.5,y=-.35]{}
\end{tikzpicture}=\begin{tikzpicture}[baseline=(current  bounding  box.center), scale=1]
\def\eps{0.5};
\draw[very thick] (-2,0) -- (.25,0);
\draw[thick, rounded corners=2pt] (-2,0.5) -- (-1+1.25,0.5);
\Text[x=-1.5,y=-.35]{}
\end{tikzpicture}\,,\label{eq:stateuni3}\\
&\qquad\begin{tikzpicture}[baseline=(current  bounding  box.center), scale=1]
\def\eps{0.5};
\draw[very thick] (-2,0) -- (.25,0);
\draw[thick, rounded corners=2pt] (-1.5,0) -- (-1,0.5) -- (-.75,0.5) -- (-1.5+1.25,0);
\draw[thick] (-1.5+1.25,0) -- (-1+1.25,0.5);
\draw[thick] (-2,0.5) -- (-1.5,0);
\draw[ thick, fill=myorange, rounded corners=2pt] (-1.5-0.35,0.2-0.25) rectangle (-1.5+0.35,0.2+0.2);
\draw[thick] (-1.5+0.1,0.15+.18)-- (-1.35+0.1,0.15+.18) -- (-1.35+0.1,0+.18);
\Text[x=-1.5,y=-.35]{}
\draw[ thick, fill=mygreen, rounded corners=2pt] (-1.5-0.35+1.25,0.2-0.25) rectangle (-1.5+0.35+1.25,0.2+0.2);
%\draw[thick] (-1.5+0.1+1.25,0.15+.18)-- (-1.35+0.1+1.25,0.15+.18) -- (-1.35+0.1+1.25,0+.18);
\draw[thick] (-1.75+1.25,0.17) -- (-1.75+1.25,0.02)-- (-1.6+1.25,0.02) ;
\Text[x=-1.5,y=-.35]{}
\end{tikzpicture}=\begin{tikzpicture}[baseline=(current  bounding  box.center), scale=1]
\def\eps{0.5};
\draw[very thick] (-2,0) -- (.25,0);
\draw[thick, rounded corners=2pt] (-2,0.5) -- (-1+1.25,0.5);
\Text[x=-1.5,y=-.35]{}
\end{tikzpicture}\,.
\label{eq:stateuni4}
\end{align}
Using this property, we can easily find one left and one right eigenvector of $\mathcal{M}_n$. In particular, note that

\begin{equation}
\begin{tikzpicture}[baseline=(current  bounding  box.center), scale=0.5]
    \draw[thick,rounded corners=1pt] (7.5,0.5)--(7,0.5)--(7,1)--(8.5,1)--(8.5,2)--(-1,2)--(-1,1);

\foreach \i in {0,3}
{
\draw[thick,rounded corners=1pt] (1.5+\i,0.5)--(1+\i,0.5)--(1+\i,1)--(3.5+\i,1)--(3.5+\i,0.5)--(3+\i,0.5);
}

\draw[thick,rounded corners=1pt] (0,-0.5)--(1.5,-0.5);
\draw[thick,rounded corners=1pt] (3,-0.5)--(4.5,-0.5);
\draw[thick,rounded corners=1pt] (6,-0.5)--(7.5,-0.5);
\draw[thick,rounded corners=1pt] (0,0.5)--(0.5,0.5)--(0.5,1)--(-1,1);

\foreach \i in {0,3,6}
{
\path (\i,0) pic[rotate=-90,scale=1.5] {state_dagfolded};
\path (\i+1.5,0) pic[rotate=90,scale=1.5] {statefolded};
}

\foreach \i in {0,3,6}
{
\draw[thick, rounded corners=1pt] (\i+-0.3,-1.5)--(\i+-0.3,-2)--(\i+1.8,-2)--(\i+1.8,-1.5);
}
\end{tikzpicture}
=
\begin{tikzpicture}[baseline=(current  bounding  box.center), scale=0.5]

\draw[thick, rounded corners=2pt] (20.5,1) rectangle (12-1,2);

\foreach \i in {0,3,6}
{
\draw[thick, rounded corners=1pt] (12+\i+-0.3,0)--(12+\i+-0.3,-2)--(12+\i+1.8,-2)--(12+\i+1.8,0);
}
%\draw [thick,decorate,decoration={brace}]
%(8,-1.75)--(-0.5,-1.75);
%\node[scale=1] at (3.75,-2.5) {$m$};

\end{tikzpicture} \, .
\end{equation}
Therefore, 

\begin{equation}
    \ket{L}=
    \begin{tikzpicture}[baseline=(current  bounding  box.center), scale=0.5]
\foreach \i in {0,3,6}
{
\draw[thick, rounded corners=1pt] (12+\i+-0.3,0)--(12+\i+-0.3,-2)--(12+\i+1.8,-2)--(12+\i+1.8,0);
}
%\draw [thick,decorate,decoration={brace}]
%(8,-1.75)--(-0.5,-1.75);
%\node[scale=1] at (3.75,-2.5) {$m$};

\end{tikzpicture} \, ,
\end{equation}
with corresponding eigenvalue $\lambda=d^2$ (we have one closed loop and in folded notation the wires are doubled). Completely analogously, 

\begin{equation}
    \bra{R}= \,\,\,\,\,\,\, 
    \begin{tikzpicture}[baseline=(current  bounding  box.center), scale=0.5]
\draw[thick, rounded corners=1pt] (1.8,1.5)--(1.8,2.5)--(2.7,2.5)--(2.7,1.5);
\draw[thick, rounded corners=1pt] (4.8,1.5)--(4.8,2.5)--(5.7,2.5)--(5.7,1.5);
\draw[thick, rounded corners=1pt] (-0.3,1.5)--(-0.3,3)--(7.8,3)--(7.8,1.5);
\end{tikzpicture}\, ,
\end{equation}
which again satisfies $\lambda=d^2$. 

At this point we make the following conjecture based on extensive numerical inspection of small-$n$ cases
\begin{conjecture}
\label{app:conj}
The matrix $\mathcal M_n$ fulfils the following two properties
\begin{itemize}
\item[(i)] For all $W$
\be
\|\mathcal M_n\|_\infty = d^2\,.
\ee
\item[(ii)] For almost all $W$ 
\be
\|{\mathcal M}_n - d^2\frac{ \ket{R} \bra{L}}{\braket{L|R}} \|_\infty < d^2\,.
\ee
\end{itemize}
\end{conjecture}
As shown in the discussion at the end of the appendix, the first statement can be rephrased in terms of a trace inequality similar to the so called matrix Chebyshev inequalities~\cite{bourin2006matrix} (see also Ref.~\cite{cope2009trace}) which appears to be always fulfilled but for which we did not yet find a proof. The second statement is analogous to the discussions in Sec.~\ref{section:dual_unitary} and \ref{sec:solvableIS}. The idea is again that $\ket{L}$ and $\bra{R}$ the sole eigenvectors corresponding to $d^2$ which are protected by unitarity, hence are generically the only ones present. Numerically, we find that the point (ii) holds whenever $W$ entangles non-trivially the two qubits it acts on.

As a consequence of Conjecture~\ref{app:conj} we have that for almost all $W$ and for large enough $x$
\begin{equation}
   \mathcal{M}_n^x \simeq d^{2x} \frac{ \ket{R} \bra{L}}{\braket{L|R}}.
\end{equation}
It is simple to apply the usual loop-counting arguments to compute $\braket{L|R}=\chi^2$. Moreover, recalling the normalisation of the $\ket{\mcirc}$ state \eqref{eq:STMfixedpoint}, $\braket{\Omega|R}=1=\braket{L|\Omega}$. Therefore,
\begin{equation}
    \textrm{Tr}(\tilde{\cal B}^n)\simeq \frac{\chi^{2n}}{d^{2n(\ell-2t)}} \frac{d^{2(\ell-2t)}}{\chi^2}=\chi^{2(n-1)}d^{2(1-n)(\ell-2t)},
\end{equation}
whereupon
\begin{equation}
S^{(n)}_{m}(A, t)|_{\rm SWAP} \simeq (\ell-2t)\log d^2-\log \chi^2,
\end{equation}
as advertised.

\subsection{Simplified form of Conjecture~\ref{app:conj}}
To simplify the expression of Conjecture~\ref{app:conj} we note that $\cal M$ can be written as the tensor product of two matrices constructed with unfolded gates, i.e.
\be
{\cal M}_n = {\cal N}_n \otimes {\cal N}^*_n
\ee
with 
\begin{equation}
\label{eq:Nmap}
\mathcal{N}_n=
    \begin{tikzpicture}[baseline=(current  bounding  box.center), scale=0.5]

\draw[thick,rounded corners=1pt] (7.5,0.5)--(7,0.5)--(7,1)--(8.5,1)--(8.5,2)--(-1,2)--(-1,1);

\foreach \i in {0,3}
{
\draw[thick,rounded corners=1pt] (1.5+\i,0.5)--(1+\i,0.5)--(1+\i,1)--(3.5+\i,1)--(3.5+\i,0.5)--(3+\i,0.5);
}

\draw[thick,rounded corners=1pt] (0,-0.5)--(1.5,-0.5);
\draw[thick,rounded corners=1pt] (3,-0.5)--(4.5,-0.5);
\draw[thick,rounded corners=1pt] (6,-0.5)--(7.5,-0.5);
\draw[thick,rounded corners=1pt] (0,0.5)--(0.5,0.5)--(0.5,1)--(-1,1);

\foreach \i in {0,3,6}
{
\path (\i,0) pic[rotate=90,scale=1.5] {statenolegs};
\path (\i+1.5,0) pic[rotate=-90,scale=1.5] {state_dagnolegs};
}

\draw [thick,decorate,decoration={brace}]
(8,-1.75)--(-0.5,-1.75);
\node[scale=1] at (3.75,-2.5) {$n$};

\end{tikzpicture}.
\end{equation}
The expectation value of $\mathcal N_n$ on a state 
\be
\ket{a} = \begin{tikzpicture}[baseline=(current  bounding  box.center), scale=0.55]
\foreach \i in {0,1,2}{
\draw[thick] (-8.5+3*\i,-2) -- (-8.5+3*\i,-1.5);
\draw[thick] (-7.5+3*\i,-2) -- (-7.5+3*\i,-1.5);
}
\draw[thick, fill=black, rounded corners=2pt] (-8.75,-2.25) rectangle (-1.25,-2);
\end{tikzpicture}
\ee
can be represented in the following trace form by rotating each red square in \eqref{eq:Nmap} above the green square on its left    
\be
\braket{a|\mathcal N_n |a}={\rm tr}[(W^\dag)^{\otimes n} (a \otimes_{e} \Pi_{n}) W^{\otimes n} (a^\dag \otimes_e \1)]
\ee
where we introduced the operator corresponding to the `un-vectorisation' of $\ket{a}$
\be
a = \begin{tikzpicture}[baseline=(current  bounding  box.center), scale=0.55]
\foreach \i in {0,1,2}{
\draw[thick] (-8.5+3*\i,-2) -- (-8.5+3*\i,-1.5);
\draw[thick, line cap=round] (-7.5+3*\i,-2) .. controls (-8.35+3*\i,-2.5) .. (-8.5+3*\i,-3);
}
\draw[thick, fill=black, rounded corners=2pt] (-8.75,-2.25) rectangle (-1.25,-2);
\end{tikzpicture}
\ee
the periodic shift $\Pi_{n}$ in a chain of $n$ qudits, and the tensor product between even and odd sites $\otimes_e$, i.e. 
\be
(a\otimes b)\otimes_e( c\otimes d) = a\otimes c\otimes b\otimes d\,.
\ee
Now we note that a sufficient condition for the validity of Conjecture~\ref{app:conj} (i) is the following trace inequality
\be
|{\rm tr}[(W^\dag)^{\otimes n} (a \otimes_{e} \Pi_{n}) W^{\otimes n} (a^\dag \otimes_e \1)]| \leq {\rm tr}[a a^\dag \otimes_{e} \Pi_{n}] =  {\rm tr}[a a^\dag] {\rm tr}[\Pi_{n}]] =  {\rm tr}[a a^\dag]  d\,. 
\ee
We verified that this inequality is satisfied for $n=1,2,3,4$ and several thousands of $W$s sampled from the circular unitary ensemble of random matrices, but we did not yet find a proof.

\section{Overlaps}
\label{app:overlaps}
Here, we compute ${}_x\!\!\braket{y|\phi_{1}}_x$, with $\ket{y}_x$ an ORS from Eq.~\eqref{eq:ors} and $\ket{\phi_{1,x}}$ a trapezium of dual-unitary gates defined in Eq.~\eqref{eq:define_T}. 

Begin by assuming an initial Bell state ($\chi=1$). Our first task will be to evaluate ${}_x\!\!\braket{\tilde{y}|\phi_{1}}_x$, with $\ket{\tilde{y}}_x$ the non-orthonormal rainbow states defined Eq.~\eqref{eq:rainbow_state}. Consider $0\leq y \leq \ell-2t$. For these rainbow states the overlap takes the form
\begin{equation} \label{eq:overlap_1} {}_x\!\!\braket{\tilde y|\phi_{1}}_x=\frac{1}{d^{2(\ell-2t)+x-y}}
    \begin{tikzpicture}[baseline=(current  bounding  box.center), scale=0.5]

\foreach \i in {0,...,2}
{\path (2.8+1.4*\i,0) pic[rotate=45] {Ur_dagfolded};
\path (7+1.4*\i,0) pic[rotate=135] {Urfolded};
}
\foreach \i in {0,...,3}
{\path (1.4+1.4*\i,-1.4) pic[rotate=45] {Ur_dagfolded};
\path (7+1.4*\i,-1.4) pic[rotate=135] {Urfolded};
}

\draw[thick] (1.4,-0.7)--(2.1,0);
\draw[thick] (0.7,-1.4)--(0,-1.4)--(0,-2.1);
\draw[thick] (11.2,-0.7)--(10.5,0);
\draw[thick] (11.9,-1.4)--(12.6,-1.4)--(12.6,-2.1);

\foreach \i in {2,...,7}
{\path (1.4*\i,0.7) pic[rotate=0] {arrowhead};
}

\draw[dashed,thick,orange] (6.3,1.3)--(6.3,-2.4);
\draw[dashed,thick,orange] (0,1.3)--(0,-1.3);

\draw [thick,decorate,decoration={brace}]
(0,1.3)--(6.2,1.3);
\draw [thick,decorate,decoration={brace}]
(6.4,1.3)--(10.3,1.3);
\draw [thick,decorate,decoration={brace}]
(10.5,1)--(13,-1.5);

\node[scale=1] at (3.1,2) {$x$};
\node[scale=1] at (8.3,2) {$x-(\ell-2t)$};
\node[scale=1] at (12.9,0.5) {$\ell-2t$};

\foreach \i in {0,...,2}
{
\draw[thick] (5.6-1.4*\i,-2.1) parabola bend (6.3,-2.3-0.2*\i) (7+1.4*\i,-2.1);
}

%\foreach \i in {0,1,8,9}
%\node[scale=1] at (1.4*\i,-2.4) {?};
\end{tikzpicture} \, .
\end{equation}

We have stripped out the normalisation from each Bell pair and included it in the prefactor. How the remaining open wires are connected --- whether by a left-to-right loop to the appropriate site on the opposite side of the network or via a $\ket{\mcirc}$ wire into the lower plane --- depends on the value of $y$. The important feature is that, for $0\leq y \leq \ell-2t$, the central part of the rainbow state always consists of the same left-to-right loops.

Applying the dual-unitarity rules \eqref{eq:du1}--\eqref{eq:du2} as usual, this network simplifies to the form
\begin{equation} \label{eq:overlap_2}
    \begin{tikzpicture}[baseline=(current  bounding  box.center), scale=0.5]

\draw [thick,decorate,decoration={brace}]
(12.8,-3)--(11,-3);

\node[scale=1] at (11.7,-3.6) {$\ell-2t$};

\foreach \i in {0,...,2}
{
\draw[thick, rounded corners=1pt] (5.6-1.4*\i,-2.1) parabola bend (6.3,-2.3-0.2*\i) (7+1.4*\i,-2.1);
}

\draw[thick, rounded corners=1pt] (5.6,0.7)--(5.6,0.4)--(6,0)--(6.6,0)--(7,0.4)--(7,0.7);
\draw[thick, rounded corners=1pt] (4.2,0.7)--(4.2,0.4)--(4.6,0)--(5.2,0)--(5.6,-0.4)--(5.6,-1)--(6,-1.4)--(6.6,-1.4)--(7,-1)--(7,-0.4)--(7.4,0)--(8,0)--(8.4,0.4)--(8.4,0.7);
\draw[thick, rounded corners=1pt] 
(2.8,0.7)--(2.8,0.4)--(3.2,0)--(3.8,0)--(4.2,-0.4)--(4.2,-1)--(4.6,-1.4)--(5.2,-1.4)--(5.6,-1.8)--(5.6,-2.1);
\draw[thick, rounded corners=1pt]
(7,-2.1)--(7,-1.8)--(7.4,-1.4)--(8,-1.4)--(8.4,-1)--(8.4,-0.4)--(8.8,0)--(9.4,0)--(9.8,0.4)--(9.8,0.7);
\draw[thick,color=myred, rounded corners=1pt] (4.2,-2.1)--(4.2,-1.8)--(3.8,-1.4)--(3.2,-1.4)--(2.8,-1)--(2.8,-0.4)--(2.4,0)--(2.1,0)--(1.4,-0.7)--(1.4,-1)--(1,-1.4)--(0,-1.4)--(0,-2.1);
\draw[thick,color=myred, rounded corners=1pt] (2.8,-2.1)--(2.8,-1.8)--(2.4,-1.4)--(1.8,-1.4)--(1.4,-1.8)--(1.4,-2.2);
\draw[thick,color=myred, rounded corners=1pt] (8.4,-2.1)--(8.4,-1.8)--(8.8,-1.4)--(9.4,-1.4)--(9.8,-1)--(9.8,-0.4)--(10.2,0)--(10.5,0)--(11.2,-0.7)--(11.2,-1)--(11.6,-1.4)--(12.6,-1.4)--(12.6,-2.1);
\draw[thick,color=myred, rounded corners=1pt]
(9.8,-2.1)--(9.8,-1.8)--(10.2,-1.4)--(10.8,-1.4)--(11.2,-1.8)--(11.2,-2.1);

%\foreach \i in {0,1,8,9}
%\node[scale=1] at (1.4*\i,-2.4) {?};

\foreach \i in {2,...,7}
{\path (1.4*\i,0.7) pic[rotate=0] {arrowhead};
}
\end{tikzpicture} \, .
\end{equation}
Once again, the black loops do not contribute because $\ket{\mcirc}$ is normalised. First consider $y=0$. This manifestly gives $\ell-2t$ loops. On the other hand, for $y=1$ there are $\ell-2t-1$ loops because the outermost wire is contracted with $\ket{\mcirc}$. Extending this logic to general $y$ we find
\begin{equation}
    {}_x\!\!\braket{\tilde{y}|\phi_{1}}_x = d^{-x-y} \,\,\,\,\,\,\,  \text{if} \,\,\,\,\,\,\, y < \ell-2t. 
\end{equation}

Another important observation of Eq. \eqref{eq:overlap_2} is that, since the black loops simply give a factor of 1, the number of them does not matter. This means that a network of the same form as Eq. \eqref{eq:overlap_1} but with a bigger or smaller width gives the same result. 

Now consider $\ell-2t < y \leq x$.  The outer part of the each rainbow state consists of $\ket{\mcirc}$ states, so our network takes the form

\begin{equation} \label{eq:overlap3} {}_x\!\!\braket{\tilde y|\phi_{1}}_x=\frac{1}{d^{2(\ell-2t)+x-y}}
\begin{tikzpicture}[baseline=(current  bounding  box.center), scale=0.5]
\foreach \i in {0,...,2}
{\path (2.8+1.4*\i,0) pic[rotate=45] {Ur_dagfolded};
\path (7+1.4*\i,0) pic[rotate=135] {Urfolded};
}
\foreach \i in {0,...,3}
{\path (1.4+1.4*\i,-1.4) pic[rotate=45] {Ur_dagfolded};
\path (7+1.4*\i,-1.4) pic[rotate=135] {Urfolded};
}
\draw[thick] (1.4,-0.7)--(2.1,0);
\draw[thick] (0.7,-1.4)--(0,-1.4)--(0,-2.1);
\draw[thick] (11.2,-0.7)--(10.5,0);
\draw[thick] (11.9,-1.4)--(12.6,-1.4)--(12.6,-2.1);
\foreach \i in {2,...,7}
{\path (1.4*\i,0.7) pic[rotate=0] {arrowhead};
}
\draw[dashed,thick,orange] (6.3,1.3)--(6.3,-2.4);
\draw[dashed,thick,orange] (0,1.3)--(0,-1.3);
\draw [thick,decorate,decoration={brace}]
(0,1.3)--(6.2,1.3);
\draw [thick,decorate,decoration={brace}]
(6.4,1.3)--(10.3,1.3);
\draw [thick,decorate,decoration={brace}]
(10.5,1)--(13,-1.5);
\node[scale=1] at (3.1,2) {$x$};
\node[scale=1] at (8.3,2) {$x-(\ell-2t)$};
\node[scale=1] at (12.9,0.5) {$\ell-2t$};
%\foreach \i in {2,...,7}
%\node[scale=1] at (1.4*\i,-2.4) {?};
\foreach \i in {0,1,8,9}
{\path (1.4*\i,-2.1) pic {arrowhead};}
\end{tikzpicture}\,.
\end{equation}
Once again, how the open wires are connected depends on the particular rainbow state considered. Using dual-unitarity,

\begin{equation}
    \begin{tikzpicture}[baseline=(current  bounding  box.center), scale=0.5]

\foreach \i in {0,...,2}
{\path (2.8+1.4*\i,0) pic[rotate=45] {Ur_dagfolded};
\path (7+1.4*\i,0) pic[rotate=135] {Urfolded};
}
\foreach \i in {1,...,3}
{\path (1.4+1.4*\i,-1.4) pic[rotate=45] {Ur_dagfolded};
\path (5.6+1.4*\i,-1.4) pic[rotate=135] {Urfolded};
}

\foreach \i in {2,...,7}
{\path (1.4*\i,0.7) pic[rotate=0] {arrowhead};
}

\draw [thick,decorate,decoration={brace}]
(2.3,1.3)--(10.3,1.3);
\node[scale=1] at (6.3,2) {$2x-2(\ell-2t)$};

\draw[thick,color=myred] (2.4,0)--(2.1,0)--(1.4,-0.7)--(1.4,-1)--(1,-1.4)--(0,-1.4)--(0,-2.1);
\draw[thick,color=myred] (2.4,-1.4)--(1.8,-1.4)--(1.4,-1.8)--(1.4,-2.2);
\draw[thick,color=myred] (10.2,0)--(10.5,0)--(11.2,-0.7)--(11.2,-1)--(11.6,-1.4)--(12.6,-1.4)--(12.6,-2.1);
\draw[thick,color=myred]
(10.2,-1.4)--(10.8,-1.4)--(11.2,-1.8)--(11.2,-2.1);

\foreach \i in {0,1,8,9}
{\path (1.4*\i,-2.1) pic {arrowhead};}

%\foreach \i in {2,...,7}
%\node[scale=1] at (1.4*\i,-2.4) {?};
\end{tikzpicture} \, .
\end{equation}
Now we make the important observation that
\begin{equation} \label{eq:overlap_5} \begin{gathered}
    \begin{tikzpicture}[baseline=(current  bounding  box.center), scale=0.5]

\foreach \i in {0,...,2}
{\path (2.8+1.4*\i,0) pic[rotate=45] {Ur_dagfolded};
\path (7+1.4*\i,0) pic[rotate=135] {Urfolded};
}
\foreach \i in {1,...,3}
{\path (1.4+1.4*\i,-1.4) pic[rotate=45] {Ur_dagfolded};
\path (5.6+1.4*\i,-1.4) pic[rotate=135] {Urfolded};
}

\draw [thick,decorate,decoration={brace}]
(2.3,1.3)--(10.3,1.3);
\node[scale=1] at (6.3,2) {$2x-2(\ell-2t)$};

\draw[thick,color=myred] (2.4,0)--(2.1,0)--(1.4,-0.7)--(1.4,-1)--(1,-1.4)--(0,-1.4)--(0,-2.1);
\draw[thick,color=myred] (2.4,-1.4)--(1.8,-1.4)--(1.4,-1.8)--(1.4,-2.2);
\draw[thick,color=myred] (10.2,0)--(10.5,0)--(11.2,-0.7)--(11.2,-1)--(11.6,-1.4)--(12.6,-1.4)--(12.6,-2.1);
\draw[thick,color=myred]
(10.2,-1.4)--(10.8,-1.4)--(11.2,-1.8)--(11.2,-2.1);

\foreach \i in {2,...,7}
{\path (1.4*\i,0.7) pic[rotate=0] {arrowhead};
}

%\foreach \i in {2,...,7}
%\node[scale=1] at (1.4*\i,-2.4) {?};

\foreach \i in {0,1,2,7,8,9}
{\path (1.4*\i,-2.1) pic {arrowhead};}

\end{tikzpicture} \\ \\ \, = \, 
\begin{tikzpicture}[baseline=(current  bounding  box.center), scale=0.5]

\foreach \i in {1,...,2}
{\path (2.8+1.4*\i,0) pic[rotate=45] {Ur_dagfolded};
\path (5.6+1.4*\i,0) pic[rotate=135] {Urfolded};
}
\foreach \i in {1,...,2}
{\path (2.8+1.4*\i,-1.4) pic[rotate=45] {Ur_dagfolded};
\path (5.6+1.4*\i,-1.4) pic[rotate=135] {Urfolded};
}

\draw[thick,color=myred] (2.8,0.7)--(2.8,0.4)--(2.4,0)--(2.1,0)--(1.4,-0.7)--(1.4,-1)--(1,-1.4)--(0,-1.4)--(0,-2.1);
\draw[thick,color=myred] (3.8,0)--(3.2,0)--(2.8,-0.4)--(2.8,-1)--(2.4,-1.4)--(1.8,-1.4)--(1.4,-1.8)--(1.4,-2.2);
\draw[thick,color=myred] (9.8,0.7)--(9.8,0.4)--(10.2,0)--(10.5,0)--(11.2,-0.7)--(11.2,-1)--(11.6,-1.4)--(12.6,-1.4)--(12.6,-2.1);
\draw[thick,color=myred]
(8.8,0)--(9.4,0)--(9.8,-0.4)--(9.8,-1)--(10.2,-1.4)--(10.8,-1.4)--(11.2,-1.8)--(11.2,-2.1);
\draw[thick,color=myred] (2.8,-2.1)--(2.8,-1.8)--(3.2,-1.4)--(3.8,-1.4);
\draw[thick,color=myred]
(8.8,-1.4)--(9.4,-1.4)--(9.8,-1.8)--(9.8,-2.1);

\foreach \i in {2,...,7}
{\path (1.4*\i,0.7) pic[rotate=0] {arrowhead};
}

\foreach \i in {0,1,2,7,8,9}
{\path (1.4*\i,-2.1) pic {arrowhead};}

%\foreach \i in {3,...,6}
%\node[scale=1] at (1.4*\i,-2.4) {?};

\end{tikzpicture} \, = \, 
\begin{tikzpicture}[baseline=(current  bounding  box.center), scale=0.5]

\foreach \i in {1,...,2}
{\path (2.8+1.4*\i,0) pic[rotate=45] {Ur_dagfolded};
\path (5.6+1.4*\i,0) pic[rotate=135] {Urfolded};
}
\foreach \i in {1,...,2}
{\path (2.8+1.4*\i,-1.4) pic[rotate=45] {Ur_dagfolded};
\path (5.6+1.4*\i,-1.4) pic[rotate=135] {Urfolded};
}

\draw[thick,color=red] (3.8,0)--(3.2,0)--(2.8,-0.4)--(2.8,-1)--(2.4,-1.4)--(1.8,-1.4)--(1.4,-1.8)--(1.4,-2.2);
\draw[thick,color=red]
(8.8,0)--(9.4,0)--(9.8,-0.4)--(9.8,-1)--(10.2,-1.4)--(10.8,-1.4)--(11.2,-1.8)--(11.2,-2.1);
\draw[thick,color=red] (2.8,-2.1)--(2.8,-1.8)--(3.2,-1.4)--(3.8,-1.4);
\draw[thick,color=red]
(8.8,-1.4)--(9.4,-1.4)--(9.8,-1.8)--(9.8,-2.1);

%\foreach \i in {3,...,6}
%\node[scale=1] at (1.4*\i,-2.4) {?};

\foreach \i in {3,...,6}
{\path (1.4*\i,0.7) pic[rotate=0] {arrowhead};
}

\foreach \i in {1,2,7,8}
{\path (1.4*\i,-2.1) pic {arrowhead};}
\end{tikzpicture} \, .
\end{gathered}
\end{equation}

As we noted above, if all the open wires in the final network of Eq. \eqref{eq:overlap_5} are connected by left-to-right lines, the result does not depend on the width of the central rectangular region. This is because black loops in Eq. \eqref{eq:overlap_2} make no contribution. It follows that ${}_x\!\!\braket{\tilde{y}|\phi_{1}}_x=d\,{}_x\!\!\braket{\widetilde{y+1}|\phi_{1}}_x$ for $\ell-2t < i \leq x$. Indeed, including the appropriate normalisation
\begin{equation}
   {}_x\!\!\braket{\tilde{y}|\phi_{1}}_x = d^{-2(\ell-2t)-x+y} \,\,\,\,\,\,\,  \text{if} \,\,\,\,\,\,\, \ell-2t < y \leq x. 
\end{equation}
To emphasise
\begin{equation}
   {}_x\!\!\braket{\tilde{y}|\phi_{1}}_x=
    \begin{cases}
      d^{-x-y} , & \text{if}\ y \leq \ell-2t \\
      d^{-2(\ell-2t)-x+y} , & \text{if}\ \ell-2t < y \leq x .  

    \end{cases}
  \end{equation}

Recalling our definition of the orthonormalised ORSs in Eq. \eqref{eq:ors}, we quickly arrive at Eq. \eqref{eq:ors_cont}, repeated here for convenience.

\begin{equation} \label{eq:appors_cont}
   {}_x\!\!\braket{y|\phi_{1}}_x = 
    \begin{cases}
      d^{-x-y-1} \sqrt{d^2-1}, & \text{if}\ y< \ell-2t\\
      0, & \text{if}\ \ell-2t\leq y<x\\
      d^{-2(\ell-2t)} , & \text{if}\ y=x.\\
    \end{cases}
  \end{equation}

We can generalise this discussion to consider more general initial states \eqref{eq:psi0}. Considering the leading contribution in the limit $\ell-2t\equiv k\to \infty$ we find 
\begin{equation} \ket{\phi_{1}}_x\simeq\frac{\chi^2}{d^{2k}}
    \begin{tikzpicture}[baseline=(current  bounding  box.center), scale=0.5]

\foreach \i in {-1,...,4}
\draw[thick] (1.4+1.4*\i,-1.3)--(1.4+1.4*\i,-2.5);

\foreach \i in {0}
{\path (2.8+1.4*\i,0) pic[rotate=45] {Ur_dagfolded};
\path (4.2+1.4*\i,0) pic[rotate=135] {Urfolded};
}
\foreach \i in {0,...,1}
{\path (1.4+1.4*\i,-1.4) pic[rotate=45] {Ur_dagfolded};
\path (4.2+1.4*\i,-1.4) pic[rotate=135] {Urfolded};
}

\draw[thick] (1.4,-0.7)--(2.1,0);
\draw[thick] (0.7,-1.4)--(0,-1.4)--(0,-2.1);
\draw[thick] (5.6,-0.7)--(4.9,0);
\draw[thick] (6.3,-1.4)--(7,-1.4)--(7,-2.1);

\foreach \i in {2,...,3}
{\path (1.4*\i,0.7) pic[rotate=0] {arrowhead=0.2};
}

\draw[dashed,thick,orange] (3.5,1.3)--(3.5,-2.4);
\draw[dashed,thick,orange] (0,1.3)--(0,-1.3);

\draw [thick,decorate,decoration={brace}]
(0,1.3)--(3.4,1.3);
\draw [thick,decorate,decoration={brace}]
(3.6,1.3)--(5,1.3);
\draw [thick,decorate,decoration={brace}]
(5.6,1)--(8.1,-1.5);

\node[scale=1] at (1.75,2) {$x$};
\node[scale=1] at (4.4,2) {$x-k$};
\node[scale=1] at (8,0.5) {$k$};

\def\epsx{5.5}
\def\epsy{-0.2}
\path (\epsx+0,\epsy+0) pic[rotate=135,scale=1.25] {statefolded};
\path (\epsx+1.4,\epsy-1.4) pic[rotate=135,scale=1.25] {statefolded};
\path (\epsx-8.25+10.65,\epsy+0.15+-2.2) pic[rotate=180] {arrowhead=0.2};
\path (\epsx-8.25++7.6,\epsy+0.15+0.85) pic[rotate=180] {arrowhead=0.2};

\def\epsx{2.8}
\def\epsy{0.05}
\path (\epsx+-2.6,\epsy+-1.6) pic[rotate=225,scale=1.25] {state_dagfolded};
\path (\epsx+-1.2,\epsy+-0.2) pic[rotate=225,scale=1.25] {state_dagfolded};
\path (-0.5+\epsx,.85+\epsy) pic[rotate=180] {arrowhead=0.2};
\path (-0.5-3.1+\epsx,.85-3.1+\epsy) pic[rotate=180] {arrowhead=0.2};

\end{tikzpicture}\,.
\end{equation}
Proceeding in close analogy with the arguments above for the Bell state but now also exploiting unitarity of $W$, we deduce that

\begin{equation}
   {}_x\!\!\braket{\tilde{y}|\phi_{1}}_x \simeq
    \begin{cases}
      d^{-x-y} \textrm{tr}(\tau^{k-y}) , & \text{if}\ y \leq k \\
      \chi^2 d^{-2k-x+y} , & \text{if}\ k< y \leq x, 
    \end{cases}
  \end{equation}
with $\tau$ the state transfer matrix (STM) defined in Eq. \eqref{eq:STM}. It follows from Eq. \eqref{eq:ors} that
\begin{align} \label{eq:apporscontgen}
   {}_x\!\!\braket{y|\phi_{1}}_x &\simeq 
    \begin{cases}
      \frac{1}{\sqrt{d^2-1}}d^{-x-y-1}[d^2\textrm{tr}(\tau^{k-y})-\textrm{tr}(\tau^{k-y-1})], & \text{if}\ y< k\\
      0, & \text{if}\ k\leq y<x\\
      \chi^2 d^{-2k } , & \text{if}\ y=x\\
    \end{cases}\\
    &\simeq 
    \begin{cases}
      d^{-x-y-1}\sqrt{d^2-1}  ,\quad\qquad\qquad\qquad\qquad\qquad & \text{if}\ y\ll k\\
      \chi^2 d^{-2k } , & \text{if}\ y=x\\
    \end{cases},
  \end{align}
where in the second step we used \eqref{eq:infinitepowertau}.

\twocolumngrid 

\bibliography{./bibliography}

\end{document}